%% file: main_combined.tex
\pdfoutput=1
\documentclass[
aps, 
prd, 
notitlepage, 
twocolumn,
superscriptaddress,
floatfix,
letterpaper,
nofootinbib,
longbibliography]{revtex4-2}

\usepackage{graphicx} 
\graphicspath{ {fig/} }

\usepackage{
amsmath, 
amsthm, 
amssymb,
slashed,
mathtools,
tabu
}
\usepackage{pifont}

\usepackage[svgnames, dvipsnames]{xcolor} 
\usepackage[colorlinks,citecolor=RoyalBlue, urlcolor=RoyalBlue, linkcolor=RoyalBlue]{hyperref}

\usepackage{babel}
\usepackage{array}
\usepackage{longtable}
\usepackage{float}
\usepackage{mathtools}

\usepackage{dcolumn}
\usepackage{bm}
\usepackage{comment}
\usepackage{float}
\usepackage{soul}
\usepackage{cancel}

\newcommand{\mathsout}[1]
{\bgroup\mathchoice
  {\sbox0{$\displaystyle{#1}$}%
    \usebox0\hspace{-\wd0}%
    \rule[0.5\ht0-0.5\dp0-.5pt]{\wd0}{1pt}}%
  {\sbox0{$\textstyle{#1}$}%
    \usebox0\hspace{-\wd0}%
    \rule[0.5\ht0-0.5\dp0-.5pt]{\wd0}{1pt}}%
  {\sbox0{$\scriptstyle{#1}$}%
    \usebox0\hspace{-\wd0}%
    \rule[0.5\ht0-0.5\dp0-.5pt]{\wd0}{1pt}}%
  {\sbox0{$\scriptscriptstyle{#1}$}%
    \usebox0\hspace{-\wd0}%
    \rule[0.5\ht0-0.5\dp0-.5pt]{\wd0}{1pt}}%
\egroup}

\newcommand{\CJ}[1]{{\color{green}{CJ: #1}}}

\newcommand{\xCornell}{Department of Physics, Cornell University, Ithaca, NY, 14850, USA}
\newcommand{\xEwha}{Department of Physics, Ewha Womans University, Seoul, South Korea}

\newcommand{\para}{}
\let\para\relax

\newcommand{\ii}{\hspace{1pt}\mathrm{i}\hspace{1pt}}

\newcommand{\bea}{\begin{eqnarray}}
\newcommand{\eea}{\end{eqnarray}}
\def\be{\begin{equation}}
\def\ee{\end{equation}}

\newcommand{\re}{\mathrm{e}}

\usepackage[export]{adjustbox}

\definecolor{col1}{HTML}{0f4d92}
\definecolor{col2}{HTML}{7c253a}
\input{zkm_math_common}



\usepackage{tikz}
\usetikzlibrary{calc} 
\usetikzlibrary{positioning}
\usetikzlibrary{shapes.geometric}
\usetikzlibrary{quantikz}

\input{util-macros}

\usepackage{multirow}

\makeatletter
\renewcommand\paragraph{\@startsection{paragraph}{4}{\z@}%
    {-2.5ex \@plus -1ex \@minus -.2ex}%
    {0.25ex \@plus .2ex}%
    {\normalfont\normalsize\bfseries\raggedright}}
\makeatother

\usepackage{
amsmath, 
amsthm, 
amssymb,
slashed,
mathtools,
tabu
}
\usepackage{pifont}

\usepackage[svgnames, dvipsnames]{xcolor} 
\usepackage[colorlinks,citecolor=RoyalBlue, urlcolor=RoyalBlue, linkcolor=RoyalBlue]{hyperref}

\usepackage{babel}
\usepackage{array}
\usepackage{longtable}
\usepackage{float}
\usepackage{mathtools}

\usepackage{dcolumn}
\usepackage{bm}
\usepackage{comment}
\usepackage{float}
\usepackage{soul}
\usepackage{cancel}

\usepackage[toc,page,titletoc]{appendix}
\usepackage[capitalise,nameinlink]{cleveref}

\def\be{\begin{equation}}
\def\ee{\end{equation}}

\newcommand{\D}{\mathcal{D}}

\newcommand{\1}{\mathbf{1}}

\begin{document}
\author{Zlatko~K.~Minev}
\affiliation{IBM Quantum, T.J. Watson Research Center, Yorktown Heights, NY 10598 USA}
\author{Khadijeh~Najafi$^{\dag\dag}$}
\affiliation{IBM Quantum, T.J. Watson Research Center, Yorktown Heights, NY 10598 USA}
\affiliation{MIT-IBM Watson AI Lab,  Cambridge MA, 02142, USA}
\author{Swarnadeep~Majumder$^{\dag\dag}$}
\affiliation{IBM Quantum, T.J. Watson Research Center, Yorktown Heights, NY 10598 USA} 
\author{Juven~Wang}
\affiliation{Center of Mathematical Sciences and Applications, Harvard University, MA 02138, USA}
\affiliation{London Institute for Mathematical Sciences, Royal Institution, W1S 4BS, UK}

\author{Ady~Stern}
\affiliation{Department of Condensed Matter Physics, Weizmann Institute of Science, Rehovot 76100, Israel}
\author{Eun-Ah~Kim}
\email{eun-ah.kim@cornell.edu}
\affiliation{\xCornell}
\affiliation{\xEwha}
\author{Chao-Ming Jian}
\email{chao-ming.jian@cornell.edu}
\affiliation{\xCornell}
\author{Guanyu~Zhu}
\affiliation{IBM Quantum, T.J. Watson Research Center, Yorktown Heights, NY 10598 USA}

\def\thefootnote{$\dagger\dagger$}\footnotetext{These authors contributed equally to this work.}

\title{
Realizing string-net condensation:\\ Fibonacci anyon braiding for universal gates and sampling chromatic polynomials
}

\begin{abstract}
{The remarkable complexity of the vacuum state of a topologically-ordered many-body quantum system 
encodes the character and intricate braiding interactions of its emergent particles, the anyons.}
Quintessential predictions exploiting this complexity use the Fibonacci string-net condensate (Fib-SNC) and its Fibonacci anyons to go beyond classical computing. 
Sampling the Fib-SNC wavefunction is expected to yield estimates of the chromatic polynomial of graph objects, a classical task that is provably hard. 
At the same time, exchanging anyons of Fib-SNC is expected to allow fault-tolerant universal quantum computation. 
\pdfoutput=1
Nevertheless, the physical realization of Fib-SNC and its anyons remains elusive. Here, we introduce a scalable dynamical string-net preparation (DSNP) approach, suitable even for near-term quantum processors, which dynamically prepares Fib-SNC and its anyons through reconfigurable graphs. Using a superconducting quantum processor, we couple the DSNP approach with composite error-mitigation on deep circuits to successfully create,  measure, and braid anyons of Fib-SNC in a scalable manner.  We certify the creation of anyons by measuring their `anyon charge', finding an average experimental accuracy of $94\%$. 
Furthermore, we validate that exchanging these anyons yields the { expected} golden ratio~$\phi$ with~$98\%$ average accuracy and~$8\%$ measurement uncertainty.
Finally, we sample the Fib-SNC to estimate the chromatic polynomial at~$\phi+2$ for {several} graphs.  
Our results establish the proof of principle for using Fib-SNC and its anyons for fault-tolerant universal quantum computation and {for aiming at} a classically-hard problem.
\end{abstract}

\maketitle

\para 
In principle, complex quantum many-body states can efficiently encode solutions to provably hard computational problems. However, protocols to generate such states, coveted even for intermediate-scale systems, without relying on random gates remain elusive~\cite{Boixo2018NaturePhys}. Surprisingly, a state of topological quantum matter considered for fault-tolerant universal quantum computation could also harbor such solutions.
Specifically, a fascinating connection between a string-net condensate wave function supporting Fibonacci anyons~\cite{Levin2005Phys.Rev.B,Levin2005Rev.Mod.Phys.} and a classically hard problem of evaluating chromatic polynomials has been noted~\cite{Fidkowski2009Commun.Math.Phys., Fendley2005Phys.Rev.B, Fendley2009Geom.Topol., Fendley2010Adv.Theor.Math.Phys.}.

\begin{figure*}[h!]
    \centering
    \includegraphics[width=2\columnwidth]{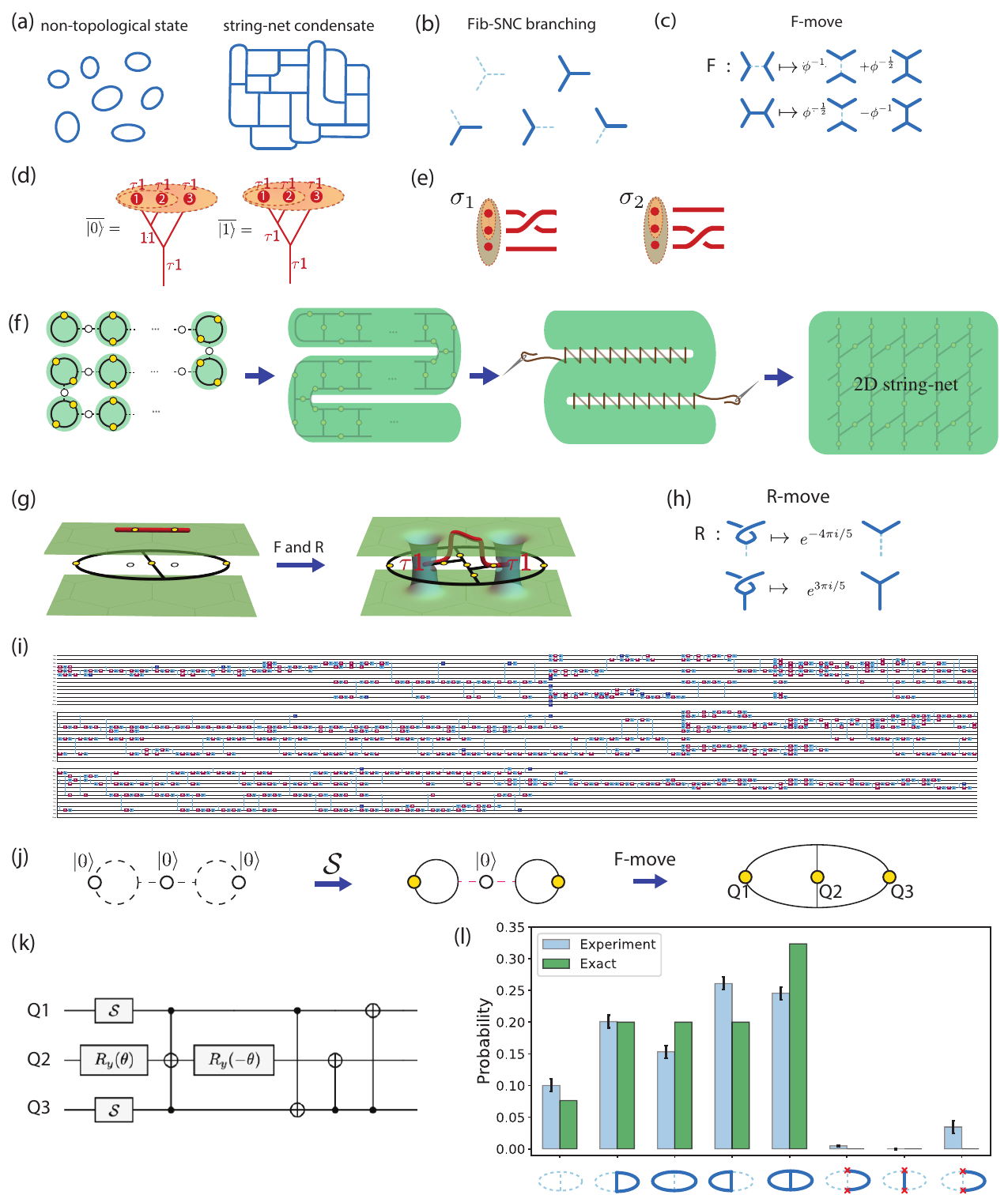}
    \caption{(next page)} 
\label{fig:1}
\end{figure*}

\para
String-net condensates (see Fig.~\ref{fig:1}a) are many-body vacuum states, encompassing essentially all parity- and time-reversal invariant topological phases \cite{Levin2005Phys.Rev.B}.
A string-net condensate is a complex superposition of `string nets', which can be visualized as trivalent graphs representing spins in excited state~$\ket{1}$ or ground state~$\ket{0}$ and subject to local geometric rules. 
The simplest condensate whose geometric rules allow a string to `branch' into two strings is the Fibonacci string-net condensate (Fib-SNC) (Fig. \ref{fig:1}b) \cite{Levin2005Phys.Rev.B,Levin2005Rev.Mod.Phys.}.
This simple `branching rule' for the Fib-SNC nevertheless leads to a remarkably complex state when combined with `$F$-move' rules (see Fig.~\ref{fig:1}c), which mandate intricate relationships among the superposition amplitudes, dictated with the golden ratio~$\phi$.
In this state, the modulus-squared amplitude of a string-net is determined by \cite{Fidkowski2009Commun.Math.Phys., Fendley2005Phys.Rev.B, Fendley2009Geom.Topol., Fendley2010Adv.Theor.Math.Phys.} the chromatic polynomial \cite{Read1968JournalofCombinatorialTheory} of its dual graph evaluated at $\phi+2$. 
Evaluating the chromatic polynomial is generally $\#$P-hard\footnote{For counting problems, $\#$P is the analog of the more familiar class NP for decision problems.} and even its approximation is computationally hard  \cite{Jaeger1990Math.Proc.Camb.Philos.Soc., Vertigan2005SIAMJ.Comput., Golderb:2008908, Goldberg:2012uh, Goldberg2014SIAMJ.Comput.}. 
Hence, realizing the Fib-SNC could open a door to a new class of classically-hard problems.

\para 
Another complexity of Fib-SNC is its role as a vacuum state supporting emergent anyons, capable of fault-tolerant universal quantum computation~\cite{Levin2005Phys.Rev.B,Levin2005Rev.Mod.Phys.}. 
The underlying universal nature of these anyons is captured by a long-wavelength effective theory treatment \cite{Levin2005Phys.Rev.B}, which combines two time-reversed copies of a topological quantum field theory (TQFT)\footnote{The level-1 Chern-Simons theory with the exceptional gauge group ${\rm G}_2$, or equivalently the integer-spin sector of the ${\rm SU}(2)$ Chern-Simons theory at level 3 \cite{Witten:1988hf, Bonderson2008,RevModPhys.80.1083}.}.
In principle, even a single copy allows for universal quantum computation \cite{Bonesteel2005Phys.Rev.Lett.}, but no microscopic blueprint exists for manifesting just a single copy. 
Each copy resembles the TQFT proposed for filling-factor $12/5$ quantum Hall state~\cite{Read1999Phys.Rev.B} and supports an anyon type $\tau$, whose multi-anyon Hilbert space dimension follows the Fibonacci sequence.
This arises from a `fusion rule' (see Fig.~\ref{fig:1}d) where two $\tau$ anyons brought together can either annihilate (resulting in ${\bf 1}$) or fuse into a single $\tau$. 
The `doubling' leads to  three anyons types, $\tau{\bf1}$, ${\bf1}\tau$, and $\tau\tau$~\cite{Koenig2010AnnalsofPhysics}, with  Fib-SNC providing a microscopic blueprint. 
A triplet of any of these anyons can encode a logical qubit, capitalizing on the two possible fusion outcomes of the $\tau$. For instance, 
 realizing Fib-SNC and creating two pairs of $\tau{\bf 1}$ allows for encoding logical information to the fusion outcome of first two $\tau{\bf 1}$ anyons
 (see Fig.~\ref{fig:1}d)~\cite{Schotte2022Phys.Rev.X}. 
Exchanging the anyons,  braiding their space-time trajectories, enacts non-Clifford logical gates, a primary requisite for universal quantum computation (see Fig.~\ref{fig:1}e).

\para
Unfortunately, faithfully realizing the Fib-SNC and its anyons has been unreachable,
despite successful implementation 
of topological states with Abelian anyons \cite{Satzinger2021Science, Semeghini2021Science} and even non-Abelian Ising \cite{Andersen2023Nature,Xu2023} and $D_4$ anyons \cite{Iqbal2024Nature}, whose braiding is restricted to Clifford gates at best.
Conventionally, the Fib-SNC is viewed as the ground state of a static Hamiltonian on a hexagonal lattice marked by high-order, 12-spin interactions \cite{Levin2005Phys.Rev.B}, which are exceptionally difficult for current capabilities. 
Nevertheless, a recent experiment showed promise \cite{Xu2024Nat.Phys.}. Yet, the formidable circuit depths necessary for $F$-moves forced the use of approximations. Moreover, the need to control 12 qubits for the smallest plaquette makes exploring the condensation of graph configurations practically infeasible. 

\para Our approach to creating minimalistic  Fib-SNC is the scalable dynamic string-net preparation (DSNP) strategy (see Fig.~\ref{fig:1}f). As explained below, we implement this strategy to create $\tau{\bf1}$ anyons, confirm their anyon charges, and braid them to extract the golden ratio. 
Furthermore, DSNP allows us to make 
 the first steps towards scaling up the Fib-SNC to estimate the chromatic polynomial { at golden ratio $\phi+2$.}

\para
DSNP leverages the inherent flexibility of graphs for efficient dynamical preparation of the Fib-SNC (see  Fig.~\ref{fig:1}f).\footnote{This is in contrast to the proposal of operating on a rigid lattice~\cite{Xu2024Nat.Phys.,Liu2022PRXQuantum}. A similar graph-centric perspective has proven productive for preparing states with Ising anyons~\cite{Lensky2023Annals}.} 
A single physical qubit can represent the smallest isolated string-net, or `bead', when prepared in a valid superposition through the modular-$\mathcal{S}$ gate:
\begin{equation}
\mathcal{S} = \frac{1}{\sqrt{1+\phi^2}}
    \begin{pmatrix}
    1&\phi\\
    \phi&-1
    \end{pmatrix}\;.
\label{eq:S}
\end{equation} 
The next step in building a larger scale Fib-SNC is to insert a qubit initialized in the $|0\rangle$ state between beads. 
Using \(F\)-moves (see Fig.~\ref{fig:1}c), the beads can be entangled 
into a strip of plaquettes. 
This strip can then be folded and sewn into a two-dimensional Fib-SNC
through additional  \(F\)-moves. 
The dynamic nature of this process allows for the optimization of resources for specific aims. 
{ The depth of the circuit grows linearly with the system size, the best scaling expected for a unitary circuit preparation of a topologically ordered state \cite{BrayviHastingsVerstraete2006}, but with the smallest prefactor to our knowledge compared to previous proposals such as \cite{Liu2022PRXQuantum}.\footnote{To prepare the Fib-SNC with $N\times N$ plaquettes, Ref. \textcite{Liu2022PRXQuantum} estimated $\approx120N$ layers of parallel CNOTS. With DSNP, the required depth scales as $2N$ 5-qubit $F$-moves. Using a conservative estimation of 40 CNOTs per 5-qubit $F$-moves \cite{Liu2022PRXQuantum}, the depth of the DSNP scales as $80N$ CNOT layers, with all other steps being parallelizable.} }

\para
{ Creation of anyons must change the topology of the many-body state. 
 In order to establish a protocol that allows explicit association between the circuit implementation and the evolving topology of the many-body state, we introduce a three-dimensional (3D) graphical representation where each copy of the TQFT is depicted as a two-dimensional surface (see Fig.~\ref{fig:1}g).
While the anyon-free Fib-SNC can be visualized entirely through two-dimensional (2D) graphs, the creation of anyons whose `anyon charge' labels which of the two copies of TQFT is affected 
necessitates keeping track of the two copies. 
Anyons connect the two surfaces through `worm holes' at the locations of anyons.
}
Furthermore, to create anyons while allowing for 
the detection and correction of local errors, we follow the `tail anyon' strategy~\cite{Schotte2022Phys.Rev.X} { that traps an end of an open string to the `tail qubit' located on a dangling edge inside a plaquette.} The $\tau{\bf1}$ or ${\bf1}\tau$ pair-creation can now be visualized as bringing in an open string from above or below the two surfaces. Fig.~\ref{fig:1}g illustrates inserting the strings from above into the Fib-SNC state shared between the two surfaces, which requires
undoing the over-crossing using the `$R$-move' shown in Fig.~\ref{fig:1}h.

\para {As a flexible state preparation strategy built on graphs, DSNP allows the preparation of Fib-SNC with an arbitrary number of plaquettes. The smallest of such only requires three qubits forming two plaquettes, which can be prepared as shown in Fig.~\ref{fig:1}j using a circuit with two $\mathcal{S}$ gates and a three-qubit $F$-move (Fig.~\ref{fig:1}k}). Fig.~\ref{fig:1}l shows the experimental result of implementing this Fib-SNC 
on the 27-qubit IBM Falcon processor \textit{ibm\_peekskill}. 
Using dynamical decoupling and readout-error mitigation \cite{Berg2022TREX}, but without other error mitigation, we sample the probability distribution of computational bitstrings using 8,192 experimental shots. The x-axis labels represent bitstrings as their corresponding graph configurations, with dashed (solid) lines indicating qubits in the $|0\rangle$ ($|1\rangle$) state and red $\times$'s denoting broken strings.
{Full tomography reconstruction of the experimental state yields a fidelity of \(0.87 \pm 0.01\) to the ideal state, which is not high in the absolute sense. However, the state shows much higher degree of 95\%  adherence to the branching rule.}

\begin{figure*}[h!]
    \centering
    \includegraphics[page=1,width=1\columnwidth]{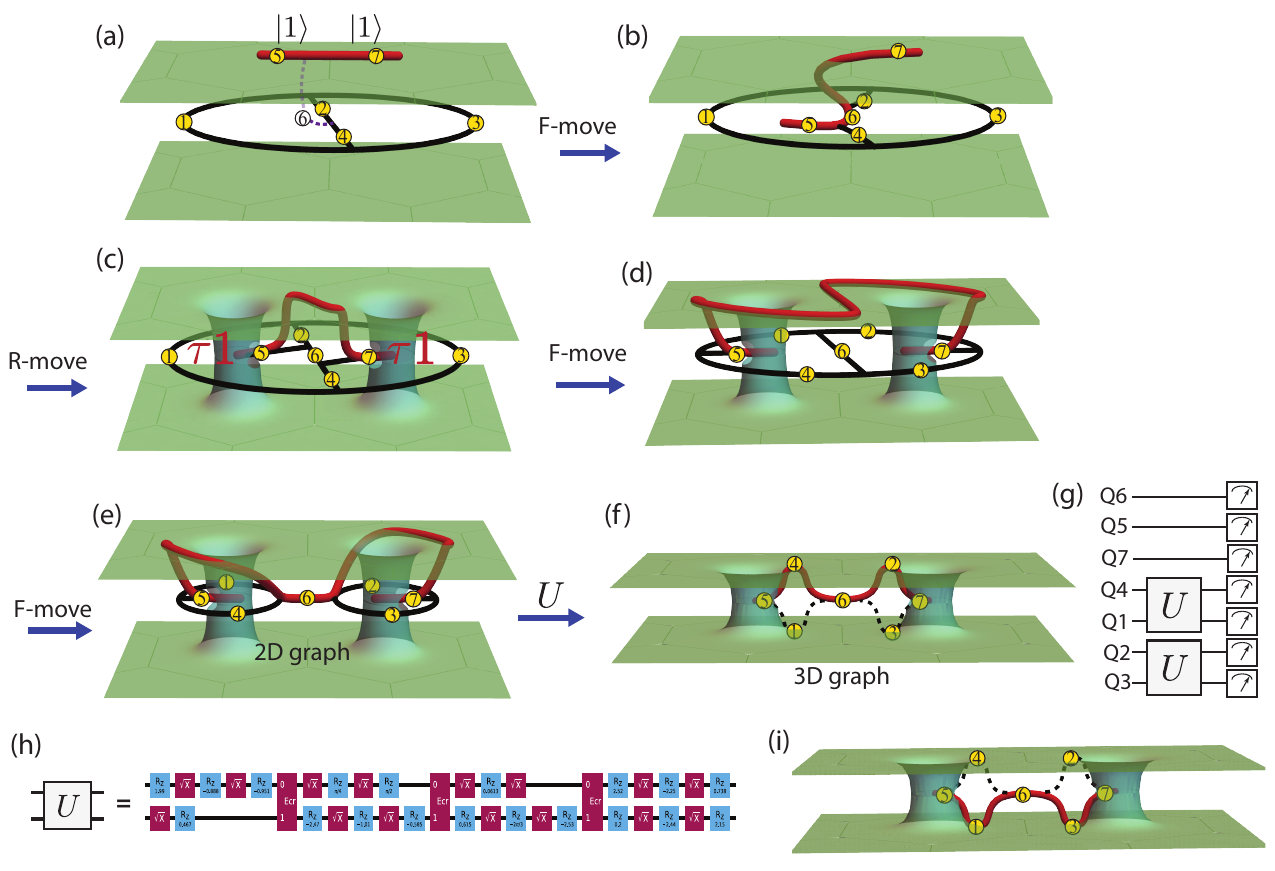}
    \begin{tikzpicture}
        \node[anchor=north west] at (0,0) {\includegraphics[page=1]{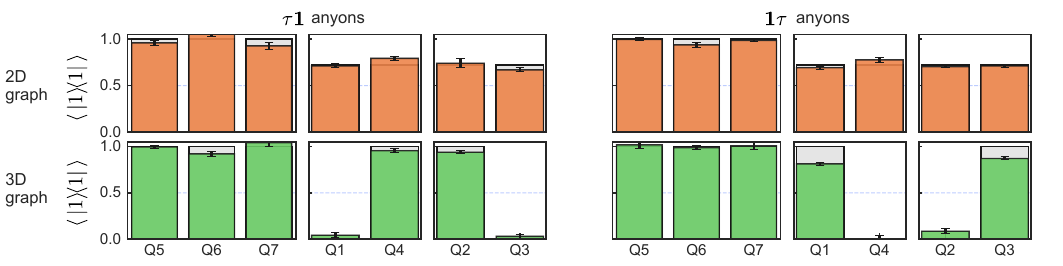}};
        \node[anchor=north west, xshift=0.0cm, yshift=-0.2cm, font=\sffamily\large] at (0,0) {(j)};
    \end{tikzpicture}
    \caption{ 
    \textbf{Fibonacci anyons: creating anyon pairs and certifying their anyon charges.}
    (a--c) {Building} a minimal two-plaquette string-net and creating a pair of $\tau \mathbf{1}$ anyons:
    {
    (a)  Qubits Q1--Q4 are initialized in $\ket{0}$, and envisioned to encode an empty two-plaque graph (black thick lines). Q5 and Q7 are initialized in $\ket{1}$, and visualized as supporting an open string (thick red line) in the upper 2D sheet (top green). Q6, prepared in  $\ket{0}$, encodes an ancillary vacuum-string segment (thin dashed line) connecting the plaquettes to the open string.
    (b) An $F$-move (blue arrow) involving Q2 and Q4--Q7 transforms the graph into a 3D configuration where edge Q6--Q7 rises above edge Q6--Q2, creating a non-planar geometry.
    (c) An $R$-move on Q6 
    yields a pair of $\tau \mathbf{1}$ anyons localized on the left and right plaquettes, with Q5 and Q7 acting as their tail qubits. The resulting open string threads through the upper plane, with its two ends supporting the $\tau \mathbf{1}$ anyons that pierce the two wormholes in the sheets (recall Fig.~1g). One further applies $F$-moves flipping edge Q2 and Q4 to reach the configuration in (d).
   (d--e) The anyon type is diagnosed by a charge measurement performed using a further $F$-move flipping edge Q6 to deform the graph into two connected plaquettes (e), joined at Q6.
   (f) Unitaries $U$ act on pairs (Q4, Q1) and (Q2, Q3), effectively lifting the unoccupied Q1 and Q3 strings away from the string path and placing Q4 and Q2 directly in it.
   }
    (g) 
    The circuit for the anyon type certification. 
    (h) 
    Compiled a circuit with three 2-qubit ECR gates (See SM Sec. F) and a few single-qubit gates for the 2-qubit unitary $U$.
    (i) 
    The contrast experiment prepares a pair of $\mathbf{1}\tau $ anyons by replacing the $R$-move with its complex conjugate. Thus, in the final measurement stage, the open string goes through the bottom plane with qubits Q1 and Q3 in its path. 
    (j) 
    Theory (grey bars) vs. experiment (colored bars) for the $\tau \mathbf{1}$ (left) and $\mathbf{1}\tau$ (right) anyon pairs measured in the 2D graph in panel (e) before applying $U$(upper) and the 3D graph in panel (f) after applying $U$(lower).
    The expectation values 
    $\big\langle \, |1\rangle\!\langle1| \,\big\rangle$
    are  indistinguishable in the 
   2D graph. 
    However, after the unitary places the qubits in the 3D graph, the one-state expectation values of {unpinned qubits Q1--4 certify the anyon type as the measurement reveals the path 
    of the open string. }
    }
\label{fig:2}
\end{figure*}

\para { 
Now we create a pair of Fibonacci anyons and certify their anyon types in the above two-plaquette Fib-SNC state. 
To create the $\tau{\bf 1}$ pair, we introduce an open string (red) of qubits (Q5 and Q7) 
above the two-plaquette Fib-SNC (Fig.~\ref{fig:2}a). With an unoccupied string (Q6 initialized in state $|0\rangle$) as the bridge, we entangle the open string with the Fib-SNC via $F$-moves and a $R$-move as illustrated in Fig.~\ref{fig:2}a--c (see {\bf Method} for details). These moves restore the planarity of the graph and effectively create two wormholes connected by an open string through the upper-copy TQFT. Now, both plaquettes each host a $\tau 1$ anyon at the tail. Although the qubits except the tail qubits now respect the local rules of Fib-SNC, the two copies of the TQFT share a complex superposition through the wormholes. 
To create the other anyon type, the ${\bf 1}\tau$ anyons, the open string should be inserted from underneath the two-plaquette Fib-SNC rather than from above.
Practically, this amounts to using the conjugate $R^*$-move instead of the $R$-move.
}

\para{
The canonical way to certify the anyon type would be to measure the five-qubit plaquette operators for each plaquette in the two-dimensional graph in Fig~\ref{fig:2}c. However, to combat the noise obscuring the certification in such extended measurements, we introduce an alternative approach that reduces this certification to independent single-qubit measurements. We first deform the graph so that the open string is pinned in the middle to be shared between the two TQFT's, as shown in Fig~\ref{fig:2}e.
Now, each plaquette can be independently measured, while the two tail qubits (Q5 and Q7) 
and the qubit bridging the plaquettes (Q6) are fixed to be in the $|1\rangle$ state, irrespective of the anyon type $\tau{\bf1}$ or ${\bf1}\tau$. At this point, all the qubits are still shared between the two TQFT's and we referred to this state as ``2D graph''.
We then lift the remaining four qubits off the shared space through a basis-changing unitary represented as $U$. In the end, the open string passes through all but two of the qubits. In particular, measurements of lifted qubits (Q1, Q4, Q2, Q3)  in the final ``3D graph'' shown in Fig~\ref{fig:2}f for $\tau{\bf 1}$'s amount to measuring the open string itself with a definite parity associated with the anyon types (see {\bf Method} for details on the implementation of Fig.~\ref{fig:2}e-f). 
}

\para To experimentally realize the anyon pair preparation and the anyon charge measurements, we need high-accuracy circuits about 150 two-qubit-gate-layers deep. We use a 133-qubit IBM Heron processor \textit{ibm\_torino}, featuring fast gates and reduced cross-talk, with median single- and two-qubit gate fidelities of $3.6\times10^{-4}$ and $4.6\times10^{-3}$, respectively (see SM Sec.~F). To address experimental noise, we employ a composite error suppression and mitigation strategy, including real-time qubit selection, dynamical decoupling, twirling \cite{Bennett1995Twirling, Knill2004Twirling}, zero-noise extrapolation \cite{Temme2016ZNEPEC, Li2017ZNE, Cai2022EMReview}, and twirled readout-error mitigation \cite{Berg2022TREX} (see SM Sec.~G).
\para {
We now create pairs of each anyon type and certify their type 
through single qubit measurements over $8.8 \times 10^6$ experimental realizations across 1,100 quantum circuit instances (see SM Sec.~H).
In Fig.~\ref{fig:2}j shows the resulting measurement outcome statistics for each of the qubits corresponding to the two anyon types. We measured 2D graphs like Fig.~\ref{fig:2}e and 3D graphs like Fig.~\ref{fig:2}f. The measurement results shown in Fig.~\ref{fig:2}j confirm the prediction with high precision. Specifically, three pinned qubits Q5--7 are consistently measured in the $|1\rangle$ state at all times.
 In the 2D graph, although single-qubit measurements hide the anyon type even for the remaining four qubits (Q1--4), the measured expectation value of $\big\langle \, \kb{1}{1} \,\big\rangle$ of 
$0.73 \pm 0.04$ as shown in the upper histograms in Fig.~\ref{fig:2}j is consistent with the theoretically predicted value of  
 \(\frac{\phi^2}{\phi^2+1} \approx 0.72\).
 However, these four qubits (Q1--4) show a dramatic contrast between the two anyon types in 3D graphs, as shown in the lower histograms in Fig.~\ref{fig:2}j. Since these four qubits are ``lifted off'' the shared plane to belong to top or bottom TQFT, the open string traverses the top TQFT with Q4 and Q2 ($\tau \mathbf{1}$) or the bottom TQFT with Q1 and Q3 ($ \mathbf{1}\tau$) depending on the anyon type. (see {\bf Method} for more details)
}

\begin{figure*}[h!]
    \centering
    \includegraphics[page=1,width=1\columnwidth]{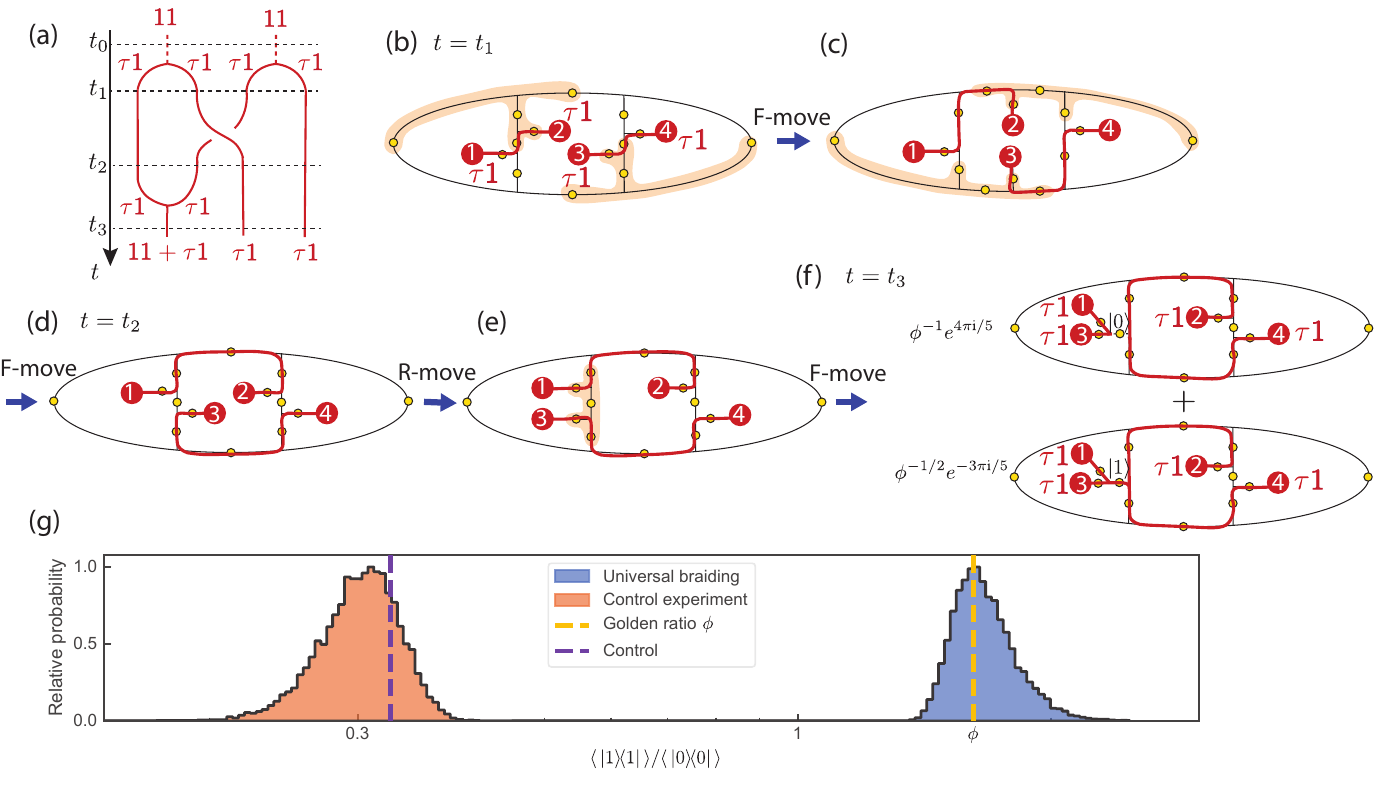}
    \caption{ 
    \textbf{Braiding of two $\tau {\bf 1}$ anyons { to implement} a non-Clifford gate on a topological logical state.}
    {
    (a) Worldlines depicting the creation of four $\tau{\bf 1}$ anyons from the vacuum ${\bf 1}{\bf 1}$, followed by the braiding of anyons 2 and 3, and concluding with a fusion-based measurement to determine the logical gate implemented by the braiding process.
    (b) Generalization of the protocol from Fig.~\ref{fig:2}, where four $\tau{\bf 1}$ anyons (red dots labeled as 1--4) are initialized on three plaquettes. Labels for the qubits (yellow dots) are suppressed. 
    (c--d) Braiding is achieved through four five-qubit $F$-moves, which permute anyons 2 and 3. The groups of five qubits undergoing the $F$-moves are indicated by the orange patches.  
    (e) An $R$-move flips anyon 3 from the center to the left plaquette, forming a new configuration for fusion.
    (f) A final $F$-move fuses anyons 1 and 3, resulting in a coherent superposition of two fusion outcomes. 
    (g) Experimental distributions of the measured ratio $\langle 1|1\rangle / \langle 0|0\rangle$ on a logarithmic scale, derived via bootstrap resampling for both the braiding and control experiments. The analysis accounts for error mitigation and statistical uncertainty (see SM Sec.~H). Vertical dashed lines indicate theoretical predictions: the golden ratio $\phi$ (yellow) and the control value (purple). Asymmetry in the distributions reflects the non-linear transformation of the ratio observable.
    }
    }
\label{fig:3}
\end{figure*}

\para  {Fully two-dimensional braiding must involve three or more plaquettes and two pairs of $\tau \mathbf{1}$ anyons. DSNP prescribes a scalable strategy for creating plaquette strips of arbitrary lengths. In Fig.~\ref{fig:3}, we demonstrate two-dimensional braiding in a scalable and error-correctable manner using the minimalistic three-plaquette strip and verify the braiding outcome through the fusion of a pair of anyons (see Fig.~\ref{fig:3}a for the schematics). }
Repeating the anyon pair preparation, we prepare two anyon pairs spread over three plaquettes as depicted in Fig.~\ref{fig:3}b. This amounts to time steps $t_0$--$t_1$ in Fig.~\ref{fig:3}a. 
Initially, the logical qubit encoded to the triplet of $\tau \mathbf{1}$ anyons (1,2,3) is in the $\overline{\ket{0}}$ state since the anyon 1 and anyon 2 are created from vacuum. 
{Now, we braid $\tau \mathbf{1}$ anyons 2 and 3 using a sequence of exact $F$-moves executing the time steps $t_1$--$t_2$ in Fig.~\ref{fig:3}a.}
Such braiding is predicted to execute 
a non-Clifford gate $\sigma_2$ (see Fig.~\ref{fig:1}e) on the logical qubit,
rotating the logical state to 
\begin{align}\label{eq:logical_qubit}
\sigma_2\overline{\ket{0}}=
\phi^{-1} e^{4\pi \ii /5} \overline{\ket{0}} + \phi^{-1/2} e^{-3\pi \ii /5} \overline{\ket{1}}\;.
\end{align}
{We certify the predicted non-Clifford gate by fusing anyon 1 and anyon 3.}
{ For this, we bring anyon 1 and anyon 3 together to share a single root edge using an $R$-move and an  $F$-move(see Fig.~\ref{fig:3}e).}
Now a measurement in the physical computational basis of the root edge onto either $\ket{0}$ or $\ket{1}$ projects the logical qubit to $\overline{\ket{0}}$ or $\overline{\ket{1}}$, respectively. Hence, if the braiding implements the correct logical gate in Eq.~\eqref{eq:logical_qubit}, the golden ratio can be measured through $\big\langle \, |1\rangle\!\langle1| \,\big\rangle/\big\langle \, |0\rangle\!\langle0|  \,\big\rangle = \phi$.

\para
As in the previous experiment, we implement this sequence on \textit{ibm\_torino} using the composite mitigation strategy, but with double the number of twirls and shots per twirl due to the increased circuit complexity. We find 
$\big\langle \, |1\rangle\!\langle1| \,\big\rangle/\big\langle \, |0\rangle\!\langle0| \,\big\rangle = 1.65 \pm 0.14$,  
within $2\%$ of the golden ratio $\phi$. Fig.~\ref{fig:3}g shows the distribution of bootstrap resampling, providing confidence intervals (see SM Sec.~H).
In a control experiment, we intentionally introduce bit-flip errors during to break two strings, generating unwanted excitations (see SM Sec.~H). This modification alters the bitstring distribution. We now measure~$
\big\langle \, |1\rangle\!\langle1| \,\big\rangle/\big\langle \, |0\rangle\!\langle0| \,\big\rangle = 0.30 \pm 0.025
$, consistent with the theoretical prediction of~0.328 for the modified circuit.

\begin{figure*}[h!]
    \centering
    \includegraphics[page=1,width=1\columnwidth]{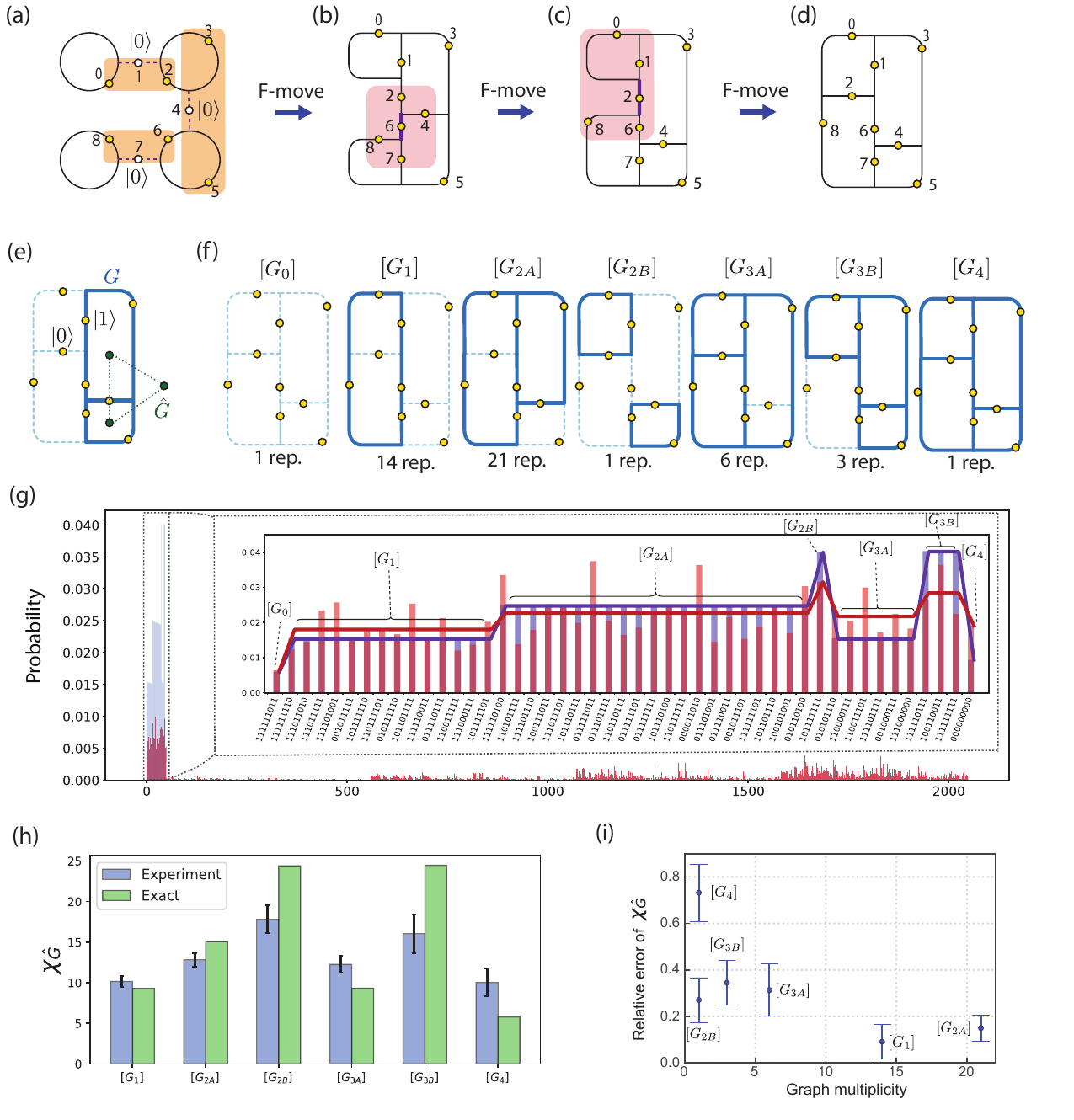}
    \caption{
    \textbf{Estimating chromatic polynomials.}
    (a-b) Four decoupled beads are
    prepared by generalizing the protocol of Fig.~\ref{fig:1}j. Three parallel $F$-moves act on the three shaded groups of qubits (orange boxes), yielding a folded strip with 4 plaquettes. 
    (c--d) Two 5-qubit $F$-moves applied to qubits in the shaded boxes deform the graph into a $2\times 2$ lattice of plaquettes: a 2D Fib-SNC. 
    (e)  The dual graph of each graph $G$ is denoted as $\hat{G}$ (green dotted lines).
    (f) For the $2\times 2$ lattice, there are 7 isomorphism classes of graphs formed by the edges in state $|1\rangle$. All the graphs are topologically equivalent (or isomorphic) within each isomorphism class. The number of representative graphs (multiplicity) is listed for each isomorphism class. The relative probability with respect to empty configuration $G_0$ is defined as $\tilde{P}([G])=P([G])/P([G_0])$.
    (g) Large panel:  Probability distribution over all $2^{11}$ bit-strings, including the two ancilla qubits (blue: theory; red: experiment). 
    Theoretically, non-zero bitstrings (47 in total) are ordered on the left.
    These satisfy the branching rule, while the remaining bitstrings on the right do not. 
    Inset: zoom-in to bitstrings obeying the branching rule.  
    The theoretical distribution reflects 7 isomorphism classes.  
    Thick red line: the measured probability 
 averaged over each isomorphism class.  
    (h) Extracting the chromatic polynomial values for graphs dual to the given string-net isomorphism class (blue: experiment; green: theory).  Error bars obtained from the standard deviation of the graph representatives in each class.  
    (i) The relative error and multiplicity of each isomorphism class of graphs. A class with a larger multiplicity tends to have smaller relative errors. 
    }
\label{fig:4}
\end{figure*}

\para {Now we move onto the most ambitious pursuit of this paper, taking the first step towards a new class of classically hard problems.
In Fig.~\ref{fig:4}, we realize a two-dimensional, four-plaquette Fib-SNC vacuum and sample it to estimate the chromatic polynomials for all possible trivalent graph embedding. 
We perform all experiments on \textit{ibm\_torino}. 
Due to the outstanding challenge of mitigating noise for sampling rather than expectation values \cite{Cai2022EMReview}, we only mitigate readout errors 
and not gate errors. 
We use DSNP as illustrated in Fig.~\ref{fig:4}a--d, starting with a four-bead strand and evolving each bead into one of the four-plaquettes of the resulting Fib-SNC vacuum (see Fig.~\ref{fig:4}d).
To reduce the circuit depth of $F$-moves we use 2 ancilla qubits(see SM Sec.~H), in addition to the 9 qubits participating in Fib-SNC.
}

\para 
{
It has long been predicted that the normalized probability weight of a subgraph $G$ in Fibonacci string-net condensate evaluates the chromatic polynomial of a dual graph $\hat{G}$ (see Fig.~\ref{fig:4}e) at $k$$=$$\phi$$+$$2$
~\cite{Fidkowski2009Commun.Math.Phys., Fendley2005Phys.Rev.B, Fendley2009Geom.Topol., Fendley2010Adv.Theor.Math.Phys.}, i.e.
\begin{equation}\label{eq:chromatic}
    \frac{P(G)}{P(G_0)}=\frac{1}{\phi+2}\chi(\hat{G},\phi+2)\;,
\end{equation}
where $P(G)$ and $P(G_0)$ are probability weight of a subgraph $G$ and the  
empty configuration $G_0$, respectively.
While 
the chromatic polynomial $\chi(\hat{G},k)$ for a positive integer $k$ counts the number of ways to $k$-color the graph $\hat{G}$\cite{Read1968JournalofCombinatorialTheory}, a recurrence relation defining the polynomial allows for extension of polynomial to non-integer valued $k$ such as $\phi+2$.
As a complex combinatorial problem, the evaluation or estimation of the chromatic polynomial is a classically hard problem  \cite{Jaeger1990Math.Proc.Camb.Philos.Soc., Vertigan2005SIAMJ.Comput., Golderb:2008908, Goldberg:2012uh, Goldberg2014SIAMJ.Comput.} despite the simplicity of the defining recurrence relation.\footnote{The proof of Ref.~\cite{Golderb:2008908,Goldberg:2012uh} is carried out for rational $k$, while one may expect that the same conclusion holds for irrationals.} 
This implies that the exact theoretical evaluation of the Fib-SNC amplitude requires an exponential-time classical algorithm in general(see SM Sec. C3). 
Hence the experimental realization of the Fibonacci string-net condensate may offer a new route for seeking quantum advantage. 

}

\para
{
Although the absence of an error-mitigation scheme poses a challenge in sampling a general state that is not highly concentrated, we can exploit the topological structure of Fib-SNC.
Firstly, valid bit-string configurations that satisfy the branching rules form a relatively small subset of all possible bitstrings. Secondly, these valid bitstrings further group into topologically equivalent isomorphism classes. Specifically, for the four plaquette Fib-SNC we implemented, there are 6 classes as shown in Fig~\ref{fig:4}(f) with different multiplicity among the bitstrings that correspond to the class.} Figure~\ref{fig:4}{g} shows the result of sampling this Fib-SNC vacuum 
for $30 \times 10^6$ realizations. With two ancilla qubits introduced to reduce the circuit depth, the probability distribution is shown over $2^{11}$ possible bit strings obtained on \textit{ibm\_torino}.
{Leveraging that the Fib-SNC amplitudes can be calculated for the present scale Fib-SNC, we benchmark experimentally sampled results against the exact predictions. The topological nature of Fib-SNC predicts amplitudes of bitstrings to be non-zero only for 47 branching rule respecting bitstrings, with the same amplitude within given isomorphism class (shown in blue in Fig.~\ref{fig:4}h).}

\para The experimentally obtained probability distribution shows robust suppression of branching rule violating, forbidden bit-strings (Fig.~\ref{fig:4}g). 
{ Moreover, class averages of the allowed bitstrings offer the estimates of the chromatic polynomials:
\begin{eqnarray}
\chi([\hat{G}_1],k)&=&k^2-k \\
\chi([\hat{G}_{2A}],k)&=&k^3-3k^2+2k \\
\chi([\hat{G}_{2B}],k)&=&k^3-2k^2+k \\
\chi([\hat{G}_{3A}],k)&=&k^4-6k^3+11k^2-6k \\
\chi([\hat{G}_{3B}],k)&=&k^4-5k^3+8k^2-4k \\
\chi([\hat{G}_{4}],k)&=&k^5-9k^4+29k^3-39k^2+18k, 
\end{eqnarray}
at $k=\phi+2$.}
For this, we estimate the relative probability $P(G)/P(G_0)$ in Eq.~\eqref{eq:chromatic}  by~$\overline{C}([G])/\overline{C}(G_0)$, where $\overline{C}([G])$ represents the average count of all bitstrings corresponding to graphs topologically equivalent to $G$\footnote{For a larger scale estimation, a graph class with higher multiplicity can be used as a reference in place of the empty configuration in Eq.~\eqref{eq:chromatic} (see SM Sec. C5).}. 
We show the resulting estimates of the chromatic polynomial in Fig.~\ref{fig:4}h, where uncertainty ranges are computed using the standard deviation within each equivalent class.
While the absence of error mitigation limits the accuracy of the estimates, Fig.~\ref{fig:4}i shows the multiplicity within each class countering errors. Specifically, the larger the multiplicity, the more accurate the estimates are. 
In particular,
the experimental estimate based on the average over $[G_1]$-class bitstrings yields $1.82$ for the golden ratio $\phi$, with 13\% relative error.  

\para 
{ In summary, our work introduces and implements a new scalable approach, DSNP,  to preparing Fib-SNC and creating,  certifying, and braiding the Fibonacci anyons the Fib-SNC supports. 
While the present experiments with physical qubits are limited by noise, the prospect of sampling of Fib-SNC with a large number of plaquettes using DSNP raises new potential frontier in the pursuit of quantum advantage. The success in this new pursuit of quantum advantage will hinge on solving two open problems.} 
Firstly, noise in sampling must be countered. 
Secondly, the sampling complexity of Fib-SNC needs to be further investigated to compare it with the complexity of classical approximations. While the sampling space of valid bitstrings still grows exponentially with system size, an estimation of $\chi(\hat{G}, \phi+2)$ for a specific dual graph $\hat{G}$ requires sampling only one equivalence class among the valid bitstrings. Hence, it is 
possible that the estimation of chromatic polynomial { at $\phi+2$} from quantum sampling can be more efficient than from classical algorithms for intermediate system size (for $O(100)$ to $O(1000)$ qubits). 
{If so, Fib-SNC sampling will become an exciting avenue for near-term fault-tolerant quantum computers. }

{
\small
\textbf{Acknowledgements:} 
While preparing our manuscript, we learned of a related study by Ref.~\onlinecite{Xu2024Nat.Phys.} on Fibonacci anyons. One significant difference in our study is that we focus on scalable planar~(2D) braiding in an error-correctable manner. Secondly, we introduce the chromatic polynomial estimation through string-net sampling. We thank Sergey Bravyi and Vojtěch Havlíček for insightful discussions on the complexity of chromatic polynomials and Dimitry Maslov for advice on simplifying multi-qubit Toffoli gates. We are grateful to Abhinav Kandala, Emily Pritchett, and Sarah Sheldon for their comments on the manuscript. We also thank Antonio Mezzacapo, Javier R. Moreno, and Ian Hincks for their valuable input. 
JW is supported by the Harvard University CMSA research associate fund. AS was supported by grants from the ERC under the European Union’s Horizon 2020 research and innovation programme (Grant Agreements LEGOTOP No. 788715), the DFG (CRC/Transregio 183, EI 519/71), and the ISF Quantum Science and Technology (2074/19). 
E-AK acknowledges support by the NSF through OAC-2118310. C-MJ is supported by
Alfred P. Sloan Foundation through a Sloan Research Fellowship.
GZ is supported by the U.S. Department of Energy, Office of Science, National Quantum
Information Science Research Centers, Co-design Center
for Quantum Advantage (C2QA) under contract number
DE-SC0012704. 
}

\clearpage
\onecolumngrid

\clearpage
{\bf Figure 1 caption}: 
\textbf{Principle of the dynamical string-net preparation (DSNP) approach and experiment.}
(a) 
Typical string-net configurations of spins in $|1\rangle$ state, for a trivial state (left) and for a string-net condensate (right).
(b) Branching rule for the Fib-SNC. A dashed line represents a qubit in the $|0\rangle$ state while a solid line represents a qubit in the $|1\rangle$. Vertices can have either 0, 2, or all 3 edges excited in the Fib-SNC. 
(c) 
Five-qubit $F$-move relating allowed string-net configurations among five qubits. When one or two pairs of four outer legs are identified, this becomes four-qubit or three-qubit $F$-moves.
$\phi$ is the golden ratio, and $\pm$ signs denote superpositions. 
(d) Logical qubit encoding using a triplet of $\tau {\bf 1}$ anyons. Logical~$\overline{|0\rangle}$ and~$\overline{|1\rangle}$ differ by the fusion outcomes of the first two $\tau {\bf 1}$ anyons. The red lines in the figure represent the space-time trajectory of anyons. Bringing together anyons 1 and 2 can either annihilate both anyons to return vacuum ($\overline{|0\rangle}$) or return a single $\tau {\bf 1}$ anyon ($\overline{|1\rangle}$).
(e) Pairwise braiding among the triplet of $\tau{\bf 1}$ anyons implements a non-Clifford gate on the logical state encoded on the triplet of anyons. The two pairwise braiding implements two distinct non-Clifford gates $\sigma_1$ and $\sigma_2$ (see SM Sec. A1). 
(f) 
Schematic outline of DSNP. Yellow dots represent qubits in SNC and empty dots represent reserve qubits in $|0\rangle$ state. The decoupled beads transform into a folded strip of plaquettes upon $F$-move entangling in reserve qubit(second sub-panel). 
The strips are sewn up via $F$-moves (third sub-panel) into the 2D Fib-SNC (right) (see SM Sec. D for details). 
(g) The Fib-SNC can be visualized with a pair of two-dimensional surfaces representing two time-reversed copies of TQFT. To create two $\tau {\bf 1}$ anyons, we bring in an open string from above. $F$- and $R$-moves bring the ends of the open string to join the two copies of TQFT through wormholes, with the ends piercing the wormholes and localizing anyons.
(h) The $R$-moves (or its complex conjugate) to resolve the over-crossing (see SM Sec. A1 for more details).
(i) A deep quantum circuit for braiding two $\tau{\bf 1}$ anyons using hardware-native gates (see Fig.~3).
(j) DSNP for the smallest Fib-SNC. Left: Three qubits, each prepared in~$\ket{0}$ (white dots), represent 3 unoccupied strings (dashed lines). Middle: Two single-qubit modular~$\mathcal{S}$ gates on Q1 and Q3 create two beads (solid rings), leaving the reserve qubit in $|0\rangle$ state. The state amounts to two decoupled beads with a reserved qubit in the middle represented by an unoccupied edge (dashed line).
Right: A 3-qubit $F$-move creates the minimal Fib-SNC by entangling in the middle qubit Q2. 
(k) Quantum circuit corresponding to { two $\mathcal{S}$ gates followed by the 3-qubit $F$-move, implementing all the steps of panel (j)}.
(l)
The probability weight of different string-net configurations corresponding to the depicted string nets for the minimal Fib-SNC. { The red $\times$ marks the vertices violating the branching rule. Only 5\% of the shots violate the branching rule.}

\clearpage

\clearpage

\clearpage

\clearpage

\twocolumngrid

\noindent \textbf{Methods}

\noindent \textit{Protocol for the anyon creation and certification experiment:}
\para  
In this section, we provide more details about the experimental protocol of creating and certifying doubled Fibonacci anyons presented in Fig.~\ref{fig:2}.  
We grow the string-net condensate to create  $\tau{\bf 1}$ and ${\bf 1}\tau$ anyon pairs and measure their anyon charges. Applying the DSNP concepts from Fig.~\ref{fig:1}f--g on our minimal example, we add the minimal number of needed qubits, Q4--Q7  shown in Fig.~\ref{fig:2}a. Qubit Q4 is incorporated into the condensate by entangling it with Q2 using a controlled-NOT operation. 
Tail qubits Q5 and Q7 are prepared in \(\ket{1}\) (yellow dots) and bridge qubit Q6 in \(|0\rangle\) (white dot). Now Q1--Q4 form a Fib-SNC shared between two copies of TQFT as depicted in Fig.~\ref{fig:2}a. To bring in \(\tau \mathbf{1}\) anyons, we start by introducing 
open string Q5--Q7 (red line) above  Q2--Q4 (black line). A five-qubit $F$-move (see Fig.~\ref{fig:1}c) entangles Q5 and Q7 with the rest, creating a single connected, 3D graph (Fig.~\ref{fig:2}b). 
Then, an $R$ move (see Fig.~\ref{fig:1}h) completes the preparation of a pair of \(\tau \mathbf{1}\)s at the end of red open string that pierces the wormholes in Fig.~\ref{fig:2}c.\footnote{Operationally, using instead a conjugate $R^*$ move related to the $R$ move in Fig.~\ref{fig:1}h by complex conjugation of phases will amount to creation of \(\mathbf{1}\tau\) anyon pairs. In this case, the open string Q5--Q7 should be initially introduced underneath Q2--Q4 instead. }
At this point, all the qubits except the tail qubits once again follow the local rules of Fib-SNC, forming a complex superposition shared between the two copies of TQFT. This adherence to the local rules means this state can be error-corrected.

\para
How do we most robustly certify the creation of the anyons associated with this abstract notion of an open string?
While the qubits are participating in the superposition of string nets as in Fig.~\ref{fig:2}c, the only measurement predicted to yield a definitive answer would be  
the left and right five-qubit plaquette operators, each comprising many Pauli terms --- but this compounds noise.
However, if the qubits are `lifted' to each copy of TQFT through basis changes such that the open string goes through the qubits, the open string itself can be measured. 
To this end, we  
 dynamically reconfigure the graph to place Q6 on a bridge that is forced to be in a definite \(|1\rangle\) state (see Fig.~\ref{fig:2}d--e). 
 Now three qubits are pinned in the \(|1\rangle\) state: Q6 and the two tail qubits, Q5 and Q7. 
 However, at this point of Fig.~\ref{fig:2}e,  Q1--Q4 will be participating in the condensate, each in a superposition of \(|0\rangle\) and \(|1\rangle\). 
 At the same time, the open string is away from the qubits except at the wormholes. 
 Remarkably, a change of basis through two-qubit unitary \(U\) (see Fig.~\ref{fig:2}g and SM Sec.~B) would lift off Q1--Q4 to make the open string go through Q4 and Q2. Topologically, the two states in Fig.~\ref{fig:2}e and Fig.~\ref{fig:2}f are equivalent. Nevertheless, the microscopic qubit placements are such that measurements of Q4 and Q2 will reveal the open string itself with a definite outcome in Fig.~\ref{fig:2}f.

\para
We certify $\tau \mathbf{1}$ pair creation through the detection of the open string as shown in Fig.~\ref{fig:2}j (see SM Sec.~H).
The qubits that should be `pinned' to $\ket{1}$ state, Q5--Q7, are measured to be in the correct state with high probability irrespective of whether the remaining qubits, Q1--Q4, are placed in 2D or 3D graphs, yielding
$0.99 \pm 0.05$ on average.
However, the measurement outcome for Q1--Q4 are strikingly different 
because 3D graph has the open string going through (Q1, Q3)-pair or (Q4, Q2)-pair.
When placed in the 2D graph with the open string away from the qubits Q1--Q4, the four qubits are indistinguishable with the expectation value of the one-state projector 
$\big\langle \, |1\rangle\!\langle1| \,\big\rangle$  
of 
$0.73 \pm 0.04$ across the 8 measurements. 
This is precisely as they should be as a part of the Fib-SNC represented by the 2D graph, with the predicted expectation value of \(\frac{\phi^2}{\phi^2+1} \approx 0.72\).
On the other hand, when the four qubits are placed in the 3D graph, the open string traverses (Q4, Q2)-pair in the upper copy while the unoccupied string ``traverses" (Q1, Q3)-pair in the lower copy.
Contrastingly, when \(\mathbf{1} \tau\) anyons are prepared with the open string passing through the lower surface, measurement outcomes between (Q4, Q2)-pair and (Q1, Q3)-pair completely reverses, as expected from the open string now passing through (Q1, Q3)-pair in the lower layer.
Our certification of \(\tau \mathbf{1}\) and \(\mathbf{1} \tau\) anyon creation through 28 measurements of the open string show the average experimental discrepancy is $-0.01 \pm 0.06$.

\clearpage



\counterwithin*{figure}{part}

\stepcounter{part}

\renewcommand{\thefigure}{S\arabic{figure}}

\begin{appendix}

\widetext

\qquad \qquad \qquad \qquad \qquad \qquad \qquad \qquad \qquad  \quad  \textbf{Supplemental Materials}

\tableofcontents

\section{Theory of string-net condensates and related anyons}

\subsection{Introduction to Fibonacci string-net condensate}
\subsubsection{Fibonacci string-net condensate: Wave functions, $F$-moves, $R$-moves, and anyon types}
String-net condensation offers a general mechanism to study party- and time-reversal invariant topological phase \cite{0404617Levin:2004miLevinWen}. A string-net condensate is the topological vacuum state formed by a complex superposition of string nets, which can be visualized as (planar) trivalent graphs constrained by local branching rules. In the following, we will review Fibonacci string-net condensate (Fib-SNC), the simplest string-net condensate that allows string branching, following Ref. \onlinecite{0404617Levin:2004miLevinWen} and \onlinecite{Koenig2010AnnalsofPhysics}. We will discuss the Fib-SNC's local branching rule and the ways to encode the complex superposition in the condensate.

\begin{figure}
    \centering
    \includegraphics[width=0.95\linewidth]{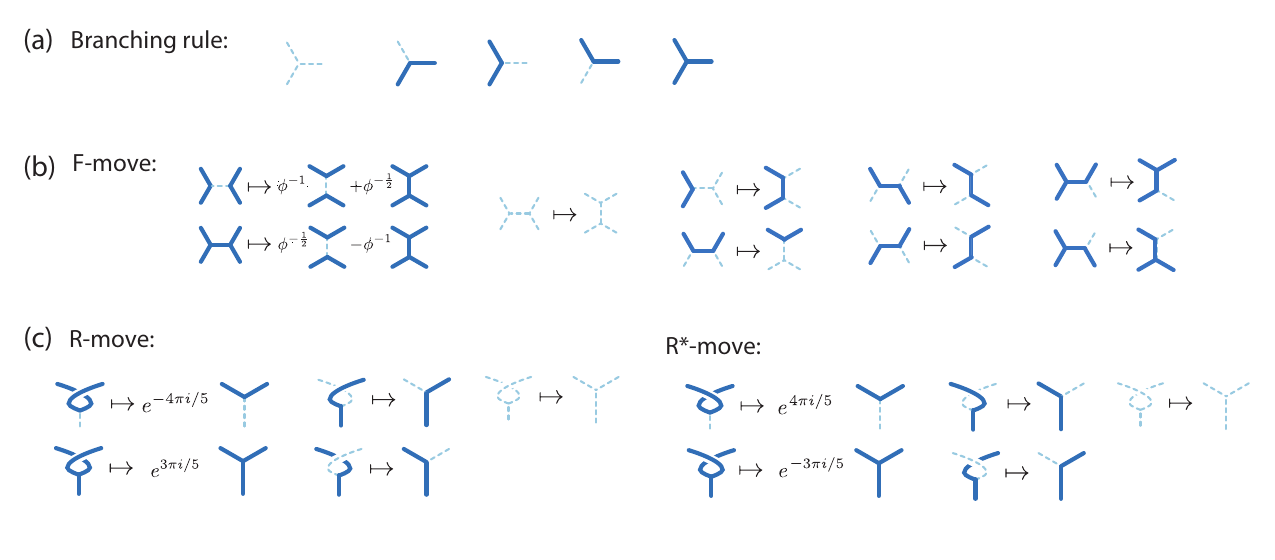}
    \caption{(a) Fib-SNC branching rule. A solid line represents an excited edge in the $|1\rangle$ state. The dashed line represents an edge in the $|0\rangle$ state. (b) $F$-moves of Fib-SNC. (c) $R$-moves and $R^*$-moves of Fib-SNC.}
    \label{fig:FRmoves}
\end{figure}

In the Fib-SNC, each edge of the trivalent graph has two states, $|0\rangle$ and $|1\rangle$, presented by a qubit. The ``excited state" $|1\rangle$ can be interpreted as a string running through the edge, and $|0\rangle$ indicates no string. The branching rule of Fib-SNC requires that at each trivalent vertex of the graph, only the configurations with 0, 2, and 3 edges excited are allowed (see Fig. \ref{fig:FRmoves}a). These configurations can be respectively interpreted as no strings, a string passing through the vertex, and a string branching into two strings. A string terminated at a vertex is not allowed in the Fib-SNC. Under this branching rule, the excited edges also form a trivalent graph. 

Next, we discuss how to encode the complex superposition of string nets in Fib-SNC. Instead of writing down the amplitude of each allowed string configuration, it suffices to specify how the amplitudes of different configurations are related to each other. First of all, two string configurations that are topologically the same share the same amplitude. A local graph deformation under the $F$-move relates the amplitude of different configurations via
\begin{align}
		\raisebox{-0.5cm}{\includegraphics[scale=.40]{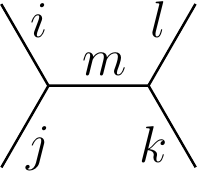}} \;\;
		& \mapsto \sum_{n} \; F^{ijm}_{kln} \;\;
		\raisebox{-.5cm}{\includegraphics[scale=.40]{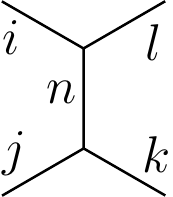}} \, \label{eq:Fmove},
\end{align}
where $i,j,k,l,m,n = 0,1$ label the states of the corresponding edges and $F^{ijm}_{kln} $ is called the $F$-symbol. All of $F$-moves of the Fib-SNC are shown in Fig. \ref{fig:FRmoves}b (or related to the shown ones by $180^\circ$ rotation). Additionally, removing a ``tadpole" (shown below) from the graph changes the amplitude by a factor
\begin{align}
	    \raisebox{-0.3cm}{\includegraphics[scale=.40]{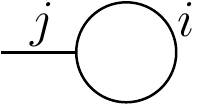}} \;
		 & \mapsto \delta_{j0} d_i \label{eq:tadpole} \,,
\end{align}
where $\delta_{j0}$ is the Kronecker delta function. $d_0 = 1$ and $d_1 = \phi$ are quantum dimensions of two string types. Recall that the $\phi = \frac{\sqrt{5}+1}{2}$ is our notation for the golden ratio throughout this work. We will comment on the mathematical structure behind the $F$-symbol and the quantum dimensions $d_i$ below. With the branching rule, $F$-moves and Eq. \eqref{eq:tadpole}, the amplitude of every string configuration is fixed.

The Fib-SNC is topologically equivalent to the combined vacuum state of a time-reversed pair of topological quantum field theory (TQFT). Each copy can be identified as the integer-spin sector of the SU(2) Chern-Simons theory at level 3 \cite{Witten:1988hfJones,Bonderson2008,0707.1889Nayak:2008zza} (or level-1 Chern-Simons theory with the exceptional gauge
group ${\rm G}_2$). Each of the copies can support a Fibonacci anyons $\tau$ (and a trivial anyon $\bf 1$). For simplicity, we will refer to this time-reversed pair of TQFT as the doubled Fibonacci TQFT.

The full list of anyons Fib-SNC (or equivalently the doubled Fibonacci TQFT) can host is given by
\begin{align}
\{\1\1, \tau \1, \1\tau,\tau\tau \},
\end{align}
which are essentially all possible combinations of anyons from the two copies of TQFT. Here, $\1\1$ represents the trivial anyon, which is equivalent to the vacuum. The fusion rules of these anyons inherit from the single copy fusion rules: 
\begin{align}\label{eq:fib_fusion_rule}
\1 \times \1 = \1,~\tau\times \1 = \1 \times \tau =\tau,~\tau\times \tau = \1+ \tau. 
\end{align}

We would like to highlight the doubled Fibonacci fusion rule
\begin{align}
    \tau\1\times \tau\1 = \1\1+ \tau\1,
\end{align}
which enables the qubit encoding discussed in Fig. 1d of the main text. To be more specific, as shown in Fig. 1d, when three $\tau\1$ anyons fuse into a single $\tau\1$, i.e. $(\tau\1 \times \tau\1)\times \tau\1 \rightarrow \tau\1$, there are two different fusion channels distinguished by whether the first two $\tau\1$'s fuse into $\1\1$ or $\tau\1$. These two fusion channels correspond to the two logical states $\overline{|0\rangle}$ and $\overline{|1\rangle}$ of a logical qubit. The pairwise braidings of the first two $\tau\1$'s and the last two $\tau\1$'s (shown in Fig. 1e) generate the non-Clifford logical gates $\sigma_1$ and $\sigma_2$ respectively:
\begin{align}
    \sigma_1 =\left(\begin{array}{cc}
        e^{-\ii 4\pi/5 } & 0 \\
        0 & e^{\ii 3\pi/5 }
    \end{array}\right),~~~~\sigma_2 =\left(\begin{array}{cc}
        \phi^{-1} e^{\ii 4\pi/5 } & \phi^{-\frac{1}{2}} e^{-\ii 3\pi/5 } \\
        \phi^{-\frac{1}{2}} e^{-\ii 3\pi/5 } & -\phi^{-1}
    \end{array}\right).
\end{align}

The Fib-SNC only involves 2D planar trivalent graphs. As will be discussed in Sec. \ref{app:anyonic_fusion} in detail, the creation of non-trivial anyons $\tau \1$, $\1\tau$, and $\tau\tau $ requires keeping tracking of each copy of the TQFTs, which can be done by extending the planar trivalent graph to 3D. The amplitude of a 3D graph can be related to the planar one with the help of additional local graph deformations called the $R$-move and $R^*$-move:
\begin{align}
		\raisebox{-0.6cm}{\includegraphics[scale=.45]{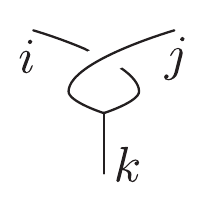}} 
		 \mapsto  \; R^{ij}_k \;\;
		\raisebox{-0.6cm}{\includegraphics[scale=.45]{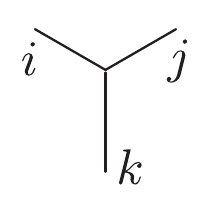}} \, ,~~~~~~~~~~
  \raisebox{-0.6cm}{\includegraphics[scale=.45]{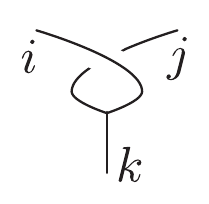}} 
		 \mapsto  \; (R^{ij}_k)^* \;\;
		\raisebox{-0.6cm}{\includegraphics[scale=.45]{fig/R-move-Y.pdf}} \, 
  \label{eq:Rmove},
\end{align}
where $R^{ij}_k$ is called the $R$-symbol. All the $R$-moves for the Fib-SNC are graphically represented in Fig. \ref{fig:FRmoves}c. The mathematical framework that underlies the branching rule, quantum dimensions, $F$-symbols, and $R$-symbols is called the modular tensor category \cite{Kitaev:2005hzj-0506438}, which we will not dive into in this discussion. The $F$-symbols and $R$-symbols obey the so-called pentagon and hexagon equations, which ensure different sequences of local deformation yield consistent amplitudes for all graphs. The quantum circuits that implement the $F$-moves and $R$-moves are given in Sec. \ref{sec:BuildingBlock}.

\subsubsection{Continuum pictures: Fattened lattice and wormhole}\label{sec:wormhole}

In practice, we work with the Fib-SNC on a given trivalent graph/lattice with a qubit on each edge. Conceptually, it is helpful to allow the string nets to extend to the continuum. To describe the Fib-SNC itself, it suffices to remain in 2D space. The way to connect a trivalent graph/lattice to continuum 2D space is to use a {\it fattened lattice picture}, which we explain below. To describe non-trivial anyons on the Fib-SNC, one needs to further extend the fatten lattice picture into a 3D {\it wormhole picture}. This wormhole picture sets the stage for the 3D string-net description of the anyonic states in Sec. \ref{app:anyonic_fusion}.

Let us first discuss the fattened lattice picture that takes Fib-SNC to a given lattice/graph to the 2D continuum space. It is the simplest to think about the fattened version of a (tailed) trivalent lattice that tessellates the 2D plane. The fattening of a general trivalent graph will follow the same idea. When all the edges of the lattice are fattened, one essentially obtains the continuum 2D plane with a puncture in every plaquette of the original lattice. The string nets initially pinned to the edges of the original lattice will be allowed to extend into the fatten lattice or, equivalently, the 2D plane with punctures where the strings cannot enter: 
\begin{equation}\label{eq:fattened_lattice}
	\raisebox{-1.3cm}{\includegraphics[scale=.50]{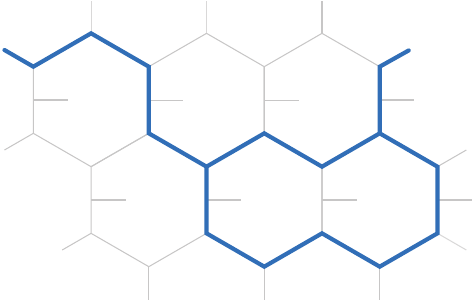}} 
 \xrightarrow{{\rm extend~to~fattened~lattice}}
			\raisebox{-1.3cm}{\includegraphics[scale=.50]{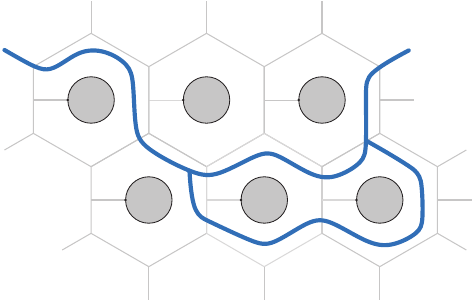}}.
\end{equation}
The topology of each string net configuration remains the same after the extension. In the fatten lattice picture, each tail edge assigns a mark point, labeled by the state of the tail edge, to its corresponding puncture. In the Fib-SNC (without anyons), the tail edges are always in the $|0\rangle$ state. They do not play an essential role in the Fib-SNC wavefunction, and their corresponding mark points are unimportant. However, the tailed edges will be excited when anyons are present \cite{Schotte2022Phys.Rev.X}, and their mark points will become non-trivial as will be shown in Sec. \ref{app:anyonic_fusion}.

The fatten lattice picture provides a compact way to represent the Fib-SNC. For this compact representation, we treat Eq. \eqref{eq:Fmove} and \eqref{eq:tadpole} as local equivalence relations between string-net configurations in the continuum. Under these equivalence relations, all the string net configurations within Fib-SNC can be reduced to a single representative:
\begin{equation}\label{eq:fattenedlattice_groundstate}
    \raisebox{-1.2cm}{\includegraphics[scale=.5]{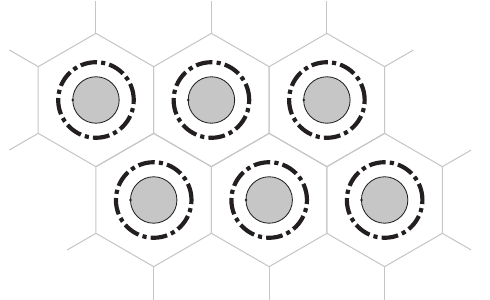}} = \sum_G \Psi(G) \raisebox{-1.3cm}{\includegraphics[scale=.50]{fig/fattenedlattice_1.pdf}}.
\end{equation}
On the right-hand side, $G$ sums over the string-net configurations in Fib-SNC, and $\Psi(G)$ is the wavefunction amplitude of the configuration $G$. Modulo the equivalence relations Eq. \eqref{eq:Fmove} and \eqref{eq:tadpole}, the right-hand side equals the left-hand side with the vacuum loop defined as 
	\begin{equation}\label{eq:vacuum_line_strnet}
		\raisebox{-0.5cm}{\includegraphics[scale=0.5]{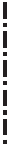}} \quad
		= 
		\frac{1}{\D} \sum_{i=0,1} d_i  \,\,\,
		\raisebox{-0.5 cm}{\includegraphics[scale=0.5]{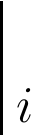}},
	\end{equation}
with ${\cal D} = \sqrt{d_0^2 + d_1^2} = \sqrt{1 + \phi^2}$. Note that a vacuum loop can be obtained from a loop in state $|0\rangle$ using the modular ${\cal S}$ gate:
\begin{align}
    {\cal S} = \frac{1}{\cal D}\left(\begin{array}{cc}
        1 & \phi \\
        \phi & -1
    \end{array}\right).
\end{align}

\begin{figure}
    \centering
    \includegraphics[width=0.9\linewidth]{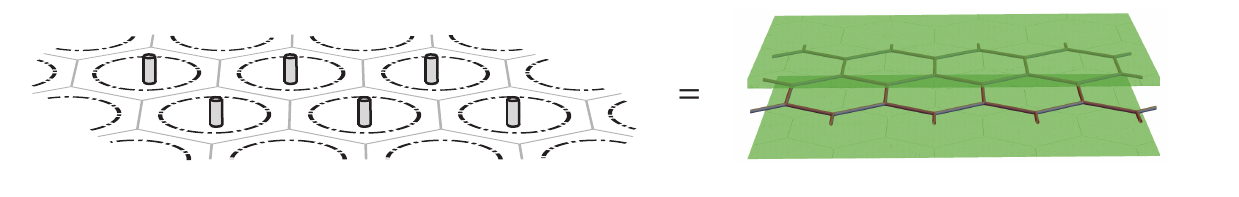}
    \caption{The Fib-SNC is equivalent to the combined vacuum of a time-reversed pair of TQFTs. Each copy of the TQFTs is represented as a green sheet on the right panel. Each grey cylinder on the left panel is a wormhole connecting the two copies. When encircled by the vacuum loop, these wormholes are effectively closed. Hence, the left and right panels are topologically equivalent.}
    \label{fig:FLWH_Translation_Extended}
\end{figure}

The Fib-SNC is equivalent to a topological vacuum state that combines a time-reversed pair of topological quantum field theories (TQFT). A physical picture that highlights this equivalence and also sets the stage to discuss anyonic states can be obtained by further extending the flattened lattice picture in the out-of-plane direction, as shown in Fig. \ref{fig:FLWH_Translation_Extended} where the two parallel sheets on the right panel represent (the vacua of) the two copies of TQFT. This extension leads to the wormhole picture. Each puncture in 2D should be extended into a wormhole (represented by a grey cylinder on the left panel of Fig. \ref{fig:FLWH_Translation_Extended}) that connects the two parallel sheets. However, in the Fib-SNC, each wormhole is effectively pinched off because it is encircled by a vacuum loop and is not connected to any other strings:
\begin{align}
        \raisebox{-1cm}{\includegraphics[scale=0.9]{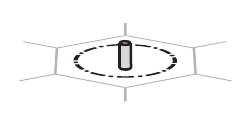}} = \raisebox{-1cm}{\includegraphics[scale=0.9]{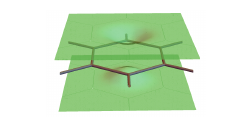}}.
\end{align}
Essentially, this pinched-off wormhole highlights the fact that the Fib-SNC is a topological vacuum state without anyons. When a non-trivial anyon is present in a plaquette, the wormhole will have extra strings attached to it, and it will have to remain open in general since the state is no longer the vacuum state:
\begin{align}
        \raisebox{-1cm}{\includegraphics[scale=0.9]{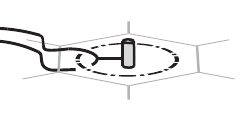}} = \raisebox{-1cm}{\includegraphics[scale=0.9]{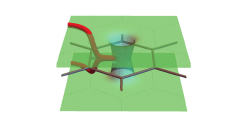}}.
\end{align}
The detailed description of this figure, especially how such a picture describes an anyon in a plaquette, will be presented in the Sec. \ref{app:anyonic_fusion}. Note that for the bi-layer wormhole picture on the right, we have omitted drawing the vacuum loops for simplicity, but one should keep in mind their presence when implicitly representing the exact wave function.

\subsection{Anyonic fusion basis states}\label{app:anyonic_fusion}

In this section, we provide more mathematical details for the wormhole pictures above, which are tightly connected to the anyonic fusion basis states. 

For the extended Levin-Wen model defined on a  tailed lattice \cite{Schotte2022Phys.Rev.X}, one can construct a basis of the string-net subspace $\mathcal{H}_{s.n.}$ called \textit{anyionic fusion basis} \cite{Koenig2010AnnalsofPhysics, Schotte2022Phys.Rev.X}, which is labeled by the fusion states of anyons from the time-reversed pair of Fibonacci TQFTs located on the plaquettes.   In order to connect the string-nets to the anyons, we also specify here that the qubit states $\ket{0}$ and $\ket{1}$ on the edges of the lattice can be interpreted as Fibonacci string labels $\mathbf{1}$ and $\tau$ respectively.  With this correspondence, the branching rules of the Fib-SNC in Fig.~\ref{fig:FRmoves} can be interpreted as the single copy fusion rules Eq.~\eqref{eq:fib_fusion_rule}.

In this work, we do not consider a compact background manifold with a finite genus, so we can focus on the simple case where the anyonic fusion basis states (also called anyonic eigenstates in the main text) are defined on a sphere $S^2$. The anyon charge from this  \textit{doubled Fibonacci TQFT} has two string labels $a$ and $b$ coming from the upper and lower copies of TQFT respectively. In addition, there needs to be a string label $\ell \in \mathcal{C}$ for the tail edge which corresponds to the fusion outcome of $a$ and $b$, i.e., $\ell = a \times b $.  For the doubled Fibonacci TQFT, each anyon charge $\bm{a}$ can be represented by three labels as  $\bm{a} \equiv (ab)_\ell \in \{ \1\1_\1,  \tau\1_\tau, \1\tau_\tau, \tau\tau_\1, \tau\tau_\tau \} $, which are all combinations satisfying the Fibonacci fusion (branching) rule Eq.~\eqref{eq:fib_fusion_rule}.  
Note that the fusion outcome of two $\tau$'s is either $\1$ or $\tau$, which leads to the splitting of the $\tau\tau$ anyon sector  into two sectors $\tau\tau_\1$ and $\tau\tau_\tau$ respectively.  Since the fusion outcome of $\tau$ and $\1$ is determined to be $\tau$, and the fusion outcome of $\1$ and $\1$ is just $\1$,   we can also suppress the subscript in the other three cases as $\1\1_\1 \equiv \1\1$,  $\tau\1_\tau \equiv \tau\1$ and $\1\tau_\tau \equiv \1\tau$.  

Now we introduce the string-net states (also called \textit{ribbon graph} states \cite{Koenig2010AnnalsofPhysics}) to represent such anyonic fusion basis on an $n$-punctured surface $\Sigma_n$.   We will start with constructing a 3D string-net defined on the thickened surface $\Sigma_n \times [-1,1]$ and reduce it to the usual planar string-net on $\Sigma_n$ which remains error-correctable.   We place the marked boundary points on the $n$ punctures in the middle of the thickened surface, i.e., $\Sigma_n \times \{0\}$.  When being far away from the puncture, the upper ribbon (string) with label $a$ stays at the top of the thickened surface $\Sigma_n \times \{1\}$, while the lower ribbon with label $b$ stays at the bottom $\Sigma_n \times \{-1\}$.  These two ribbons fuse into the third ribbon with the tail label $l$ at the middle  $\Sigma_n \times \{0\}$ and is then attached to the puncture at the marked point $p$. In addition, a vacuum loop (defined in Eq.~\eqref{eq:vacuum_line_strnet}) circulates the puncture and the upper and lower ribbons. We can visualize this 3D string-net state near the wormhole as 
	\begin{align}\label{eq:an_fus_basis_closing_off}
		\raisebox{-1.1cm}{\includegraphics[scale=.52]{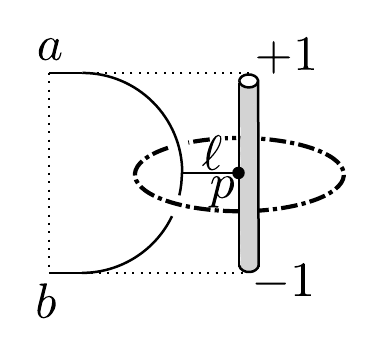}}
		\;\; .
	\end{align}
 
We first consider the simplest situation of the 2-punctured surface $\Sigma_2$ (equivalent to a cylinder/tube), where the anyonic fusion state of  $\mathcal{H}_{\Sigma_2}$ can be expressed as 
\begin{equation}\label{eq:anyonic_fusion_basis_states_2}
		\ket{k, \ell, a, b} = \; \raisebox{-.9cm}{\includegraphics[scale=.5]{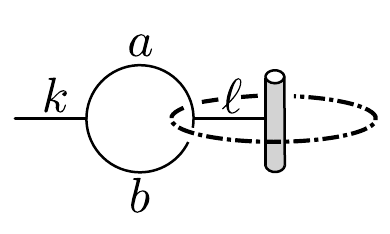}}   \;.
	\end{equation}
The crossing between the vacuum loops and the upper/lower ribbons can be resolved by applying the following relation with the $R$-symbol:
	\begin{align}\label{eq:resolve_crossing}
		\raisebox{-.6cm}{\includegraphics[scale=.40]{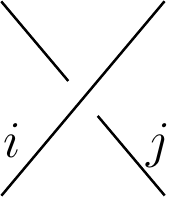}}
		\quad = \sum_{k} \frac{v_k}{v_i v_j} \, R^{ij}_k \,\,\,
		\raisebox{-.6cm}{\includegraphics[scale=.40]{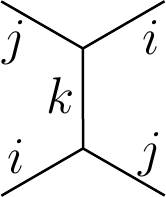}}\;,
	\end{align}
 which can itself be derived from the more elementary relations \eqref{eq:Fmove}, \eqref{eq:tadpole} and \eqref{eq:Rmove}.
By further using the F-move relation Eq.~\eqref{eq:Fmove} and the tadpole relation Eq.~\eqref{eq:tadpole},  one can  turn the 3D string-net into the superposition of planar string-nets as
	\begin{align}\label{eq:doubled_leaf_segment_red}
		\raisebox{-0.9cm}{\includegraphics[scale=.5]{fig/anyonic_fus_bas_state_2.pdf}} \;
		\ = \
		\frac{1}{\D} v_a v_b \sum_{\alpha, \beta} v_\alpha v_\beta \sum_{\gamma, \delta} d_\gamma d_\delta R^{a \alpha}_{\gamma} R^{\alpha b}_{\delta} G_{\alpha a b}^{k \delta \gamma} G_{b \alpha \ell}^{\beta a \delta} G_{a \gamma \delta}^{k \beta \alpha}
		\raisebox{-0.7cm}{\includegraphics[scale=.80]{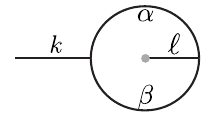}} \; ,
	\end{align} 
where $G^{i j k}_{\lambda \mu \nu} = \frac{1}{v_k v_\nu} F^{i j k}_{\lambda \mu \nu} \,$ and $v_\lambda = \sqrt{d_\lambda}$ (see Ref.~\cite{Schotte2022Phys.Rev.X} for detailed derivation). Note that the edge $\ell$ on the right is nothing but the tail edge in the extended Levin-Wen model \cite{Schotte2022Phys.Rev.X}.


We now embed the anyonic fusion state into the lattice which generalizes the fattened lattice picture in Eqs.~\eqref{eq:fattened_lattice} and \eqref{eq:fattenedlattice_groundstate} to a 3D string-net representation (equivalent to the bi-layer wormhole picture), where the strings can be located above, below or in the middle plane (with coordinate $1$, $-1$ and $0$ respectively).  In particular, we show an example of creations a pair of anyons with charge labels $(a_1 b_1)_{l_1}$ and $(a_2 b_2)_{l_2}$ created out of the topological vacuum, which is related to the experimental demonstration in the Fig. 2 of the main text:
\vspace{-1.3cm}
\begin{align}
		\hspace{-10pt}
		\ket{\vec{\ell}, \vec{a}, \vec{b}} \equiv &
		 \raisebox{-1.8cm}{\includegraphics[scale=.7]{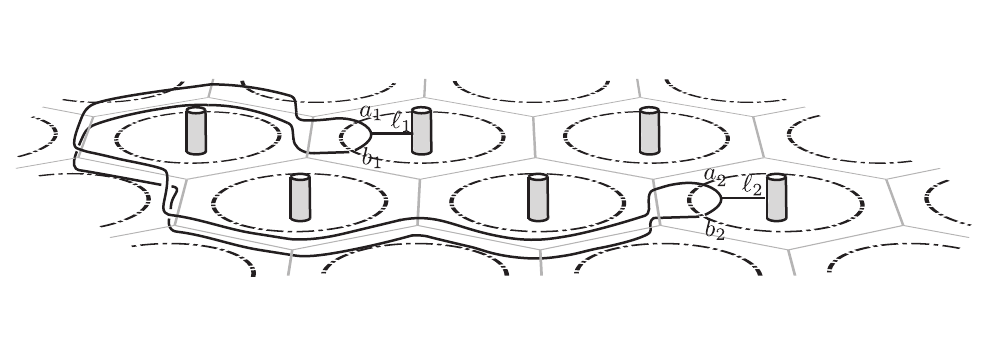}}  \hspace{-.8cm}& \label{eq:embed_into_lattice} \\  
		 =&  \sum_{\vec{\alpha}, \vec{\beta}, k} X^{\vec{a}, \vec{b},  \vec{\ell}}_{\vec{\alpha}, \vec{\beta}, k}   \hspace{.2cm}\raisebox{-1.8cm}{\includegraphics[scale=.4]{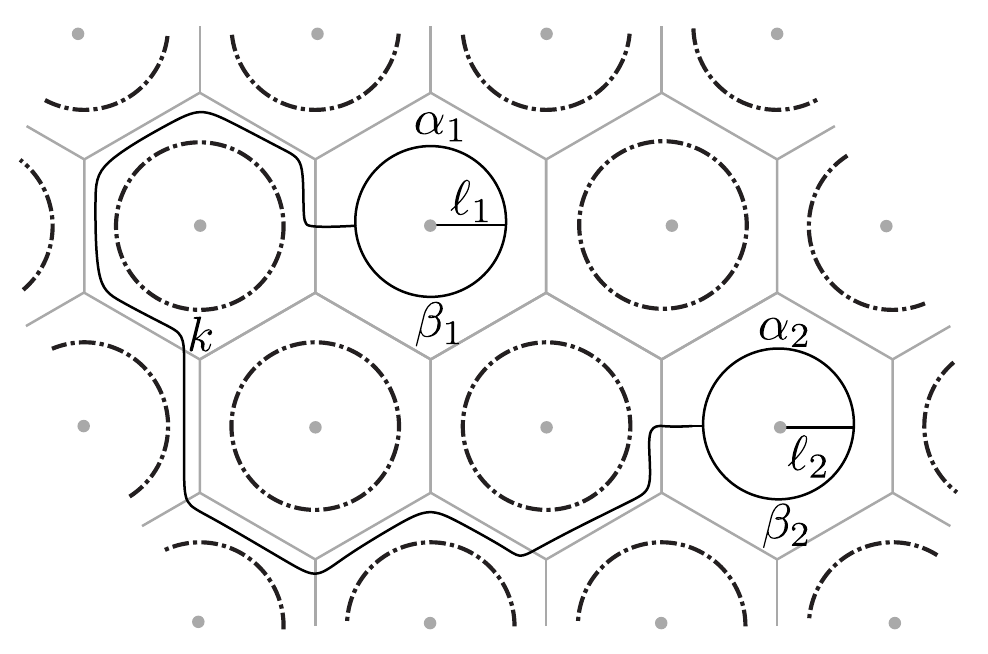}}   \quad ,  \label{eq:embed_into_lattice_tube}
	\end{align}
where the vectors $\vec{\ell}, \vec{a}, \vec{b}$ represent the sets of labels $\{\ell_j\}$, $\{a_j\}$ and $\{b_j\}$  with $j=1,2$. Due to the pair creation out of the vacuum, the strings above the middle plane near each wormhole have the same label: $a_1 =a_2$, and similar for the strings below the middle plane: $b_1 =b_2$. On the other hand, the labels for the tail edges threaded into the womholes $\ell_1$ and $\ell_2$ do not have to be the same in general.
Note that both Eqs.~\eqref{eq:embed_into_lattice} and \eqref{eq:embed_into_lattice_tube} represent the exact wavefunctions of the anyonic fusion states, while the equality shows an equivalence  relation between the 3D string-net state and the superposition of planar string-net states with a tube composed by edges $\alpha_i, \beta_i, \ell_i$ and $k$ inserted in each plaquette with anyonic excitation, where $X^{\vec{a}, \vec{b},  \vec{\ell}}_{\vec{\alpha}, \vec{\beta}, k} $ represent the coefficients and the sum is over the labels $\vec{\alpha}$, $\vec{\beta}$ and $k$.  Here, the vectors $\vec{\alpha}$ and $\vec{\beta}$ also represent the sets of labels $\{\alpha_j\}$ and $\{\beta_j\}$ for $j=1,2$. This equivalence relation and the coefficient $X^{\vec{a}, \vec{b},  \vec{\ell}}_{\vec{\alpha}, \vec{\beta}, k} $  can be derived in a similar manner as Eq.~\eqref{eq:doubled_leaf_segment_red} using the relation Eq.~\eqref{eq:anyonic_fusion_basis_states_2} to resolve the crossing and additional local relations Eq.~\eqref{eq:Fmove} and \eqref{eq:tadpole}.  

Here, we would like to clarify the meaning of Eq. \eqref{eq:embed_into_lattice}. Recall that the lefthand side of Eq. \eqref{eq:fattenedlattice_groundstate} represents the Fib-SNC in the sense that it is equivalent to the superposition of string-net configurations that forms the Fib-SNC modulo the local equivalence relations Eq. \eqref{eq:Fmove} and \eqref{eq:tadpole}. In a similar fashion, 
the righthand side of Eq. \eqref{eq:embed_into_lattice} represents the anyonic state $\ket{\vec{\ell}, \vec{a}, \vec{b}}$ in the sense that it is equivalent to the superposition of string-net configurations that form $\ket{\vec{\ell}, \vec{a}, \vec{b}}$ modulo the equivalence relations Eq. \eqref{eq:Fmove}, \eqref{eq:tadpole} and also  \eqref{eq:Rmove}. Note that we need to include the $R$- and $R^*$-moves as a part of the equivalence relations here because we are working with the 3D string-net representation of the anyonic states.  Recall that these equivalence relations can be used to resolve the crossing of the strings and hence and turn the 3D string-nets into superposition of planar string-nets.

We also note that one can also represent the anyionic fusion basis state on the usual tailed lattice (without additional tubes) by pinning all the strings in Eq.~\eqref{eq:embed_into_lattice_tube} to the edges of the underlying grey lattice as
\begin{align}
		\ket{\vec{\ell}, \vec{a}, \vec{b}}   
		 =  \sum_{\vec{\alpha}, \vec{\beta}, k} X^{\vec{a}, \vec{b},  \vec{\ell}}_{\vec{\alpha}, \vec{\beta}, k}   \hspace{.2cm}\raisebox{-1.8cm}{\includegraphics[scale=.4]{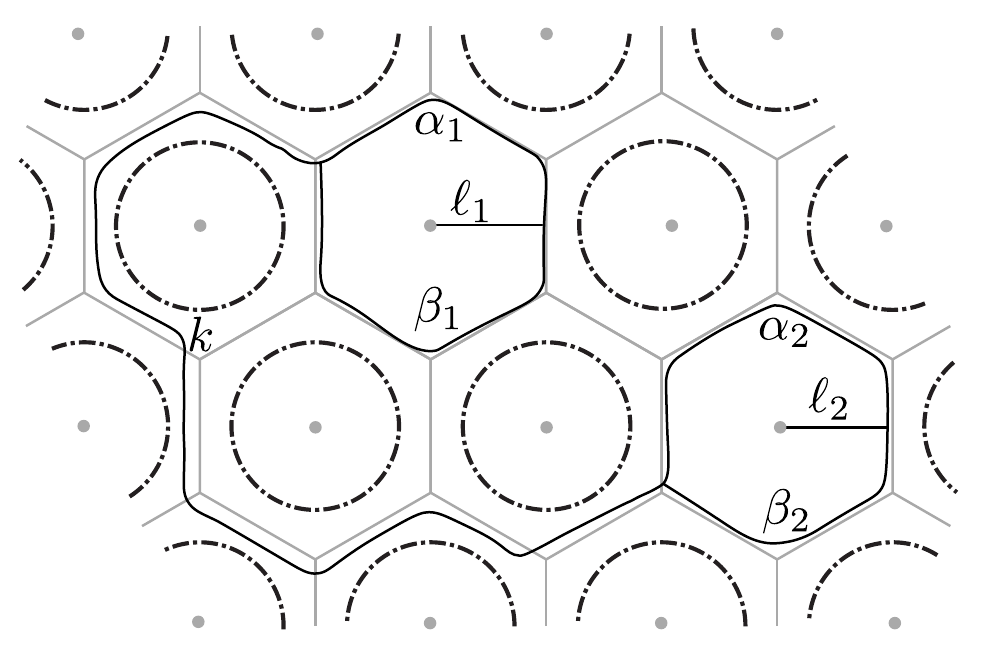}}   \quad .  
	\end{align}
In the above picture, the edges with label $\ell_1$ and $\ell_2$ become the tail edges of the lattice.  Note that one can always move the location of the tails by further applying F-moves as we have done in the experiments in the main text.  In general, one can always deform the usual tailed lattice (grey) into a lattice  with tubes in order to directly measure the anyon charges (see Ref.~\cite{Schotte2022Phys.Rev.X}).

Now we illustrate the 3D string-net states with a specific example of a pair of $\tau \mathbf{1}$ anyons as in the experiment in Fig.2 of the main text by setting $a_1=a_2=\ell_1=\ell_2=\tau$ and $b_1=b_2=\mathbf{1}$:
\begin{align}
\raisebox{-1.8cm}{\includegraphics[scale=.6]{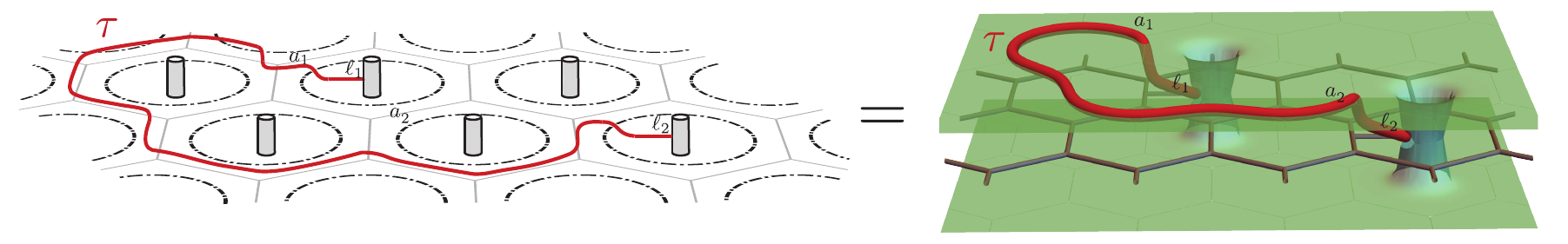}}, 
\end{align}
where a more concise bi-layer wormhole picture is shown on the right of the equality corresponding to the pictures used in the main text.  Note that the non-contractible string with label $\tau$ stays on the upper plane ($z=+1$) most of the time and only goes down to the middle plane ($z=0$) when penetrating to the wormhole.   Note that in the bi-layer wormhole picture on the right, all the vacuum loops are omitted for simplicity.

We now express all the anyonic fusion basis states near a wormhole as a superposition of planar string-nets on a tube as follows: 
\begin{align}
	\ket{\mathbf{1}\mathbf{1}} 		\label{eq:psi11} 
	& \equiv \raisebox{-0.6cm}{\includegraphics[scale=.5]{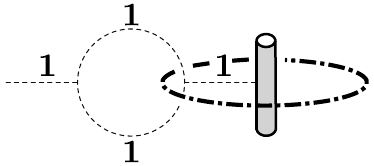}}  =\frac{1}{\D}  \raisebox{-0.48cm}{\includegraphics[scale=.6]{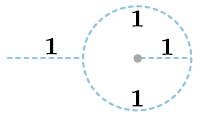}} \, 
	+ \frac{\phi}{\D} \raisebox{-0.48cm}{\includegraphics[scale=.6]{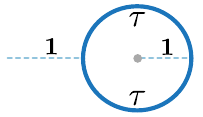}} \,  \\
\nonumber    &= \ket{00}\otimes \frac{1}{\D} \Big(\ket{00} + \phi \ket{11} \Big) \equiv \ket{00} \otimes  \widetilde{\ket{\1\1}},  \\
    \ket{\tau \1} 				\label{eq:psit1}
	& \equiv \raisebox{-0.6cm}{\includegraphics[scale=.5]{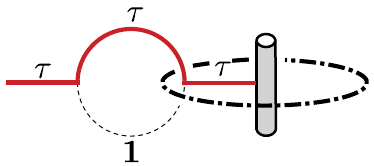}} = \frac{1}{\D}  \raisebox{-0.48cm}{\includegraphics[scale=.6]{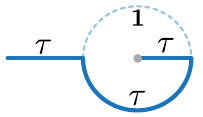}} \,  
	+  \frac{e^{-4 \pi \ii/5}}{\D}  \raisebox{-0.48cm}{\includegraphics[scale=.6]{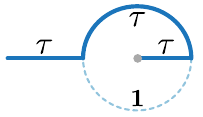}} \,    
	+ \sqrt{\phi} \frac{e^{3\pi \ii/5}}{\D}  \raisebox{-0.48cm}{\includegraphics[scale=.6]{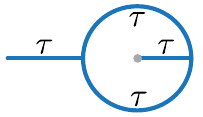}} \,   \\
	& 
	=   \ket{11} \otimes \frac{1}{\D} \Big( \ket{01} + e^{-4 \pi \ii/5} \ket{10}  + \sqrt{\phi} e^{3\pi \ii/5}  \ket{11} \Big)     \nonumber 
	 \equiv  \ket{11} \otimes \widetilde{\ket{\tau \1}}  , \nonumber \\
      \ket{\1\tau} 			\label{eq:psi1t}	
	& \equiv \raisebox{-0.6cm}{\includegraphics[scale=.5]{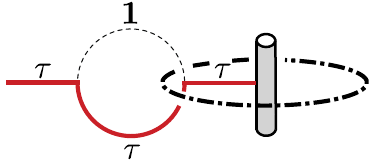}} = \frac{1}{\D}
	 \raisebox{-0.48cm}{\includegraphics[scale=.6]{fig/tube_t1tt.pdf}} \,  
	+ \frac{e^{4 \pi \ii/5}}{\D}  \raisebox{-0.48cm}{\includegraphics[scale=.6]{fig/tube_ttt1.pdf}} \,  
	+ \sqrt{\phi} \frac{e^{-3\pi \ii/5}}{\D}  \raisebox{-0.48cm}{\includegraphics[scale=.6]{fig/tube_tttt.pdf}} \,    \\
	& 
	=   \ket{11} \otimes \frac{1}{\D} \Big( \ket{01} + e^{4 \pi \ii/5} \ket{10}  + \sqrt{\phi} e^{-3\pi \ii/5}  \ket{11} \Big)  \nonumber 
	 \equiv  \ket{11} \otimes \widetilde{\ket{\mathbf{1}\tau}} ,  \nonumber\\   
	\ket{\tau\tau_\1} 			\label{eq:psitt_1}
	& \equiv \raisebox{-0.6cm}{\includegraphics[scale=.5]{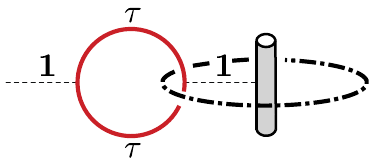}} = \frac{\phi }{\D} \raisebox{-0.48cm}{\includegraphics[scale=.6]{fig/tube_1111.pdf}} \, 
	-  \dfrac{1}{\D}  \raisebox{-0.48cm}{\includegraphics[scale=.6]{fig/tube_1t1t.pdf}} \,   \\
\nonumber	 &=  \ket{00} \otimes \frac{1}{\D} \Big( \phi\ket{00} -  \ket{11}  \Big)  
	\equiv  \ket{00} \otimes 	
	\widetilde{\ket{\tau\tau_\1}}   ,     \\
	\ket{\tau\tau_\tau }			\label{eq:psitt_t}	
	& \equiv \raisebox{-0.6cm}{\includegraphics[scale=.5]{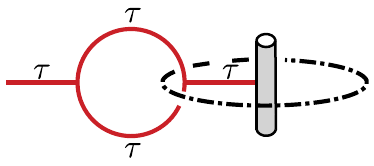}} = \frac{\sqrt{\phi}}{\D}  \raisebox{-0.48cm}{\includegraphics[scale=.6]{fig/tube_t1tt.pdf}} \, 
	+  \frac{\sqrt{\phi}}{\D}  \raisebox{-0.48cm}{\includegraphics[scale=.6]{fig/tube_ttt1.pdf}}  
	+  \frac{1}{\phi \D}   \raisebox{-0.48cm}{\includegraphics[scale=.6]{fig/tube_tttt.pdf}} \,  \\	 
	& =  \ket{11} \otimes \frac{1}{\D} \Big( \sqrt{\phi} \ket{01} + \sqrt{\phi} \ket{10} + \frac{1}{\phi} \ket{11} 
	\Big) \nonumber
	 \equiv   \ket{11} \otimes \widetilde{\ket{\tau\tau_\tau}},  \nonumber
\end{align}
which are closely related to the tube algebra (see Refs.\cite{Koenig2010AnnalsofPhysics, Schotte2022Phys.Rev.X} for details).
 In the above equations, we have also factorized the 4-qubit states into two parts as $\ket{kl} \otimes {\ket{\alpha \beta}}$ according to the labels in Eqs.~\eqref{eq:doubled_leaf_segment_red}. Besides the above five states corresponding to anyonic excitations,  there are two additional states on the tube which does not carry anyonic excitations which correspond to the nilpotents of the tube algebra \cite{Koenig2010AnnalsofPhysics, Schotte2022Phys.Rev.X}:
\begin{align}
    \ket{\tau\tau_{\1,\tau}} 		 \label{eq:psitt_1t}
    & \equiv \raisebox{-0.6cm}{\includegraphics[scale=.5]{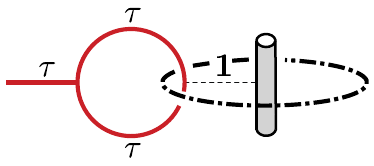}}=   \raisebox{-0.48cm}{\includegraphics[scale=.6]{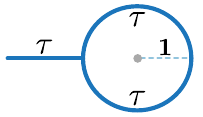}} \,  
     =  \ket{1011} ,  
    \\
    \ket{\tau\tau_{\tau,\1}}  	 \label{eq:psitt_t1}
    & \equiv \raisebox{-0.6cm}{\includegraphics[scale=.5]{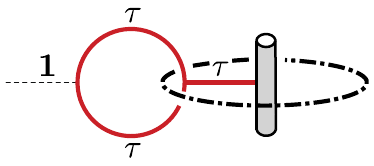}} =   \raisebox{-0.48cm}{\includegraphics[scale=.6]{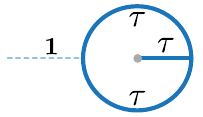}} \, 
    =    \ket{0111}   
\end{align}
Note that the above seven states span the 7-dimensional subspace satisfying the fusion (branching) rules on the tube ($\Sigma_2$).

\section{Anyon charge measurement protocol}
In this section, we discuss the details of the anyon charge measurement protocol shown in Fig.~2 of the main text. 
The measurement protocol starts with the following graph configuration:
\begin{equation}
\centering \includegraphics[scale=0.8]{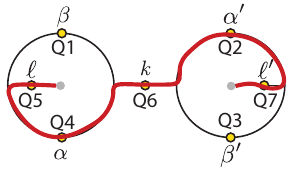}, 
\end{equation}
which is equivalent to two connected tubes overlapping at qubit Q6 with state label $k$ (the red ribbon is suppressed for clarity).   We need to measure the anyon charges created on the two plaquettes and confirm that they are indeed the same as we expected.     Note that the tail edge $l$ and $l'$ (qubit Q6 and Q7) connect to the punctures (grey circle), which trap the anyon and are equivalent to the wormholes discussed in Sec.~\ref{sec:wormhole}.   This planar graph configuration can be related to the anyonic fusion basis states in the 3D string-net picture or, equivalently the bi-layer wormhole picture using the relation in Eq.~\eqref{eq:embed_into_lattice_tube}:
\begin{equation}\label{eq:connected_tubes_3D}
\centering \includegraphics[scale=1]{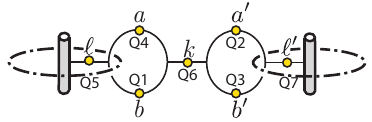}.
\end{equation}
As has been pointed out in Sec.~\ref{app:anyonic_fusion}, there are 5 types of doubled anyon charges: $\mathbf{a} $$=$$\mathbf{11}$ (``vacuum"), $\tau \mathbf{1}$, $\mathbf{1} \tau$, $\tau \tau_\mathbf{1}$ and $\tau\tau_\tau$, with the corresponding anyonic eigenstate being graphically represented in Eqs.~(\ref{eq:psi11}-\ref{eq:psitt_t}).    Since the tail edge $l$ and $l'$ (qubit Q5 and Q7)  are threaded by a $\tau$-ribbon in our case, we know that the possible anyon charge can only be $\tau \mathbf{1}\equiv \tau \mathbf{1}_\tau$, $\mathbf{1} \tau \equiv \mathbf{1} \tau_\tau$ or $\tau \tau_\tau$, where the first two anyon charges are rewritten to indicate the total charge $\tau$ on the middle ribbon resulting from the fusion of $\tau$ and $\mathbf{1}$.  In addition, the edge $k$ (qubit Q6) is also passed by a ribbon in our case.   We can hence declare that edge $k$, $l$ and $l'$ (qubit Q6, Q5, Q7) in the tubes are all in the $\ket{1}$ state (corresponding to a $\tau$ label), which is verified by projective measurement on the diagonal basis (see main text).  The corresponding anyon eigenstates on the left and right tubes can hence be factorized to $\ket{\mathbf{a}}=\ket{1,1}_{k,l}\otimes \widetilde{\ket{\mathbf{a}}}_{\alpha, \beta}$ and $\ket{\mathbf{a}'}=\ket{1,1}_{k,l}\otimes \widetilde{\ket{\mathbf{a}'}}_{\alpha', \beta'}$ respectively, where $\widetilde{\ket{\mathbf{a}}}_{\alpha, \beta}$ ($\widetilde{\ket{\mathbf{a}'}}_{\alpha', \beta'}$) represents the anyon eigenstate restricted to edge $\alpha$ and $\beta$ ($\alpha'$ and $\beta'$) and is a superposition state in the diagonal basis as can be seen from Eqs.~\eqref{eq:psit1}, \eqref{eq:psi1t}, \eqref{eq:psitt_t} and listed as below for the left tube: 
\begin{align}
\widetilde{\ket{\tau\mathbf{1}}}_{\alpha, \beta}=&   \frac{1}{\D} \Big( \ket{01}_{\alpha,\beta} + e^{-4 \pi \ii/5} \ket{10}_{\alpha, \beta}  + \sqrt{\phi} e^{3\pi \ii/5}  \ket{11}_{\alpha, \beta} \Big),  \\
\widetilde{\ket{\mathbf{1}\tau}}_{\alpha, \beta}=&   \frac{1}{\D} \Big( \ket{01}_{\alpha,\beta} + e^{4 \pi \ii/5} \ket{10}_{\alpha, \beta}  + \sqrt{\phi} e^{-3\pi \ii/5}  \ket{11}_{\alpha, \beta} \Big), \\
\widetilde{\ket{\tau\tau_\tau}}_{\alpha, \beta} =&  \frac{1}{\D} \Big( \sqrt{\phi} \ket{01}_{\alpha,\beta} + \sqrt{\phi} \ket{10}_{\alpha,\beta} + \frac{1}{\phi} \ket{11}_{\alpha,\beta} 
	\Big), 
\end{align}
where $\D=\sqrt{1+\phi^2}$.
From the above expressions, we can predict that when performing projective measurements on edge $\alpha$ and $\beta$ (qubit Q4 and Q1)  of both $\widetilde{\ket{\tau\mathbf{1}}}_{\alpha, \beta}$ and $\widetilde{\ket{\mathbf{1}\tau}}_{\alpha, \beta}$ states, the probabilities of observing the $\ket{1}$ state are 
\begin{equation}
P_1^{(\alpha)}=P_1^{(\beta)} = \frac{1+\phi}{2+\phi} = \frac{\phi^2}{\phi^2+1} \approx 0.724. 
\end{equation}
The experiment presented in the main text (Fig. 2) confirms this prediction.   Note that the projective measurement directly in the diagonal basis cannot tell the difference between $\widetilde{\ket{\tau\mathbf{1}}}_{\alpha, \beta}$ and $\widetilde{\ket{\mathbf{1}\tau}}_{\alpha, \beta}$, since they only differ by the phase factor in the diagonal basis.

The 4-dimensional Hilbert space on edge $\alpha$ and $\beta$ (qubit Q1 and Q4) of the left tube is spanned by the three orthorgonal anyonic eigenstates $\widetilde{\ket{\tau\mathbf{1}}}_{\alpha, \beta}$, $\widetilde{\ket{\mathbf{1}\tau}}_{\alpha, \beta}$ and $\widetilde{\ket{\tau\tau_\tau}}_{\alpha, \beta}$, as well as the diagonal state $\ket{00}_{\alpha,\beta}$ which violates the vertex fusion (branching) rule, and similar for the right tube.  In order to facilitate the anyon charge measurement, we apply the following 2-qubit unitary on edges with state labels $\alpha$ and $\beta$ on the left tube:
\begin{align}\label{eq:U_transformation}
U=&\ket{10}_{a,b}\widetilde{\bra{\tau\mathbf{1}}}_{\alpha,\beta}+\ket{01}_{a,b}\widetilde{\bra{\mathbf{1}\tau}}_{\alpha, \beta}+\ket{11}_{a,b}\widetilde{\bra{\tau\tau_\tau}}_{\alpha, \beta}+\ket{00}_{a,b}\bra{00}_{\alpha, \beta} \\
=& \frac{1}{\sqrt{\phi^2+1}}\begin{pmatrix}
    \sqrt{\phi^2+1} & 0 & 0 & 0 \\
    0 & 1 &  e^{-4\pi\ii/5 } & \sqrt{\phi} e^{3\pi\ii/5 }\\
    0 & 1 & e^{4\pi\ii/5 } & \sqrt{\phi} e^{-3\pi\ii/5 } \\
    0 & \sqrt{\phi} & \sqrt{\phi} & \frac{1}{\phi}
    \end{pmatrix},
\end{align}
which performs a basis transformation to the diagonal basis $\{\ket{10}_{a,b}, \ket{01}_{a,b}, \ket{11}_{a,b}, \ket{00}_{a,b}\}$, namely
\begin{equation}\label{eq:basis_transform}
    U :\widetilde{\ket{\tau\mathbf{1}}}_{\alpha, \beta} \mapsto \ket{10}_{a,b},  \quad \widetilde{\ket{\mathbf{1}\tau}}_{\alpha, \beta} \mapsto \ket{01}_{a,b},  \quad \widetilde{\ket{\tau\tau_\tau}}_{\alpha, \beta} \mapsto \ket{11}_{a,b},  \quad \ket{00}_{\alpha, \beta} \mapsto \ket{00}_{a,b}.
\end{equation}
Note that these diagonal basis states are nothing but the anyonic fusion basis states in the 3D string-net picture presented in Eq.~\eqref{eq:connected_tubes_3D}, where the state labels $\alpha, \beta$ on qubit Q1 and Q4  are transformed to the state labels $a$ and $b$. One then applies an identical unitary for the right tube, which can be obtained by replacing the labels ($\alpha, \beta, a,b$) in Eq.~\eqref{eq:U_transformation} with ($\alpha', \beta', a',b'$).  We graphically  represent this basis transformation from the planar tube basis to the 3D string-net basis as:
\begin{equation}\label{eq:U_basis_transform}
   U \otimes U : \left\{\raisebox{-1.0cm}{\includegraphics[scale=0.8]{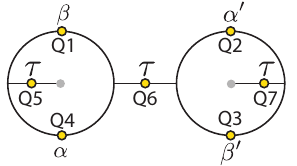}} \right\}  \longmapsto \left\{\raisebox{-0.9cm}{\includegraphics[scale=1]{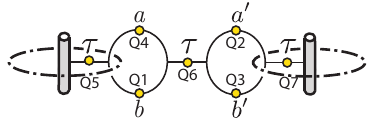}} \right\}
\end{equation}
The graphic representation of the basis transformation in Eq.~\eqref{eq:basis_transform} for the right tube is shown below:
\begin{align}
U:  \frac{1}{\D} \left( \raisebox{-0.48cm}{\includegraphics[scale=.6]{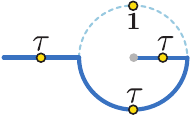}} 
	+  e^{-4 \pi \ii/5} \raisebox{-0.48cm}{\includegraphics[scale=.6]{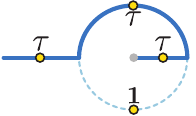}}  
	+ \sqrt{\phi} e^{3\pi \ii/5}  \raisebox{-0.48cm}{\includegraphics[scale=.6]{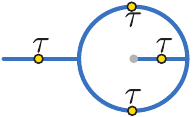}} \right) &\longmapsto \raisebox{-0.6cm}{\includegraphics[scale=.5]{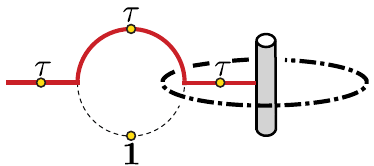}} \quad , \label{eq:basis_transform1} \\
  \frac{1}{\D} \left( \raisebox{-0.48cm}{\includegraphics[scale=.6]{fig/tube_right1.pdf}} 
	+  e^{4 \pi \ii/5} \raisebox{-0.48cm}{\includegraphics[scale=.6]{fig/tube_right2.pdf}}  
	+ \sqrt{\phi} e^{-3\pi \ii/5}  \raisebox{-0.48cm}{\includegraphics[scale=.6]{fig/tube_right3.pdf}} \right) &\longmapsto \raisebox{-0.6cm}{\includegraphics[scale=.5]{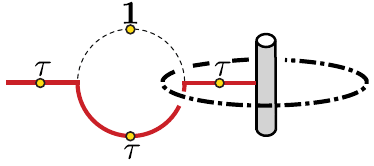}} \quad , \label{eq:basis_transform2} \\
  \frac{1}{\D} \left( \sqrt{\phi} \raisebox{-0.48cm}{\includegraphics[scale=.6]{fig/tube_right1.pdf}} 
	+  \sqrt{\phi} \raisebox{-0.48cm}{\includegraphics[scale=.6]{fig/tube_right2.pdf}}  
	+ \frac{1}{\phi}  \raisebox{-0.48cm}{\includegraphics[scale=.6]{fig/tube_right3.pdf}} \right) &\longmapsto \raisebox{-0.6cm}{\includegraphics[scale=.5]{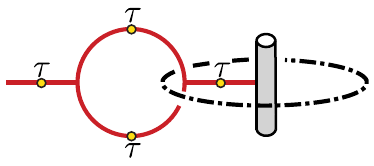}} \quad , \label{eq:basis_transform3} \\
  \raisebox{-0.48cm}{\includegraphics[scale=.6]{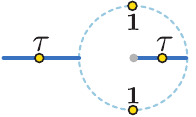}} \quad &\longmapsto \raisebox{-0.6cm}{\includegraphics[scale=.5]{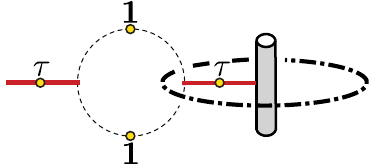}}.   \label{eq:basis_transform4}
\end{align}
Note that the states on both sides of Eq.~\eqref{eq:basis_transform4} violate the fusion (branching) rule and are hence outside the string-net subspace. Nevertheless, the basis transformation $U$ is trivial (identity) outside the string-net space. A similar expression exists for the left tube up to a left-right reflection.  The 3D string-net pictures on the right side of Eqs.~\eqref{eq:basis_transform1} and \eqref{eq:basis_transform2} is equivalent to the bi-layer wormhole pictures shown in Fig.2(f,i) in the main text.   

Here, we comment on the subtle difference between Eq.~\eqref{eq:U_basis_transform} and Eq.~\eqref{eq:embed_into_lattice_tube}.  The latter uses a ``$=$" to indicate the equivalence of the same quantum state, while the former uses ``$\longmapsto$" to indicate a linear map, or equivalently the unitary transformation, between different quantum states.  The subtlety lies in the fact that the qubits are located in the same place, i.e., the underlying grey lattice, on both the left and right sides of the equality in  Eq.~\eqref{eq:embed_into_lattice_tube}. The two sides of the equation Eq.~\eqref{eq:embed_into_lattice_tube} are hence equivalent string-net representations (3D and 2D, respectively) for the same quantum state.  On the other hand,  the qubits Q1, Q2, Q3 and Q4 are only on residing on the 2D plane on the left side of Eq.~\eqref{eq:U_basis_transform}, and are moved above and below the middle plan on the right side of the equality, which suggests Eq.~\eqref{eq:U_basis_transform} should be interpreted as a linear map between two different Hilbert spaces defined on a planar qubit lattice and a 3D qubit lattice respectively.  This same subtlety also appears in Eq.~\eqref{eq:basis_transform1} to Eq.~\eqref{eq:basis_transform4} which uses ``$\mapsto$" to indicate linear map between different states and Eq.~\eqref{eq:psi11} to Eq.~\eqref{eq:psitt_t1} which uses ``$=$" to indicate equivalence relation between the same state.

Now one can perform a projective measurement in the diagonal basis on qubits Q1 and Q4 (Q2 and Q3) of the left tube (right tube) and the measurement results in $\ket{10}_{a,b},$
$\ket{01}_{a,b}$, $\ket{11}_{a,b}$ \big( $\ket{10}_{a',b'},$
$\ket{01}_{a',b'}$, $\ket{11}_{a',b'}$\big) correspond to the anyon charge $\tau \mathbf{1}$, $\mathbf{1} \tau$ and $\tau\tau_\tau$ respectively, while only the former two are expected in our experiment presented in the main text (Fig. 2).

\section{Chromatic polynomials and string-net sampling}

\subsection{Introduction of the chromatic polynomials}
The chromatic $\chi(G, k)$ is a graph polynomial polynomial of variable $k$ defined on graph $G=(E,V)$ \cite{Read1968JournalofCombinatorialTheory}. For positive integer $k \in \mathbb{N}^+$, the chromatic polynomial $\chi(G, k)$  counts the number of proper vertex colorings with $k$ colors on a graph $G$.  Here, proper coloring means the two vertices connected by an edge cannot have the same color.  As an example, for a triangle $K_3$,  one has $\chi(K_3, k)=k(k-1)(k-2)$, which is a polynomial for the variable $k$. 

There exists a local recurrence relation called the \textit{deletion-contraction relation} for the chromatic polynomial \cite{Read1968JournalofCombinatorialTheory}:
\begin{equation}\label{eq:deletion-contraction}
    \chi(G, k) = \chi(G-uv, k) - \chi(G/uv, k), 
\end{equation}
which can also be represented graphically as
\begin{equation}\label{eq:deletion-contraction_graph}
		\raisebox{-0.3cm}{\includegraphics[scale=1]{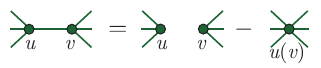}}. 
	\end{equation}
The above relation states that the number of colorings of a graph $G$ with a specified edge $uv$ is the difference between the number of colorings of $G-uv$ (the graph $G$ with the edge $uv$ being deleted) and $G/uv$ (G with the two vertices $u$ and $v$ being contracted to a point).  This relation follows the simple logical identity: Different = All $-$ Same, i.e., the number of colorings when the two vertices on $u$ and $v$ are different equals the number with no constraint on $u$ and $v$ subtracting the number where $u$ and $v$ have the same color.  

Now one can also extend the definition of the chromatic polynomial $\chi(G, k)$ from integer $k$ to $k \in \mathbb{C}$ on the entire complex plane $\mathbb{C}$  through the recurrence relation Eq.~\eqref{eq:deletion-contraction} and the identity  $\chi(G, k)= k^{|V|}$ for the edgeless graph with $|V|$ vertices.\footnote{For integer $k$,  this identity holds since each vertex can have $k$ different colors and do not interfere with each other due to the absence of edges in this graph.  For non-integer $k$, we assume this identity still holds in order to define the chromatic polynomial.}

The chromatic polynomial is a topological invariant, or more specifically, a graph invariant, which means isomorphic (topologically equivalent) graphs must have the same chromatic polynomial. Nevertheless, non-isomorphic (topologically inequivalent) graphs can still have the same chromatic polynomial. Two graphs $G_1$ and $G_2$ are said to be \textit{chromatically equivalent} if their chromatic polynomials are the same, i.e., $\chi(G_1, k) = \chi(G_2, k)$ for any $k \in \mathbb{C}$.

\subsection{Evaluation of the string-net wavefunctions via the chromatic polynomials}

The Fib-SNC ground state can be expressed as a condensation (superposition) of all closed-string network configurations (also called string diagrams) satisfying the branching (fusion) rule in Fig.~\ref{fig:FRmoves}(a):
\begin{equation}\label{eq:string-net_wavefunction}
\ket{\Psi}= \Psi(G_1)\raisebox{-1.23cm}{\includegraphics[scale=0.7]{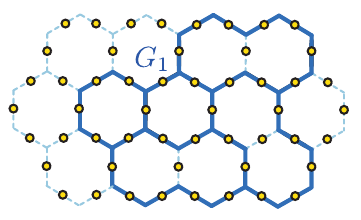}}  + \Psi(G_2)\raisebox{-1.23cm}{\includegraphics[scale=0.7] {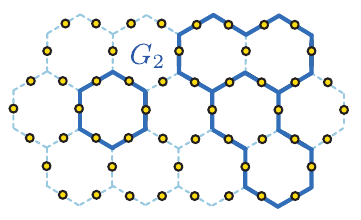}}+\Psi(G_3)\raisebox{-1.23cm}{\includegraphics[scale=0.7]{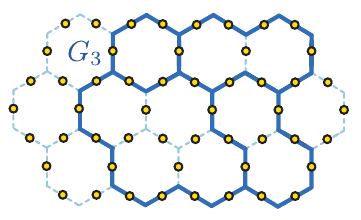}}+\cdots.
\end{equation}
The amplitude in front of each string diagram associated with the trivalent subgraph $G_i$ is denoted by $\Psi(G_i)$, which is the string-net wavefunction amplitude in the diagonal (bit-string) basis. 

\begin{figure}[t]
\centering
\includegraphics[width=0.7\linewidth]{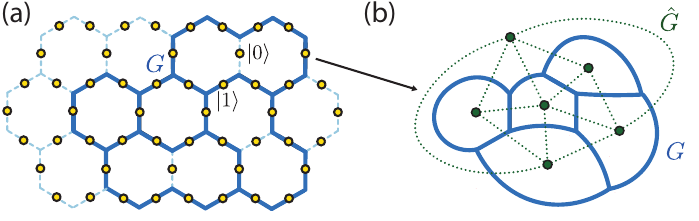}
\caption{(a) A string diagram corresponding to the subgraph $G$. (b) The corresponding dual graph $\hat{G}$.}
\label{fig:dual_chromatic_polynomial}
\end{figure}

We now consider the evaluation of the string-net wavefunction amplitude $\Psi(G)$ on a sphere for a particular string diagram $G$, as illustrated in Fig.~\ref{fig:dual_chromatic_polynomial}a.  We also define the relative wavefunction amplitude $\widetilde{\Psi}(G) = \Psi(G)/\Psi(\text{vac})$, where $\Psi(\text{vac})$ is the amplitude for the empty configuration (``vacuum"). We call the calculation of $\widetilde{\Psi}(G)$ an evaluation of the diagram $G$. 

As an example, the simplest non-trivial graph $G$ is a loop diagram $\bigcirc$.  We evaluate the loop diagram (with label~$\tau$) through the following relation for the Fibonacci string-net [a special case of Eq.~\eqref{eq:tadpole}]:
\begin{equation}\label{eq:loop}
\raisebox{-0.5cm}{\includegraphics[scale=0.8]{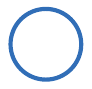}} = d_\tau=\phi,
\end{equation}
where $d_\tau$ represents the quantum dimension for the Fibonacci category.
We can re-express the above relation as $\widetilde{\Psi}(\bigcirc)=\Psi(\bigcirc)/\Psi(\text{vac})=\phi$.

It has been shown in Refs.~\onlinecite{Fidkowski2009Commun.Math.Phys., Fendley2005Phys.Rev.B, Fendley2009Geom.Topol., Fendley2010Adv.Theor.Math.Phys.} 
that the Fib-SNC wavefunction can be related to the chromatic polynomial as
\begin{equation}\label{eq:chromatic2}
    \widetilde{\Psi}(G)^2=\frac{\Psi(G)^2}{\Psi(\text{vac})^2}=\frac{1}{\phi+2}\chi(\hat{G},\phi+2),
\end{equation}
where $\hat{G}$ is the dual graph (triangulation) of $G$ on a sphere obtained by interchanging the faces (plaquettes) and vertices.  One should caution that on a sphere, there is an extra face (plaquette) outside the graph $G$ corresponding to a vertex on the dual graph $\hat{G}$.
Note that the Fibonacci string-net wavefunction $ \widetilde{\Psi}(G)$ is real-valued.  Therefore, the square of the wavefunction is just the probability distribution $\Psi(G)^2=|\Psi(G)|^2=P(G)$, which can then be measured from sampling the distribution of bitstrings where the 1's are supported on the graph $G$.   

For the simplest example of the loop diagram $G=\bigcirc$, we can use Eq.~\eqref{eq:loop} to compute the relative sampling probability with respect to the empty configuration as $P(\bigcirc)/P(\text{vac}) = \Psi(\bigcirc)^2/ \Psi(\text{vac})^2 = \phi^2$. Alternatively, we can compute the chromatic polynomial on the dual graph $\hat{G}$ on a sphere: 
\begin{equation}
\raisebox{-0.5cm}{\includegraphics[scale=0.8]{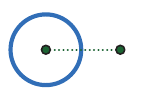}}.
\end{equation}
One can calculate the chromatic polynomial $\chi(\hat{G},\phi+2)$ defined on $\hat{G}$ using the deletion-contraction relation Eq.~\eqref{eq:deletion-contraction} or \eqref{eq:deletion-contraction_graph}: 
\begin{equation}\label{eq:minimal_chromatic}
\chi(\hat{G}, k=\phi+2)=\raisebox{-0.3cm}{\includegraphics[scale=0.8]{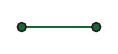}} \equiv  \raisebox{-0.3cm}{\includegraphics[scale=0.8]{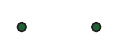}} - \raisebox{-0.3cm}{\includegraphics[scale=0.8]{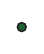}}=k^2-k=(\phi+2)^2-(\phi+2),
\end{equation}
where we have used the fact that $\chi(G',k)=k^n$ for edgeless graph $G'$ with $n$ vertices.
Together with Eq.~\eqref{eq:chromatic2}, we then obtain
\begin{equation}\label{eq:loop_probability}
\frac{P(\bigcirc)}{P(\text{vac})}=\frac{\Psi(\bigcirc)^2}{\Psi(\text{vac})^2}=\frac{1}{\phi+2}\chi(\raisebox{-0.2cm}{\includegraphics[scale=0.6]{fig/one_edge.pdf}},\phi+2)= \phi+1=\phi^2,
\end{equation}
which is consistent with the result we have obtained above from directly evaluating the loop diagram $\bigcirc$ using Eq.~\eqref{eq:loop}.

\subsection{Classical algorithm and complexity for the string-net wavefunctions}\label{sec:direct_evaluation}

We now present a classical algorithm to evaluate the string diagram and, hence, the string-net wavefunction. 

We start with the evaluation of a simple example of the smallest string-net discussed in the main text:
\begin{align}\label{eq:theta_evaluation_general}
\nonumber\raisebox{-0.6cm}{\includegraphics[scale=0.8]{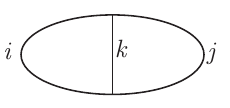}}=&\sum_l F^{ijk}_{jil}\raisebox{-0.6cm}{\includegraphics[scale=0.8]{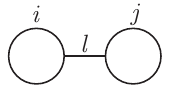}}  \\
\nonumber =& F^{ijk}_{ji\textbf{1}}\raisebox{-0.6cm}{\includegraphics[scale=0.8]{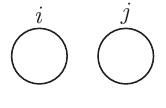}} \\
=& F^{ijk}_{ji\textbf{1}}d_i d_j = \sqrt{\frac{d_k}{d_i d_j}}\delta_{ijk} d_i d_j=\sqrt{d_i d_j d_k}\delta_{ijk},
\end{align}
where $i,j,k,l=\mathbf{1} \ \text{or} \ \tau$ for the Fibonacci string-net.
The first equality corresponds to an $F$-move, which in general creates a superposition state that the flipped edge $l$ in state $\mathbf{1}$ or $\tau$.   However, for $l=\tau$, a tadpole diagram is created, which is forbidden in the ground state, and so the coefficient for that term $F^{ijk}_{jk\tau}$ must be 0.  We, hence, only get two disconnected loops in the second equality. For the third equality, we have used the identity that the evaluation of the loop diagram equals its quantum dimension:    
\begin{equation}
    \raisebox{-0.4cm}{\includegraphics[scale=0.8]{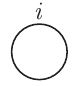}} = d_i,
\end{equation}
which generalizes Eq.~\eqref{eq:loop} and is a special case of Eq.~\eqref{eq:tadpole}.  For the Fibonacci string-net, one has $d_\mathbf{1}=1$ and $d_\tau=\phi$.  The forth equality has used the identity $F^{ijk}_{ji\textbf{1}}=\sqrt{\frac{d_k}{d_i d_j}}\delta_{ijk}$ coming from the F-symbol data, where $\delta_{ijk}$ encodes the fusion-rule data.   Now by setting the labels at $i=j=k=\tau$, we evaluate the $\theta$-diagram where all the edges are occupied by the $\tau$ strings:
\begin{equation}\label{eq:theta_evaluation}
\raisebox{-0.6cm}{\includegraphics[scale=0.8]{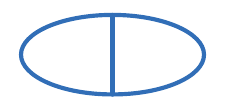}}=F^{\tau\tau\tau}_{\tau\tau\mathbf{1}} \raisebox{-0.5cm}{\includegraphics[scale=0.8]{fig/loop.pdf}} \quad \raisebox{-0.5cm}{\includegraphics[scale=0.8]{fig/loop.pdf}}=\sqrt{d_\tau^3}=\phi^{3/2}.
\end{equation}
We hence get its relative probability as $P(G)/P(
\text{vac})= \phi^3.
$

As we can see, such a classical algorithm for evaluating a generic string diagram has exponential complexity (in the worst case) scaled with the number of vertices, i.e., $\sim 2^{O(|V|)}$.  This is because whenever one applies an $F$-move on two vertices to deform the diagrams, a superposition of two string diagrams appears in the worst case.   One, hence, needs to sum over exponentially many string diagrams in the worst case.    As we will see in the next sub-section, this exponential complexity coincides with the exponential complexity of exactly evaluating the corresponding chromatic polynomial.

\subsection{Classical algorithm and complexity for the chromatic polynomials}

Due to the connection between the square of string-net wavefunction amplitudes and the chromatic polynomial in Eq.~\eqref{eq:chromatic2}, one can also compute the string-net probability distribution $P(G)$ using a classical algorithm for evaluating the chromatic polynomial.

The only existing classical algorithm for exactly evaluating the chromatic polynomial on a generic graph is the \textit{deletion-contraction algorithm}  based on recursively applying the deletion-contraction relation Eq.~\eqref{eq:deletion-contraction} or \eqref{eq:deletion-contraction_graph} until all the polynomials in the sum are defined on edgeless graphs.  

As a simple example, we consider evaluating the relative probability of the theta diagram via the chromatic polynomial defined on the dual graph $\hat{G}$ as:
\begin{equation}
    \raisebox{-0.4cm}{\includegraphics[scale=0.8]{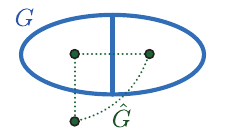}}.
\end{equation}
We now apply the deletion-contraction algorithm to evaluate the chromatic polynomial on $\hat{G}$:
\begin{align}\label{eq:deletion-contraction_algorithm}
\nonumber \chi(\hat{G}, k) \equiv  \raisebox{-0.5cm}{\includegraphics[scale=0.8]{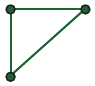}} \ =&  \ \raisebox{-0.5cm}{\includegraphics[scale=0.8]{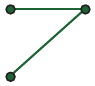}} \  - \  \raisebox{-0.3cm}{\includegraphics[scale=0.8]{fig/one_edge.pdf}} \\
\nonumber = & \ \raisebox{-0.5cm}{\includegraphics[scale=0.8]{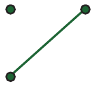}} \  - \quad  \raisebox{-0.5cm}{\includegraphics[scale=0.8]{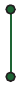}} \quad + \   \raisebox{-0.3cm}{\includegraphics[scale=0.8]{fig/one_edge_deleted.pdf}} \ - \ \raisebox{-0.3cm}{\includegraphics[scale=0.8]{fig/one_vertex.pdf}} \\
\nonumber =& \ \raisebox{-0.5cm}{\includegraphics[scale=0.8]{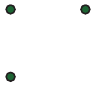}} \ - \ \raisebox{-0.3cm}{\includegraphics[scale=0.8]{fig/one_edge_deleted.pdf}} \ - \quad  \raisebox{-0.5cm}{\includegraphics[scale=0.8]{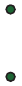}} \quad + \  \raisebox{-0.3cm}{\includegraphics[scale=0.8]{fig/one_vertex.pdf}} - \ \raisebox{-0.3cm}{\includegraphics[scale=0.8]{fig/one_edge_deleted.pdf}} \ + \ \raisebox{-0.3cm}{\includegraphics[scale=0.8]{fig/one_vertex.pdf}} \\
\nonumber =& \ \raisebox{-0.5cm}{\includegraphics[scale=0.8]{fig/triangle_edgeless.pdf}} \ - \ 3\raisebox{-0.3cm}{\includegraphics[scale=0.8]{fig/one_edge_deleted.pdf}} \ + \ 2\raisebox{-0.3cm}{\includegraphics[scale=0.8]{fig/one_vertex.pdf}} \\
=& k^3 -3 k^2 + 2k,
\end{align}
where in the last equality, we have used the formula that the chromatic polynomial for an edgeless graph $G=(V, E)$ with $|V|$ vertices is $\chi(G, k)= k^{|V|}$.  We can hence obtain the relative probability:
\begin{equation}
    \frac{P(G)}{P(\text{vac})}=  \frac{\Psi(G)^2}{\Psi(\text{vac})^2}=\frac{1}{\phi+2}\chi(\hat{G},\phi+2)=\frac{1}{\phi+2}((\phi+2)^3-3(\phi+2)^2+2(\phi+2)) = \phi^3, 
\end{equation}
which is consistent with the direct evaluation of the $\theta$-diagram using $F$-moves shown in Eq.~\eqref{eq:theta_evaluation}
in Sec.~\ref{sec:direct_evaluation}.

As we can conclude in Eq.~\eqref{eq:deletion-contraction_algorithm}, a single application of the deletion-contraction relation will split one term into two, so the total number of terms in this classical algorithm should scale exponentially with the number of edges and vertices.   Indeed, it has been found that 
the worst-case running time satisfies the same recurrence relation as the Fibonacci numbers and scales exponentially as $O(\phi^{|V|+|E|})$. Here,  $\phi$ is the golden ratio, while $|V|$ and $|E|$ are the number of vertices and edges of the graph $G=(V,E)$, respectively.

Besides this specific algorithm, the general computational complexity of exactly evaluating or approximating the chromatic polynomial has also been well-studied.  It has been known that for a generic graph $G$,  the chromatic polynomial $\chi(G, k)$ is $\#$P-hard to evaluate for $k \in \mathbb{C}$ except for the three ``easy points" $k = 0,1,2$ \cite{Jaeger1990Math.Proc.Camb.Philos.Soc.}. The $\#$P-hard complexity has also been proved when restricting to planar graphs \cite{Vertigan2005SIAMJ.Comput.}.  No classical approximation algorithm is known for chromatic polynomials except for the three easy points.  Moreover, it has been shown that no fully polynomial randomized approximation scheme (FPRAS) exists for $k$$>$$2$ \cite{Golderb:2008908,Goldberg:2012uh} (where $k$ is rational).  One may expect that FPRAS also does not exist for irrational $k$. 

On the other hand, an additive approximation of the chromatic polynomial on a planar graph can be achieved with a polynomial-time quantum algorithm \cite{Aharonov2007}.  The BQP-completeness is also proven in Ref.~\cite{Aharonov2007} for certain parameter regimes in a generalized version of the chromatic polynomial: the Tutte polynomial.  This implies that the chromatic polynomial as a combinatorial object typically appearing in classical computing problems may be intrinsically quantum.  Therefore, it is potentially possible to demonstrate quantum advantage when using a quantum computer to evaluate or approximate the chromatic polynomial.

Since the string-net wavefunction amplitude $\Psi(G)$ is real-valued, it only contains additional information about the sign other than the probability distribution $\Psi(G)$.   Due to the proportionality of $\Psi(G)$
to the chromatic polynomial, we hence know that the probability distribution $\Psi(G)$ is also $\#$P-hard to evaluate exactly.   Since the wavefunction amplitude $\Psi(G)$ is at least as hard as $P(G)$ due to the additional sign structure, we know that the exact evaluation of the wavefunction amplitude $\Psi(G)$ is also $\#$P-hard.  

Finally, the sampling of the bit-string distribution of the string-net ground state, as will be discussed in the next sub-section,  provides a way to approximate the probability distribution  $P(G)$, and hence could be a good candidate for demonstrating quantum advantage.

\subsection{String-net sampling}
\label{subsec:string-net-sampling}

Instead of classically evaluating the string-net wavefunction or the chromatic polynomials, one can estimate them by performing a quantum sampling of the string-net wavefunction on the quantum computer. We call such a process \textit{string-net sampling}, where the name resembles boson sampling.  A 2D string-net state with linear size $L$ can be prepared by a geometrically local unitary circuit with depth $O(L)$ (see Sec.~\ref{SM:preparation_protocol}). If allowing long-range (geometrically non-local) connectivity in the hardware, one can use a Multiscale Entanglement Renormalization Ansatz (MERA) circuit to prepare the state in $O(\log L)$ time.   

According to Eq.~\eqref{eq:chromatic2}, one can estimate the chromatic polynomial via the string-net sampling as
\begin{equation}\label{eq:estimate_polynomial}
    \chi(\hat{G},\phi+2)=(\phi+2)\frac{\Psi(G)^2}{\Psi(\text{vac})^2}=(\phi+2)\frac{P(G)}{P(\text{vac})}\approx (\phi+2)\frac{C(G)}{C(\text{vac})},
\end{equation}
where $C(G)$ represents the count of the bit-string corresponding to graph $G$ and $C(\text{vac})$ is the count for the empty configuration, which is also referred to as $G_0$ in the main text for the $2\times 2$ lattice. However, the number of samples of each graph $G$ is small when the system becomes large, and this estimation will suffer from large fluctuation.  Instead, we can sample the isomorphism class $[G]$, i.e., the class of graphs that are all isomorphic to the graph $G$:
\begin{equation}
 \chi(\hat{G},\phi+2) \approx (\phi+2)\frac{C([G])}{C(\text{vac})},
\end{equation}
where $\overline{C}([G])$ represents the average count over the isomorphism class $[G]$.   This estimation method is used in the main text (Fig. 4).

Nevertheless, when scaling up the system size, this estimation method could suffer from the disadvantage that the count $C(\text{vac})$ is typically very small since there is only a unique empty configuration, which leads to large fluctuation.   Instead, one can choose a simple graph with high multiplicity, i.e., more isomorphic configurations,  as a reference.   A good candidate is the loop configuration $\bigcirc$, which has very high multiplicity for arbitrary lattice size.   The relative probability is known theoretically to be $P(\bigcirc)/P(\text{vac})=\phi^2$ according to Eq.~\eqref{eq:loop_probability}.   Substituting this into Eq.~\eqref{eq:estimate_polynomial} leads to another estimation method:
\begin{equation}\label{eq:estimate_polynomial_loop}
   \chi(\hat{G},\phi+2) = \phi^2(\phi+2) \cdot  \frac{P(G)}{P(\bigcirc)} \approx  \phi^2(\phi+2) \cdot  \frac{\overline{C}([G])}{\overline{C}([\bigcirc])}, 
\end{equation}  
where $\overline{C}([\bigcirc])$ denotes the average count over the isomorphism class of the loop $\bigcirc$.  
The averaging over the isomorphism class in both the numerator and denominator of Eq.~\eqref{eq:estimate_polynomial_loop} can greatly suppress the estimation error due to the fluctuation of the sampling distribution either from the shot noise or the device noise.   Note that the total number of isomorphic classes of the subgraphs denoted by $[G]$ on a given Levin-Wen lattice $\mathcal{L}$ is much smaller than the total number of subgraphs $G$.  Therefore, the estimation method in Eq.~\eqref{eq:estimate_polynomial_loop}, which effectively samples the isomorphic class $[G]$, has a much smaller sampling overhead than directly sampling over individual graph $G$, and is hence expected to be more scalable when growing the lattice size.

\begin{figure}[hbt]
\centering
\includegraphics[width=0.7\linewidth]{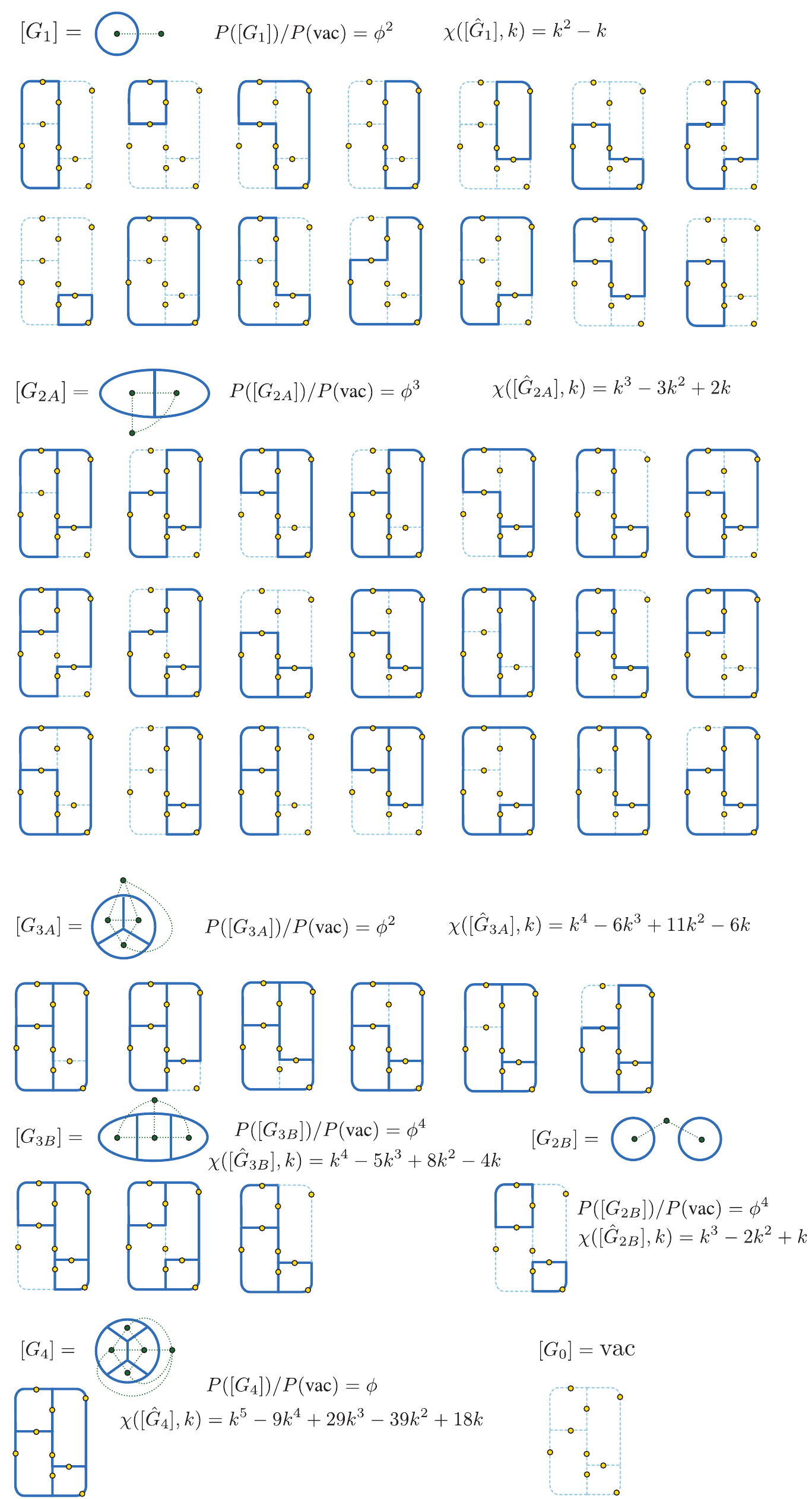}
\caption{List of all the isomorphism class of subgraphs (thick blue lines) on a 4-plaquette string-net model as well as their dual graphs (green dotted lines).  The corresponding relative probability and the chromatic polynomials are listed for each class.}
\label{fig:string_classification}
\end{figure}

For the experiment in Fig.~4 of the main text defined on a lattice with four plaquettes,  we summarise all the seven isomorphism classes (including the empty configuration) in Fig.~\ref{fig:string_classification}.  As we can see, all the isomorphism classes have different chromatic polynomials of their dual graphs.  Therefore, none of these classes are chromatically equivalent.  However, some of the classes do have the same evaluation of chromatic polynomials at $k=\phi + 2$.

\newpage

\begin{figure}[hbt]
\centering
\includegraphics[width=0.5\linewidth]{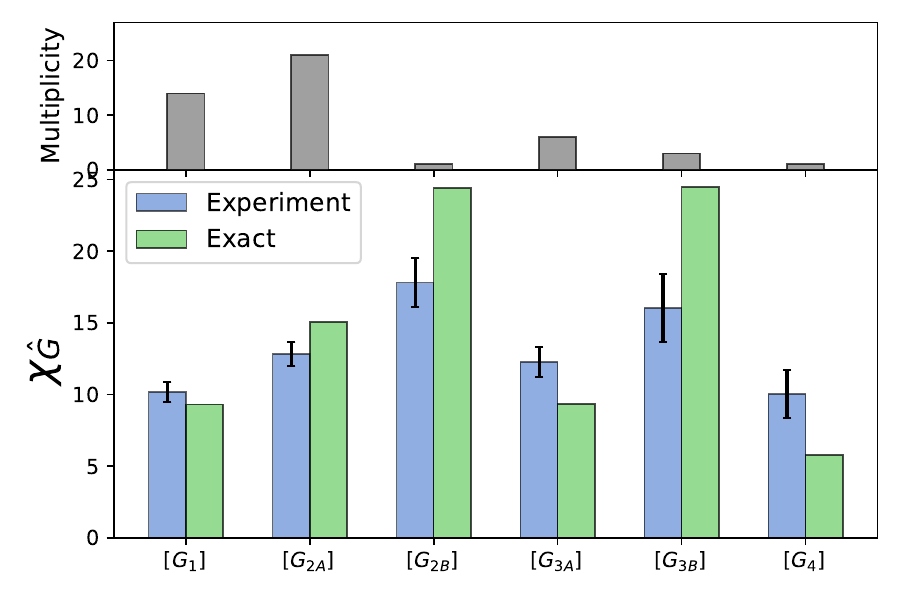}
\caption{Alternative way to estimate the chromatic polynomial of isomorphism class of a graph $[G]$ using Eq.~\eqref{eq:estimate_polynomial_loop}. Blue:  experimental estimation of the polynomial; Green: exact theoretical prediction of the chromatic polynomial;   Grey: multiplicity of each isomorphism class.}
\label{fig:Chrom_poly}
\end{figure}

Now we use the alternative estimation method in Eq.~\eqref{eq:estimate_polynomial_loop} to evaluate the chromatic polynomial for the same 4-plaquette lattice considered in the main text (Fig.~4).  The result is summarized in Fig.~\ref{fig:Chrom_poly}.  

\section{Scalable protocol for preparing the 2D Fib-SNC}\label{SM:preparation_protocol}
\begin{figure}[!h]
    \centering
    \includegraphics[width=0.95\textwidth]{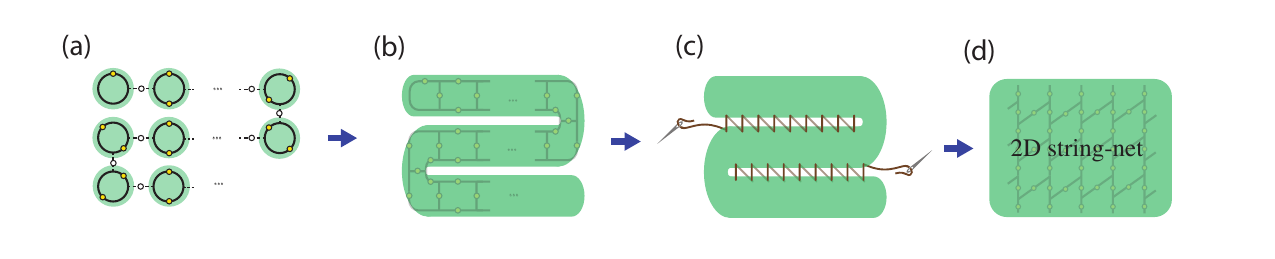}
    \caption{The stages of a scalable protocol for 2D string-net: (a) Bead strand configuration (b) A folded strip of plaquettes traversing through the targeted 2D system (c) Sewing up the gaps between the folds using unitary gates (d) Fib-SNC in 2D.}
    \label{fig:2D_protocal_cartoon}
\end{figure}

\begin{figure}[!h]
    \centering
    \includegraphics[width=0.95\textwidth]{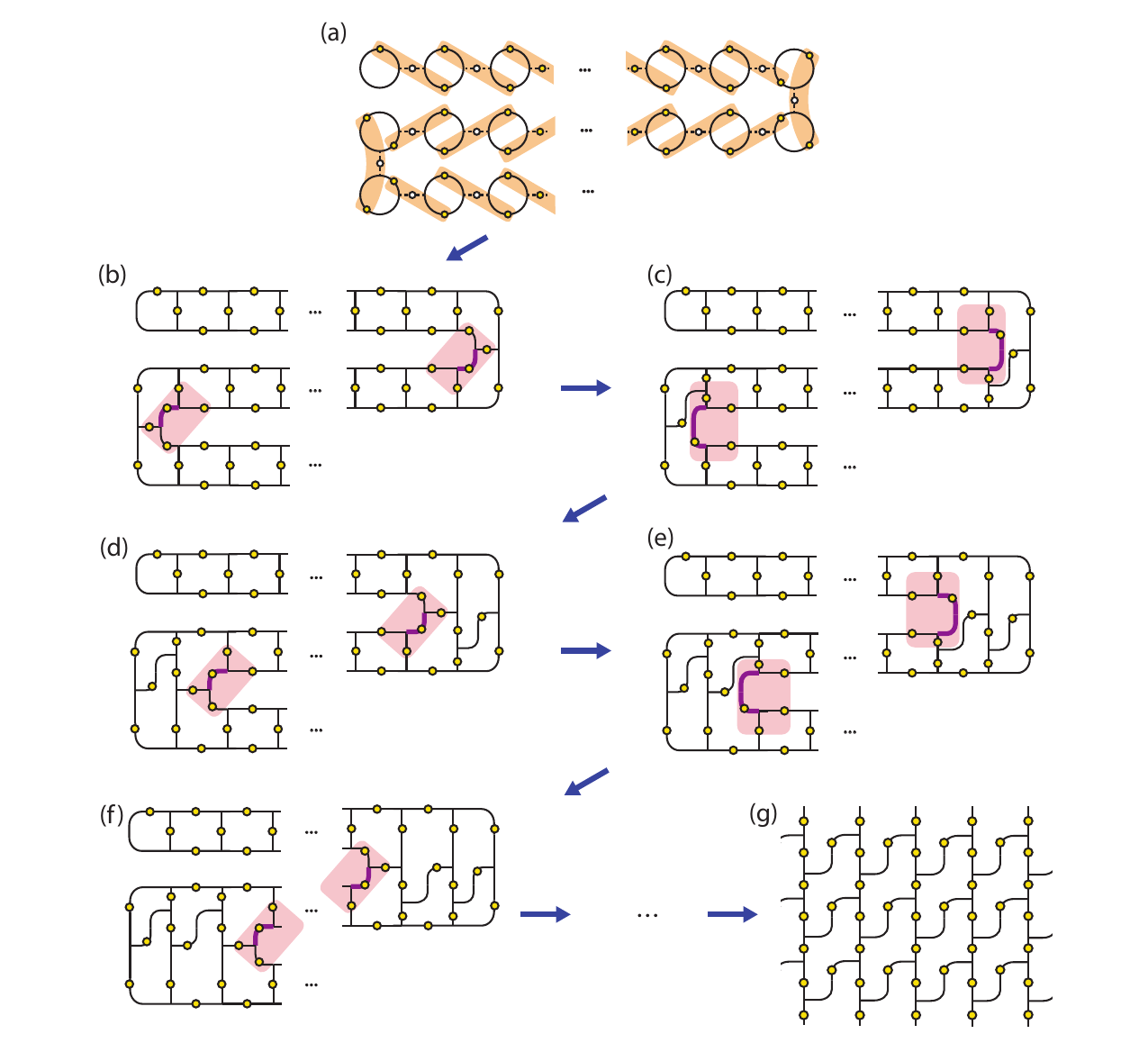}
    \caption{(a) Bead strand configuration with the location of each 3-qubit $F$-moves indicated by a orange box. (b-f) On the folded strip of plaquettes, consecutive 5-qubit $F$-moves are applied to sew up the gap between the folds. (g) Fib-SNC on a 2D lattice. }
    \label{fig:2D_protocal_full}
\end{figure}

We discuss our scalable protocol for dynamical string-net preparation (DSNP), which creates 2D Fib-SNC via dynamically deforming the trivalent graphs using unitary operations. This protocol is not tied to any specific rigid lattice. It offers a flexible alternative to the strategies designed for the honeycomb lattice \cite{Xu2024, Liu2022PRXQuantum}. Also, this DSNP naturally generalizes to string-net condensation with general branching rules (or, more precisely, the LW string-net ground state with a more general input category). In the following,  we will first outline the overall strategy of DSNP schematically (see Fig. \ref{fig:2D_protocal_cartoon}) and provide the details of the circuit implementation afterward (see Fig. \ref{fig:2D_protocal_full}). 

The outline of the DSNP strategy contains several stages, as illustrated in Fig. \ref{fig:2D_protocal_cartoon}. First, we start with the bead strand shown in Fig. \ref{fig:2D_protocal_cartoon}a. Then, we use $F$-moves to turn it into a folded strip of plaquettes (Fig. \ref{fig:2D_protocal_cartoon}b). The folded strip is arranged to traverse through the 2D space we target. Next, as shown in Fig. \ref{fig:2D_protocal_cartoon}c, we sew up the gap between the folds using $F$-moves, which brings us to the 2D Fib-SNC in Fig. \ref{fig:2D_protocal_cartoon}d. The circuit depth to prepare the strip of plaquettes (Fig. \ref{fig:2D_protocal_cartoon}b) is $\sim O(1)$ regardless of the length of the strip. The circuit depth needed to sew up the folds in (Fig. \ref{fig:2D_protocal_cartoon}c) is $\sim O(L)$, where $L$ is the linear size of each fold. Overall, our strategy can prepare a 2D Fib-SNC using a unitary circuit whose total depth scales linearly with the system's linear dimension.

Now, we describe the unitary circuit implementation of our DSNP strategy. Here, we describe this implementation in terms of the basic building blocks, such as the modular ${\cal S}$ gate, $F$-moves, etc. The quantum circuit for each building block will be provided in Sec. \ref{sec:BuildingBlock}. The starting bead strand is shown in Fig. \ref{fig:2D_protocal_full}a. All the white qubits are in the $|0\rangle$ state. On each bead (represented by the black circle), the two qubits are in the state $\frac{1}{\sqrt{1+\phi^2}}
(|00\rangle +  \phi |11\rangle)$, which can be obtained from the consecutive actions of a single-qubit ${\cal S}$ gate and a two-qubit CNOT gate acting on $|00\rangle$ just like what is done in Fig. 3 of the main text. For the bead with a single qubit, the qubit is prepared in the state  $\frac{1}{\sqrt{1+\phi^2}}
(|0\rangle +  \phi |1\rangle) = {\cal S} |0\rangle$. At this point, the total wave function of the system is essentially the state with a vacuum loop surrounding each plaquette, just like the left-hand side of Eq. \eqref{eq:fattenedlattice_groundstate}. This interpretation is not affected by the qubits between the beads because they are all still in the $|0\rangle$ state, representing the absence of any string on the corresponding edges. Recall that the equation Eq. \eqref{eq:fattenedlattice_groundstate} holds only after we use Eq. \eqref{eq:Fmove} and Eq. \eqref{eq:tadpole} as equivalence relations to reduce the Fib-SNC wave function on the right-hand side. Conceptually, the remaining steps of our DSNP strategy are to use the $F$-moves to undo this reduction and recover the full Fib-SNC wave function from the state with a vacuum loop surrounding each plaquette. For the next step of the DSNP protocol, we apply a parallel set of 3-qubit $F$-moves (indicated by the orange boxes) to turn the bead strand into a folded strip of plaquettes shown in Fig. \ref{fig:2D_protocal_full}b. To sew up the gap between the folds, one needs to apply a set of consecutive 5-qubit $F$-moves (pink boxes in Fig. \ref{fig:2D_protocal_full}b-f). For example, after applying the 5-qubit $F$-moves shown in Fig. \ref{fig:2D_protocal_full}b-c, the length of the gap between the folds is shortened by the size of one plaquette.  Fig. \ref{fig:2D_protocal_full}d-e iterates the same steps as in Fig. \ref{fig:2D_protocal_full}b-c. After iterating the same steps for $\sim O(L)$ times, resulting in a circuit depth $\sim O(L)$, the gaps between the folds are completely sewn up, and we arrive at the 2D Fib-SNC (see Fig. \ref{fig:2D_protocal_full}g). Here, for a concrete demonstration, the 2D Fib-SNC realized in Fig. \ref{fig:2D_protocal_full} resides on the honeycomb lattice. It is worth noting that the DSNP strategy can be readily generalized to any 2D trivalent graph and also to more general string-net condensates.

\pagebreak
\newpage

\section{Quantum circuits for the building blocks of DSNP}
\label{sec:BuildingBlock}

\begin{figure}[hbt] \label{Circuits_Fmoves}
    \centering
\includegraphics[width=1\textwidth]{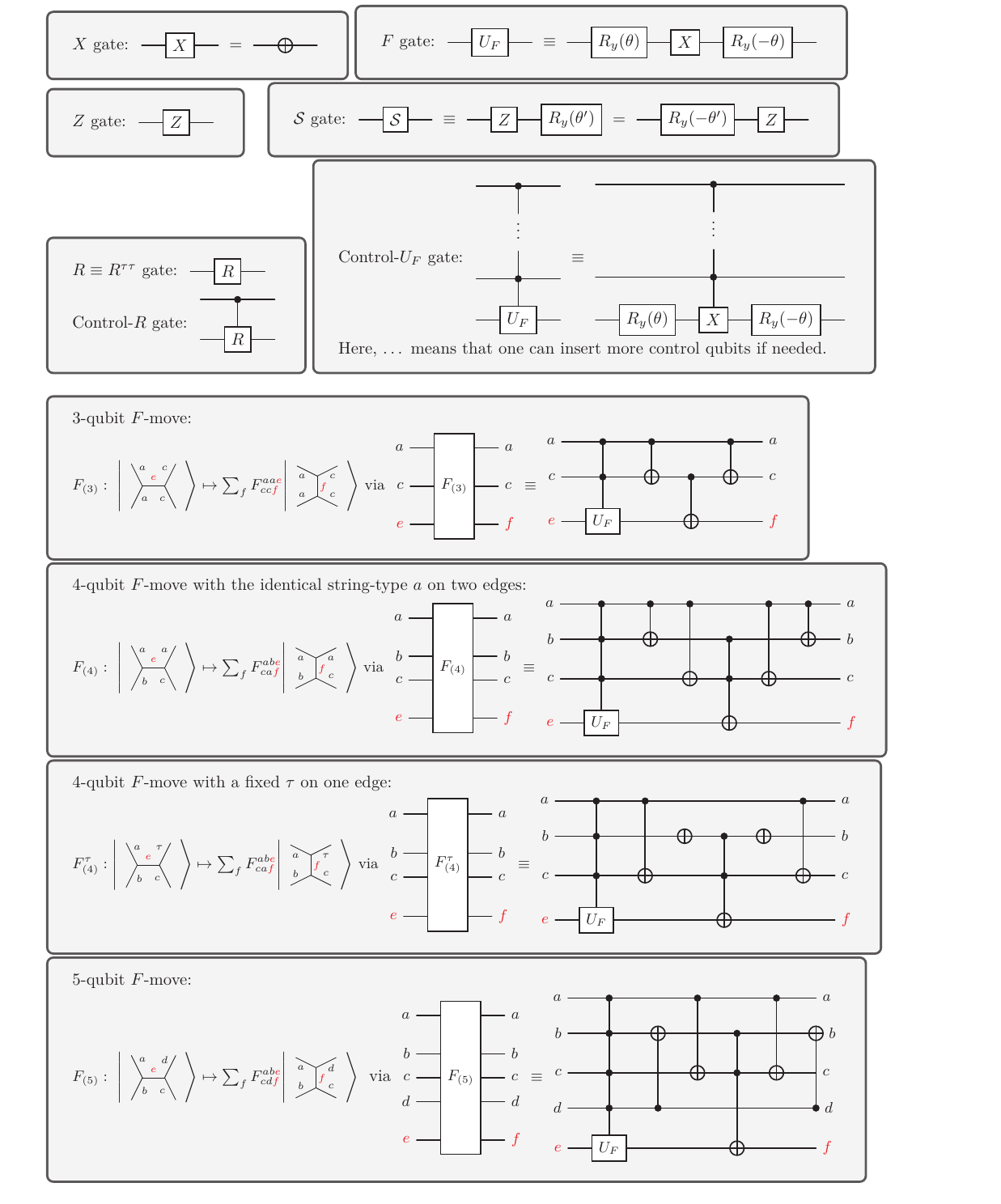}
    \caption{Quantum circuits for the building blocks of DSNP: ${\cal S}$ gate, $F$-moves, $R$-moves, and etc. \\
    Note that in terms of circuit, the input initial state is on the left side, 
    while after the gate operations, the output final state is on the right side of the circuit; 
    namely the unitary evolution is from the left to the right.
    In contrast, the unitary matrix representation of the quantum gate acts on the ket state $|$ string-net configuration $\rangle$ 
    on the right end from the left; 
    namely the unitary evolution of the ket state is from the right to the left.
    Quantum circuit here is \LaTeX\; thanks to Quantikz \cite{1809.03842Kay:2018huf}.
   }
    \label{fig:RF_Circuits}
\end{figure}

In this section, we provide the details on the quantum circuit realization of the basic building blocks of our DSNP strategy, including the ${\cal S}$ gate, $F$-moves, and $R$-move (summarized in Fig.~\ref{fig:RF_Circuits}).

First, we establish the convention for notations. As a reminder, $
\phi$ denotes the golden ratio, i.e. $\phi = \frac{\sqrt{5}+1}{2}$. Additionally, we define the angles $\theta$ and $\theta'$ by
 $\theta = \arctan \sqrt{\phi}$ and $\theta' = 2\arctan \phi$. The three Pauli operators are denoted by $2\times 2$ matrices $X,Y,Z$. The single-qubit rotation $R_y(\alpha)$ for an angle $\alpha$ is defined as 
\begin{align}
    R_y(\alpha)\equiv\exp(-\ii \alpha Y/2).
\end{align}
$R_x(\alpha)$ and $R_z(\alpha)$ are defined similarly as $R_y(\alpha)$ but with $Y$ replaced by $X$ and $Z$ respectively.

In this convention, the ${\cal S}$ gate can be decomposed as the product:
\begin{align}
    {\cal S} = \frac{1}{\cal D}\begin{pmatrix}
        1 & \phi \\
        \phi & -1
    \end{pmatrix} = R_y(\theta')Z.
\end{align}
One can also find other decompositions involving $R_x(\alpha)$ or $R_z(\alpha)$.

Next, we discuss the quantum circuits for the $R$-move, which is defined by $R$-symbol $R^{ij}_k$ shown in Eq. \eqref{eq:Rmove} and Fig. \ref{fig:FRmoves}c (also Fig. 1h of the main text). As a quantum gate, the most generic $R$-move corresponds to a three-qubit unitary gate that acts as a diagonal matrix in the computation basis: 
\begin{align}
		\ket{\raisebox{-0.5cm}{\includegraphics[scale=.35]{fig/R-move-over.pdf}}} 
		 \mapsto  \; R^{ij}_k \;\;
		\ket{\raisebox{-0.5cm}{\includegraphics[scale=.35]{fig/R-move-Y.pdf}}} \, 
\end{align}
where $i,j,k = 0,1$ labels the states of the three edges meeting at a vertex. The $R^*$-move corresponds to a three-qubit unitary gate which is the complex conjugate of the $R$-move. In our experiment, all occurrences of the $R$-move act on triplets of qubits with either $|i\rangle = |1\rangle$ or $|j\rangle = |1\rangle$. When $|i\rangle = |1\rangle$, the $R$-move reduces to a 2-qubit control gate where $|j\rangle$ acts as the control qubit and the action on the state $|k\rangle$ is a phase rotation given by
$R \equiv \begin{pmatrix}  
 \re^{\frac{4\pi}{5} \ii} & 0\\
 0 & \re^{-\frac{3\pi}{5} \ii}
\end{pmatrix}$. When  $|j\rangle = |1\rangle$, the $R$-move reduces to the same 2-qubit control gate except that $|i\rangle$ becomes the control qubit. The circuit diagram of this reduced 2-qubit version of the $R$-move is given by the control-$R$ gate present in Fig.~\ref{fig:RF_Circuits}. A similar reduction also applies to the $R^*$ move. 

Now, we discuss the circuit implementation of the $F$-move. The most general form of the $F$-move acts on 5 qubits (see the last panel of Fig.~\ref{fig:RF_Circuits}). Notice that the state $|abcd\rangle$ of the first four qubits remains unchanged under the $F$-move, but it controls the action on the last qubit. In our experiments, we often encounter situations where the $F$-move can be simplified according to the initial state of the first four qubits before the $F$-move. The simplified versions result in lower circuit depths. 
When the initial state satisfies $|a\rangle = |b\rangle$ and $|c\rangle = |d\rangle$, the $F$-move can be effectively reduced to a 3-qubit gate, which we call the 3-qubit $F$-move.  When the initial state satisfies $|a\rangle = |d\rangle$, the $F$-move reduces to a 4-qubit gate, a 4-qubit $F$-move (denoted as $F_{(4)}$ in Fig.~\ref{fig:RF_Circuits}). When one of the first four qubits is in the $|1\rangle$ state, for example, $|d\rangle = |1\rangle$, the $F$-move reduces to another 4-qubit gate (denoted as $F_{(4)}^\tau$ in Fig.~\ref{fig:RF_Circuits}).

\clearpage

\section{Experimental setup}

\subsection{Devices and error rates}

In the following, we report on the experimental setup and device error rates.  The experiments used three IBM Quantum devices, only two of which are used for presentation in the main text. See Sec.~\ref{sec:exp_flow_params} for exact details on the devices used in the experiments. Here, we provide device-level information. 
To this end, let us establish a few common notions and metrics. 

\paragraph{Readout-error metric.}

\begin{figure}[t]
    \centering 
\includegraphics{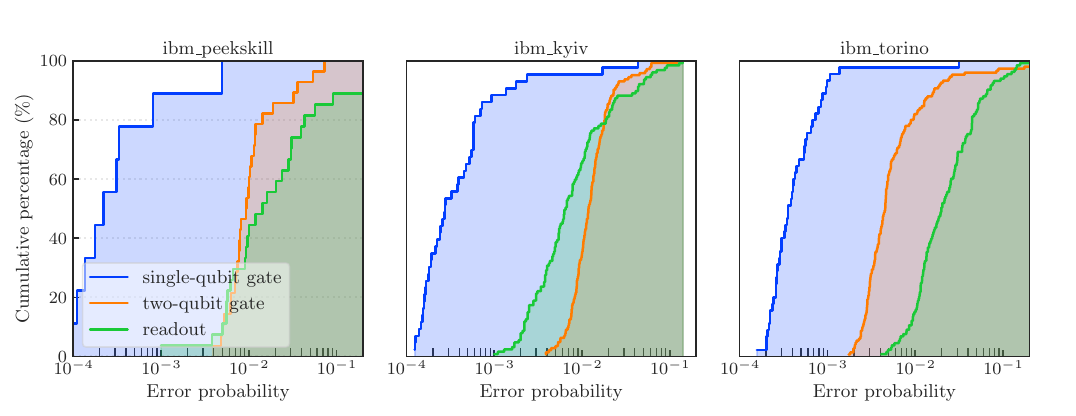} 
\caption{
Empirical cumulative distribution functions (CDFs) of error probabilities for three IBM quantum devices: \textit{ibm\_peekskill}, \textit{ibm\_kyiv}, and \textit{ibm\_torino}. The plots display error probabilities for single-qubit gates (\textcolor{blue}{blue}), two-qubit gates (\textcolor{orange}{orange}), and readout errors (\textcolor{green}{green}). Single-qubit gate errors correspond to those of the square root of a CNOT gate (sx). Two-qubit gate errors correspond to those native to the device, such as the CX, ECR, or CZ gates. The x-axis represents error probability on a logarithmic scale, while the y-axis shows the cumulative percentage of qubits or gates. The shaded areas under the CDF curves highlight the distribution of errors across the devices.
}
 \label{fig:si_device_error_rates}
\end{figure}

To characterize the imperfection of the quantum-to-classical meters in our devices, the measurement operations, we report the average readout-assignment error  \(1 - \mathcal{F}_a\) for each device (see Fig.~\ref{fig:si_device_error_rates}).
We define this quantity in the standard way for a single qubit as 
$$
\mathcal{F}_a := 1 - \frac{1}{2} \left( P(1|0) + P(0|1) \right)\;,
$$ 
where \(P(A|B)\) is the empirical probability to measure the qubit in state \(A \in \{0,1\}\) given that the qubit was nominally prepared in state \(B \in \{0,1\}\).  We note that in practice, the probability distributions are biased in superconducting devices due to the asymmetric nature of the energy relaxation processes, $T_1$ in the qubit, such that typically \(P(1|0) \ll P(0|1)\).

\paragraph{Single-qubit native, basis gates and their errors.}

One can decompose any single-qubit unitary~$U$ into a combination of~$R_Z$ rotation gates and $\sqrt{X}$ (or sx) gates. For example, one valid decomposition of a unitary $U$ is
\begin{equation}
    U = R_z(\alpha+\pi)\sqrt{X}R_z(\beta+\pi)\sqrt{X}R_z(\gamma)\;, 
\end{equation}
where $\alpha,\beta,\gamma$ are the Euler angles. This is how our circuits are compiled, down to $R_Z$ and $\sqrt{X}$ gates. The device-level implementation of our $R_Z$ rotation gates is virtual. Thus, it incurs no noise and no error. On the other hand, our $\sqrt{X}$ (or sx) gates are based on finite-time pulses. These are calibrated carefully to implement the $\sqrt{X}$ operation. However, they are ultimately imperfect at the level of $10^-4$ in our devices (see Fig.~\ref{fig:si_device_error_rates}). Their error rate characterizes the main errors due to single-qubit operations. Of course, the way they are measured does not account for all possible classical and quantum cross-talk errors that may result from the parallel application of gates or the effect of single qubit gates on spectator qubits. See below for our benchmarking setup, which uses parallel gates. 

\paragraph{Two-qubit native, basis gates and their errors.}

Our two-qubit calibrated native gates are either a controlled-NOT (CX), controlled-Z (CZ), or echo-cross resonance (ECR) gate \cite{Sheldon2016, Sundaresan2020, Jurcevic2021}. The ECR gate is native to the Falcon and Eagle IBM processors. It is maximally entangling and equivalent to a CX gate up to single-qubit pre-rotations. It's two-qubit unitary is
\begin{equation}
    \operatorname{ECR} = \frac{1}{\sqrt{2}}
            \begin{pmatrix}
                0   & 1   &  0  & i \\
                1   & 0   &  -i & 0 \\
                0   & i   &  0  & 1 \\
                -i  & 0   &  1  & 0
            \end{pmatrix}\;,
\end{equation}
in the two-qubit Hilbert space of the control and target qubits $\mathcal H_\mathrm{target} \otimes \mathcal H_\mathrm{control}$. It is equivalent to the control-echoed rotation around the ZX Pauli of $\pi/2$, that is $\operatorname{ECR} = i R_\mathrm{ZX}(-\pi/4) R_\mathrm{XI}(\pi) R_\mathrm{ZX}(\pi/4)$, where we have accounted for an $i$ global phase, and have defined $R_\mathrm{ZX}(\theta):=\exp(-i\frac{1}{2}\theta \,\mathrm{ZX})$ and $R_\mathrm{XI}(\theta)$ similarly.
All two-qubit and higher-level unitaries are compiled down to these native single-qubit and two-qubit error gates. We report the error rates of our two-qubit gates in the following.

\paragraph{Device-level error rates: overview.}

In Fig.~\ref{fig:si_device_error_rates}, we show the empirical cumulative distribution functions (CDFs) of the error probabilities for three IBM quantum devices: \textit{ibm\_peekskill} (27 qubits), \textit{ibm\_kyiv} (127 qubits), and \textit{ibm\_torino} (133 qubits). These devices exhibit typical error rates across different sizes and generations of processors, with the latest device \textit{ibm\_torino} performing the best.  Generally, single qubit gates are the fastest and least complex operation and thus tend to be the least noisy, as evident in the data. Two-qubit gates and readout gates are typically longer and more complex, and incur more error.  Naturally, all qubit parameters and error rates fluctuate over time. The data presented here is representative of the common case. 
To be concrete, we report in detail the error rates at the time of writing.

The \textit{ibm\_peekskill} device is a Falcon processor of the 8th revision (r8) family. It's single-qubit gate errors have a mean of \(8.0 \times 10^{-4}\) and a median of \(2.2 \times 10^{-4}\). The two-qubit gate errors show a mean of \(1.5 \times 10^{-2}\) and a median of \(9.4 \times 10^{-3}\). Readout errors for this device have a mean of \(4.2 \times 10^{-2}\) and a median of \(1.4 \times 10^{-2}\).

The \textit{ibm\_kyiv} device (an Eagle processor, revision 3) has single-qubit gate errors with a mean of \(1.8 \times 10^{-3}\) and a median of \(2.7 \times 10^{-4}\). Its two-qubit gate errors have a mean of \(1.5 \times 10^{-2}\) and a median of \(1.2 \times 10^{-2}\). The readout errors show a mean of \(1.5 \times 10^{-2}\) and a median of \(6.5 \times 10^{-3}\).

The \textit{ibm\_torino} device belongs to the new generation of processors called Heron (first revision r1). It features 133 fixed-frequency qubits integrated with tunable couplers for two-qubit gates. Its performance is so far seen to be on the order of 3 to 5 times improvement over the previous state-of-the-art 127-qubit Eagle processors. Typically, the Heron's crosstalk is lower, reducing the critical challenge of cross talk.  For \textit{ibm\_torino}, single-qubit gate errors have a mean of \(1.2 \times 10^{-3}\) and a median of \(3.6 \times 10^{-4}\). The two-qubit gate errors show a mean of \(2.3 \times 10^{-2}\) and a median of \(4.6 \times 10^{-3}\), significantly better than those of the other devices. The readout errors for this device have a mean of \(3.3 \times 10^{-2}\) and a median of \(2.0 \times 10^{-2}\).

\begin{figure}[t]
    \centering 
\includegraphics{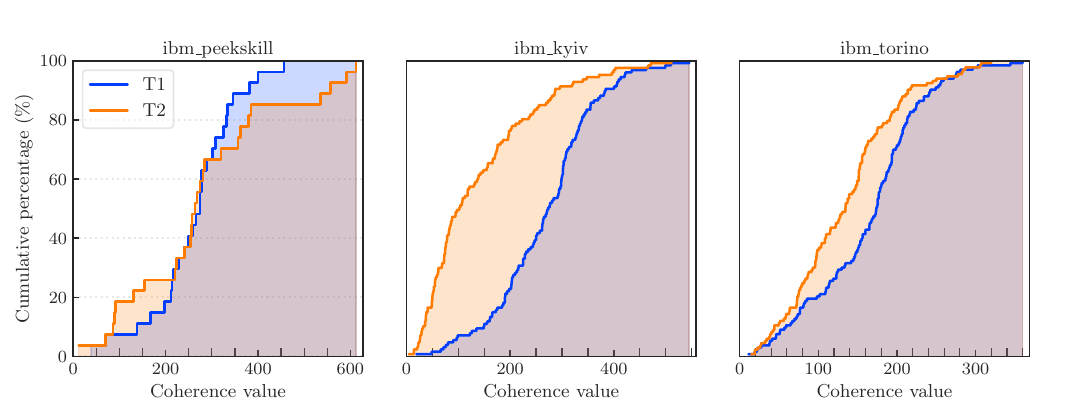} 
\caption{
Empirical cumulative distribution functions (CDFs) of coherence values (T1 and T2) for three IBM quantum devices: \textit{ibm\_peekskill}, \textit{ibm\_kyiv}, and \textit{ibm\_torino}. The plots display the cumulative percentage of qubits with given T1 (blue) and T2 (orange) coherence values. The x-axis represents the coherence value, while the y-axis shows the cumulative percentage of qubits. 
}
 \label{fig:si_device_coherence}
\end{figure}

\paragraph{Device: coherence times.}
In Fig.~\ref{fig:si_device_coherence}, we show the empirical cumulative distribution functions (CDFs) for the \( T_1 \) and \( T_2 \) (specifically, \( T_2 \) echo) coherence times for the three IBM quantum devices. These coherence times ultimately impact the gate and quantum circuit fidelity, especially when there are deep circuits and long idle periods. The empirical CDFs make it clear that the Peekskill device exhibits the largest mean and median \( T_1 \) and \( T_2 \) coherence times compared to the other devices. 
Although Torino has comparatively lower \( T_1 \) and \( T_2 \) coherence times, it compensates with significantly faster tunable-coupler two-qubit gates. The increased gate speed in Torino results in lower two-qubit gate error rates. This showcases the trade-off between coherence times and gate speeds. We find best results on Torino as discussed in the following sections.

\subsection{Real-time benchmarking of the device}

\paragraph{The importance of qubit selection for deep circuits.}

The selection of qubits in our experiments is of critical importance, particularly due to the unique challenges presented by the Fibonacci anyon model. This model requires circuits that are very deep and, thus, very sensitive to noise.  The depth of the circuits is on the order of 150 two-qubit gate layers. The precise depth depends on the specific qubit topology and the optimization techniques employed during the transpilation process. To our knowledge, these circuits appear to be among the deepest executed for such problems.

In our experiments, we observed that sub-optimal layout --- one including even one or two poorly performing qubits --- significantly degraded the results. Conversely, choosing a chain of high-fidelity qubits leads to markedly improved outcomes. This, of course, is not surprising in and of itself, but it's worth emphasizing that the sensitivity of our circuits is much more pronounced compared to typical, shallower circuits.

\paragraph{Real-time benchmarking experiments.}

To ensure optimal qubit selection, we employ a real-time benchmarking protocol to assess the properties of the qubits at the time of execution of the Fibonacci circuits. That is rather than using the reported backend properties, which are updated at a lower cadence and become outdated. We found that real-time data is very helpful for quality results. This is expected given the inherent fluctuations over time of the qubit, gate, and device properties. These are known to arise from various calibration drifts and intrinsic device noise mechanisms beyond our control.

In particular, we conduct parallel real-time benchmarking of the entire chip using standard benchmarks: T1 relaxation time, T2 Hahn Echo, local readout error characterization, and two-qubit randomized benchmarking (performed on adjacent qubits). For completeness, we briefly review the purpose of each benchmark executed. 

There are three node-based and one edge-based set of experiments. The first set of experiments measure the energy relaxation time $T_1$ of each qubit. We use 5-time delay data points, logarithmically spaced out with a rate that depends on the mean reported $T_1$ lifetime of all the qubits. We measure all qubits simultaneously.  The second set of experiments evaluate the qubit coherence times $T_2$ using the single Hahn Echo sequences. 

Due to the higher sensitivity of this experiment to cross-talk on the devices, and our aim to first estimate the isolated qubit performance, we measure all qubits in two batches. To achieve this, we color the nodes of the device graph using two colors. The two colors partition the qubits into disjoint sets. Each set is then measured independently. We use 8 time-delay points with logarithmic spacing determined by the mean $T_2$ of the device.    To characterize readout errors quickly, we prepare both the all zero and the all one state and assume a local readout noise model. We repeat this experiment several times to achieve lower additive error on the results. We thus obtain the readout error
$\mathcal{F}_a = 1 - \frac{1}{2} \left( P(1|0) + P(0|1) \right)$ from the measurements of \(P(A|B)\) for the measurement outcome \(A \in \{0,1\}\) given that the qubit was nominally prepared in state \(B \in \{0,1\}\).

Finally, we assess the two-qubit gate performance on connected qubits uses randomized benchmarking (RB) to find the average gate fidelity of our native two-qubit gates.  We color the edges of our heavy-hex device using three colors to partition all edges into three disjoint sets. We use 5 RB sequence lengths logarithmically spaced based on the mean two-qubit error rate of the device.  For an example data set, see Fig.~\ref{fig:rb_curves}. It is conspicuous from the decay and confidence intervals in this example that some of the edges should be avoided at all costs.

\begin{figure}[t]
    \centering
    (a) \textit{ibm\_peekskill}\\
    \includegraphics{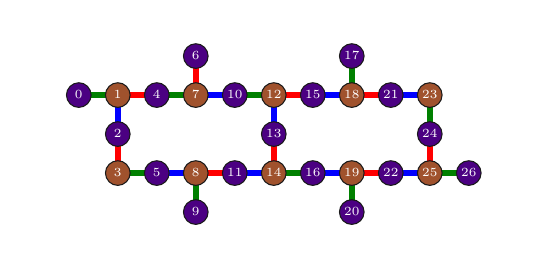}
    \\
    
    (b) \textit{ibm\_torino}\\
    \includegraphics{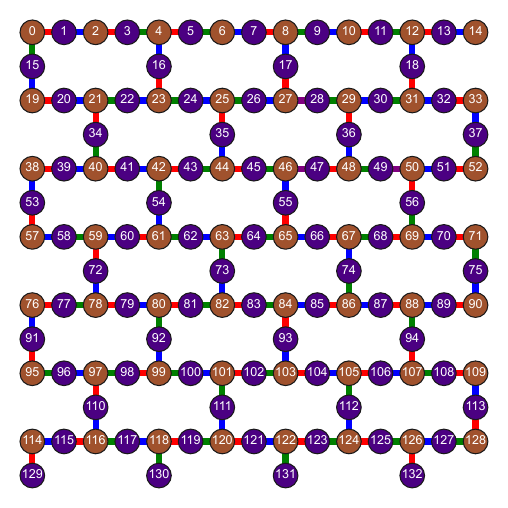}
    \caption{Example device layout and graph coloring for benchmarking for \textit{ibm\_peekskill} (a) and \textit{ibm\_torino} (b). The labels in the round nodes indicate qubit numbers. Edges between nodes indicate two-qubit gate connectivity.  The colors of the edges and nodes represent the partitioning schemes used to isolate and independently measure qubit and gate properties. Experiments are either node-based or edge-based. These include measuring \( T_1 \) relaxation time, \( T_2 \) Hahn Echo, local readout errors, and two-qubit randomized benchmarking on adjacent qubits, here using 3 layers.
    \label{fig:si_device_color}
    }
\end{figure}

\begin{figure}
    \centering
    \includegraphics[width=\textwidth]{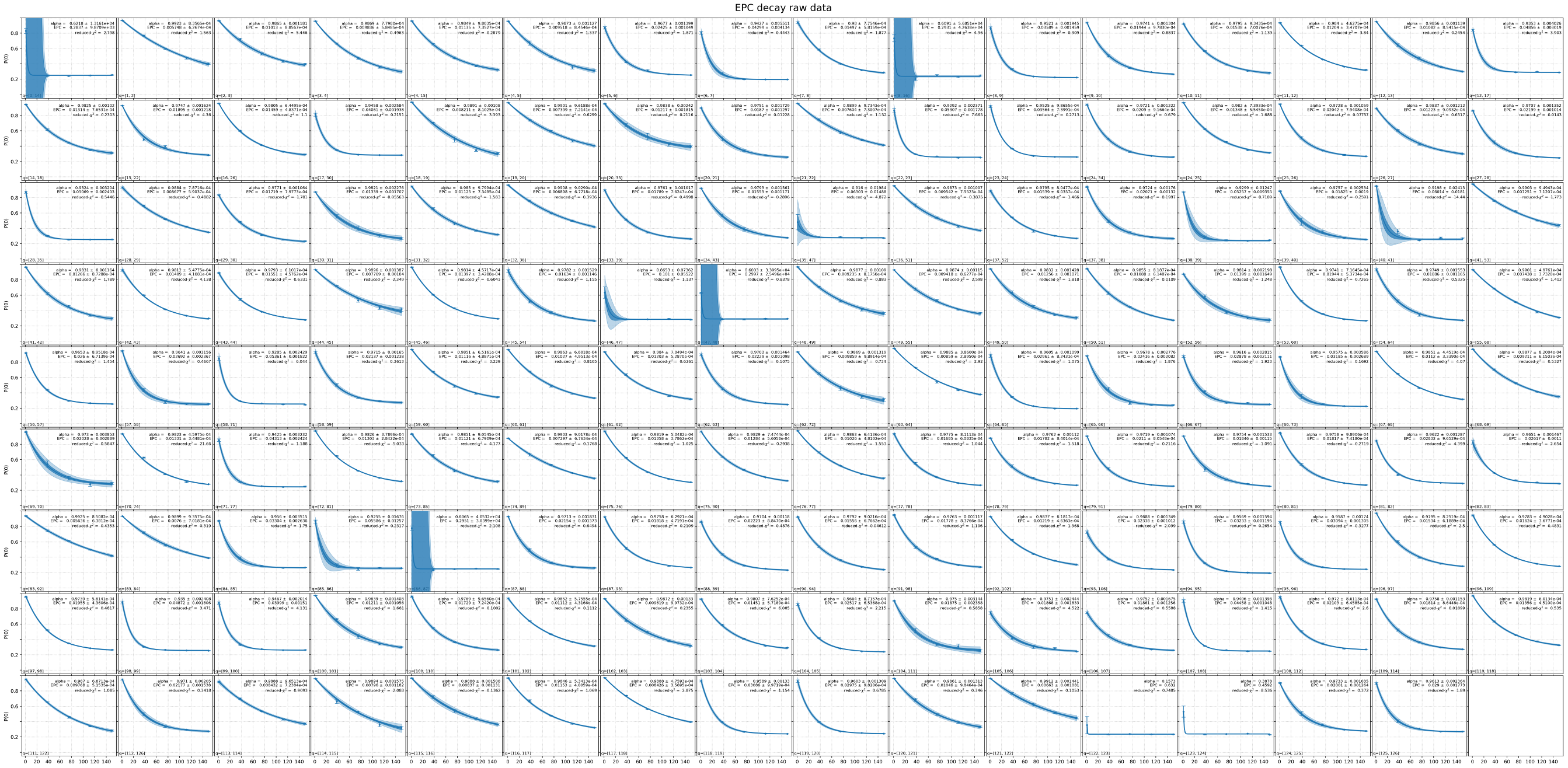}
    \caption{Two-qubit randomized benchmarking curves for each of the edges in one of the devices (\textit{ibm\_kyiv}). Each panel contains the raw data and anlysis for one of the edges in the device. The x-axis represents the length of the randomized benchmarking sequence. The raw data points are displayed with error bars as caps. The fits are exponential decay curves with offsets, shown with confidence intervals as shaded regions. Each panel reports the main fit parameters and their confidence intervals.}
    \label{fig:rb_curves}
\end{figure}

\clearpage

\section{Experiments: Error suppression and mitigation strategies}

\subsection{Qubit selection using real-time benchmark data}

\begin{figure}[th]
    \centering
    \includegraphics{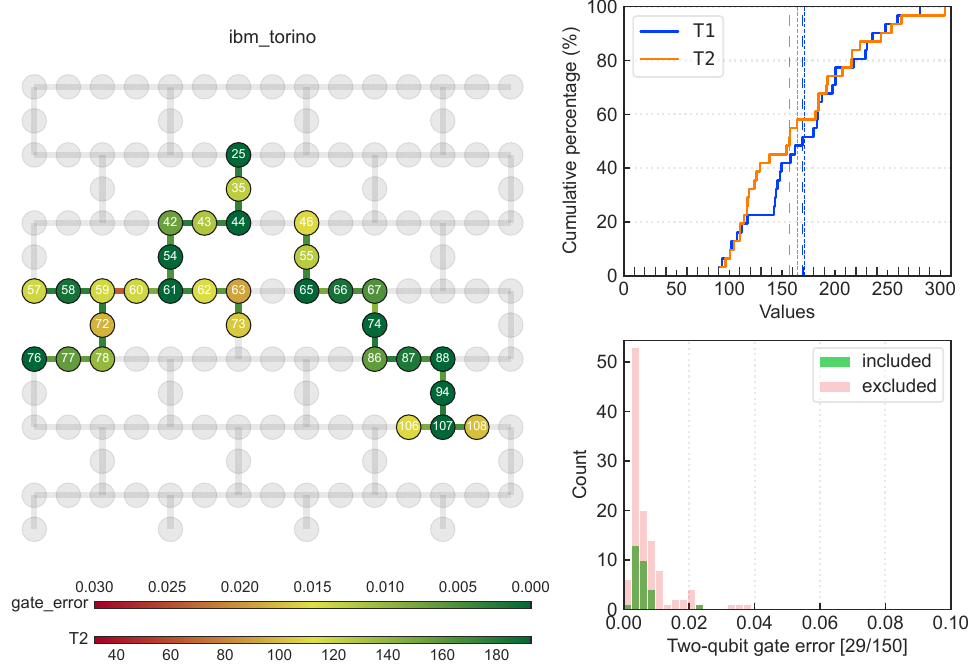} 
 \caption{
 \textbf{Qubit selection.} 
    \textit{Device Layout and Qubit Selection Criteria.} The left panel presents the layout of the \textit{ibm\_torino} quantum processor, comprising 133 qubits, depicted with transparent nodes. Highlighted in color and labeled with numbers are the qubits that meet specific selection criteria: high coherence times (\(T_1\) and \(T_2\)), low gate error, and low readout error. In this example, qubits in the selected set exhibit \(T_1\) and \(T_2\) times of at least 80 microseconds, readout error rates of no more than 10\%, connected gate error rates of no more than 4\%, and must be part of a cluster of such qubits that is at least large enough. Otherwise, they are not shown.
    \textit{Color Encoding for Qubit and Edge Performance.} The color bars at the bottom illustrate the color-coding scheme for nodes and edges. Nodes are colored based on their \(T_2\) times. Edges represent two-qubit gate errors. Greener colors denote lower error rates.
    \textit{Cumulative Distribution Functions of Coherence Times.} The top right panel displays the cumulative distribution functions (CDFs) of \(T_1\) and \(T_2\) lifetimes for the selected qubits, with \(T_1\) in blue and \(T_2\) in orange.
    \textit{Distribution of Gate Errors.} The bottom right panel shows histograms of two-qubit gate error rates. The green bars represent the error distribution of the included qubits' edges, whereas the red bars depict the distribution for all edges.
    }
    \label{fig:qubit_selection}
\end{figure}

Using a qubit selection protocol for mapping the virtual circuits we want to execute on the device, we can improve the noise performance of the experiments. We use the real-time benchmark data to this end. This allows us to identify the most suitable qubits for our experiments. By continuously monitoring qubit performance, we aim for our experimental setup to be optimized for the highest fidelity and reliability.

For example, the layout of the \textit{ibm\_torino} quantum processor is depicted in Figure~\ref{fig:qubit_selection}. This device consists of 133 qubits, represented by transparent nodes in the layout. The qubits highlighted in color and labeled with numbers are those that meet specific selection criteria, including high coherence times (\(T_1\) and \(T_2\)), low gate error, and low readout error. In this simple example, we require qubits with coherence times of at least \(T_1 = 80\) $\mu$s, at least \(T_2 = 80\) $\mu$s, and readout error rates of no more than of 10\%. The qubit must be connected to edges of error rate of no greater than 4\%. The patch of connected qubits must be at least 13 in this example. All smaller patches of connected qubits that meet the criteria are filtered out.

The color bars at the bottom of the left panel illustrate the color-coding scheme used for nodes and edges. Nodes are colored according to their \(T_2\) times, with greener shades indicating better performance (longer coherence times). Edges are colored based on two-qubit gate errors, with greener colors representing lower error rates. A red hue indicates worse performance for both nodes and edges.

The top right panel of Figure~\ref{fig:qubit_selection} shows the cumulative distribution functions (CDFs) of \(T_1\) and \(T_2\) lifetimes for the selected qubits, with \(T_1\) in blue and \(T_2\) in orange. These CDFs provide a statistical overview of the coherence performance of the selected qubits.

The bottom right panel of Figure~\ref{fig:qubit_selection}  presents histograms of two-qubit gate error rates. The green bars represent the error distribution of the edges associated with the selected qubits, whereas the red bars depict the distribution for all edges, highlighting the performance differences between the selected subset and the entire device. We note that many of the larger-error edges are selected out. Here, we have selected 29 of the total 150 edges in the device.

\subsection{Error suppression}

\paragraph{Optimized transpilation and qubit selection.}

To reduce the noise susceptibility of our quantum circuits, we map them (VF2++ algorithm) \cite{VF2} and transpile \cite{Qiskit10} them to the real-time benchmark selected qubits under certain optimizations and conditions. This process involves optimizing the circuits to minimize their depths and the number of two-qubit gates, which are typically the most error-prone. One notable technique we employ is the Maslov trick for implementing Toffoli (CCX) gates \cite{Maslov2016Phys.Rev.A}. This significantly reduces the overhead. Multiple optimizations are applied in a stochastic manner, allowing for various mappings and layouts to be assessed based on a cost function that incorporates error rates and the number of two-qubit gates. This comprehensive approach ensures the circuits are executed in the most noise-resilient manner possible.

\paragraph{Dynamical decoupling.}

To suppress idling coherent noise and some quantum cross-talk, we use dynamical decoupling (DD) \cite{Viola1998, Viola1999, Jurcevic2021, Ezzell2022}. This involves applying a sequence of curated control pulses to average out unwanted system-environment interactions. This results in lower effective decoherence rates, so long as the noise spectrum allows for the decoupling. This technique is especially beneficial for sparse circuits with significant idling times, such as those used in our experiments. Among the various DD sequences one case use, we focused on the simple and standard XY4 sequence due to its widespread use and tested efficiency in reducing decoherence secondary to low-frequency noise.
It consists of four pulses: $X$, $Y$, $X$, and $Y$, where each pulse represents a $\pi$ rotation around the respective axis. This sequence is applied during idle periods. 

\paragraph{Twirling.}

Twirling is a noise tailoring technique that randomizes the gates of a circuit while preserving the logical unitary operation. This process transforms arbitrary noise channels into stochastic Pauli channels, thereby simplifying the error accumulation in the circuit \cite{Wallman2017Twirling}. For a tutorial on twirling, see Ref.~\cite{Minev2022twirl} In our experiments, we applied Pauli twirling to the entangling layers, improving the overall fidelity of our quantum computations. 

\paragraph{Readout error mitigation.}

Readout error mitigation is needed to correct the imperfect measurements of our quantum processors. When determining the expectation values of observables, readout errors can be mitigated by twirling the readout noise channel, learning it, and then applying its inverse in post processing. This method is known as twirled extinction readout error mitigation (TREX) \cite{Berg2022TREX}. As with any error mitigation, it comes at some sampling cost. 
We employed TREX in our analyses for Figures 1, 2, and 3 of the main text. For applications requiring accurate bitstring distributions, we used the Mthree method, which involves learning and inverting a calibration matrix to correct the readout errors \cite{Nation2021M3}.

\subsection{Error mitigation}

Error mitigation \cite{Temme2016ZNEPEC, Cai2022EMReview} is crucial for extending the capabilities of current quantum hardware, allowing for more accurate and reliable results despite the presence of noise. Among the various techniques available, two state-of-the-art methods stand out: probabilistic error cancellation (PEC) \cite{Temme2016ZNEPEC, Berg2022PEC} and zero-noise extrapolation (ZNE)  \cite{Temme2016ZNEPEC, Li2017ZNE, Kim2021}. These two techniques have also been combined in into a new method known as probabilistic error amplification (PEA) \cite{Kim2023Utility}.

PEC is known for its ability to provide bias-free expectation values of observables assuming a time-stationary noise model for all different entangling layers in the circuit can be learned without error. These noise channels can be inverted and their noise inverse can be implemented using a quasi-probability circuit decomposition. PEA relies on the learning of noise models of each of the unique layers in the circuit as well. While for many unique layers, in principle it is possible to do this efficiently under certain assumptions on the noise,  this is an active frontier in the field of mitigation. Since our target circuits typically contain at least 60 unique layers,  
it becomes a challenge to readily learn noise for each of these layers within current infrastructure and experimental limitations.

We thus turn to ZNE as our mitigation method of choice in this study. ZNE is a well tested and widely used method \cite{Temme2016ZNEPEC, Li2017ZNE, TudorDZNE2020, He2020, Lowe2020, Kim2021, Urbanek2021, Pascuzzi2022, Russo2022, Cai2022EMReview,  Chen2022e, Rivero2022ZNE, Shtanko2023, Liao2023ZNEEM, Yu2023,  Hour2024} that systematically amplifies the noise in the circuits. From measurements at different levels of noise, one can extrapolate back to the zero-noise limit. The noise amplification can be implemented in various ways, each with its assumptions and trade-offs.  One has to thus be careful and to check empirically if their experimental setup and circuits are suitable. For simplicity, we employ the platform-independent gate folding method to amplify noise and examine the type of extrapolation model in the following section.

\section{Experiments: Parameters, data analysis, and confidence intervals}

\subsection{Experimental workflow and parameters}
\label{sec:exp_flow_params}

\paragraph*{Experimental workflow.}
In the preceding section, we set the stage for our experiments.
Let us briefly summarize here our experimental workflow for the deep Fibonacci circuits run on our superconducting qubit processors.

First, the selection of qubits is critical due to the circuit depth, which is approximately 120--150 two-qubit gate layers, depending on the exact layout, optimization, and qubit patch used. To select the qubits, we use a real-time benchmarking protocol to assess qubit properties at the time of execution. With this, we aim to account for fluctuations due to calibration drifts and intrinsic device noise.
We then map and transpile circuits onto a selected subgraph of the device, optimizing them to minimize depth and the number of two-qubit gates. 
When executing, we use the XY-4 dynamical decoupling sequence applied to idle periods to suppress decoherence. We also twirl the gate operations and the readout channels.
For readout error mitigation of expectation values, we used the twirled-extinction readout-error mitigation (TREX) method \cite{Berg2022TREX}. For bitstring distributions, we used the the Mthree method \cite{Nation2021M3}. To correct the decay of the signal, we used in conjunction with all these zero-noise extrapolation (ZNE). In summary, we use a composite error mitigation and suppression strategy tailored to each of the experiments.

\paragraph*{Parameters.}
The exact type of mitigation and parameters used for each of these for each of the experiments in the main text are summarized in the following table:
\begin{table}[h]
    \centering
    \begin{tabular}{ |p{4.5cm}||p{3cm}|p{3cm}|p{3cm}|p{3cm}|  }
        \hline
        \multicolumn{5}{|c|}{Experiment details} \\
        \hline
         & Figure 1 & Figure 2 & Figure 3 & Figure 4\\
        \hline
        2Q Depth (min-max) & 10 & 120 -- 150 & 124 -- 150 & 107 -- 134\\
        Number of shots per circuit & 8,192 & $2\times10^{5}$ & $4\times10^{5}$ & $3\times10^{7}$\\
        Number of twirls & N/A & 100 & 200 & 1,200\\
        Dynamical decoupling (DD) & XY-4 & XY-4 & XY-4 & XY-4\\
        Readout mitigation & TREX & TREX & TREX & M3\\
        Number of ZNE noise factors & N/A & 11 & 20 & N/A\\
        Device & \textit{ibm\_peekskill} & \textit{ibm\_torino} & \textit{ibm\_torino} & \textit{ibm\_torino} \\
        \hline
    \end{tabular}
    \caption{Summary of experimental details for each figure. The table lists the two-qubit depth, number of shots, number of twirls, dynamical decoupling (DD) sequence, readout error mitigation (REM) method, zero-noise extrapolation (ZNE) noise factors, and the device used for the experiments. The number of shots per circuit is for a single ZNE noise factor, divided among the twirls.}
    \label{tab:experiment_details}
\end{table}

\paragraph*{Figure 3 control experiment circuit setup.}
The control data in Fig.~3 of the main text corresponds to a control circuit that is identical to the golden-ratio circuit but with the deliberate addition of two X gates. These simulate errors within the operations. The first error is introduced at the beginning of the circuit, and the second near the end.
Specifically, the first error affects qubit Q2 during the initial step of the protocol within the $F$-move. This occurs after decomposing the $F$-move on qubits Q1, Q2, and Q3. The X gate acts immediately before the target bit of the Toffoli gate. The second error impacts the final $F$-move that is applied before the~$R^*$ gate. Here, a bit-flip is introduced on qubit Q4, the target of the last Toffoli gate within this step, which involves qubits Q1, Q4, and Q13.

\clearpage

\subsection{Data analysis}

\begin{figure}[t]
    \centering 
\includegraphics[
]{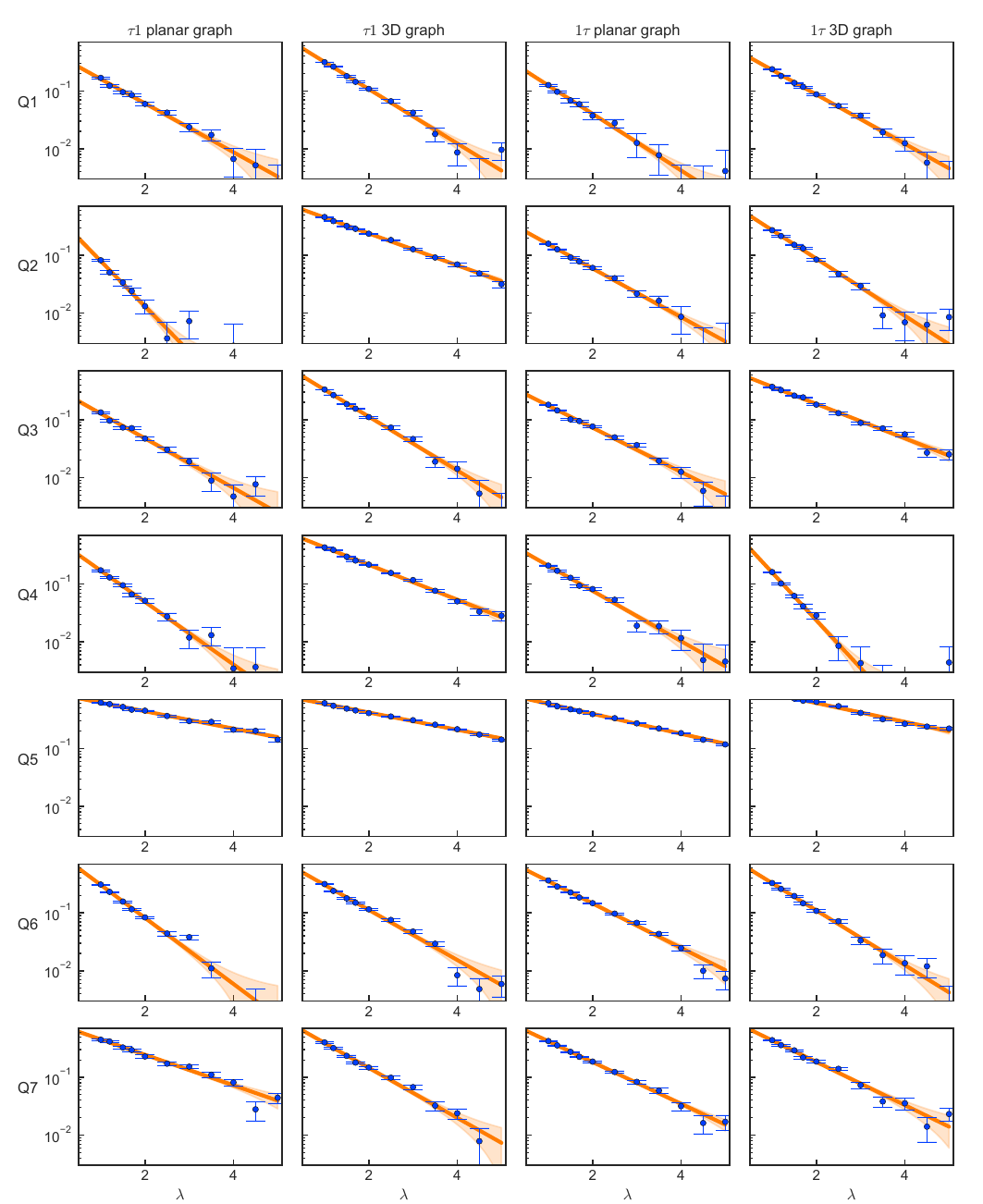}
\caption{
\textbf{Raw data and model fits for the experiments of Fig.~2 of the main text.}
Raw measured $\langle Z \rangle$ values from the four experiments shown in Fig.~2 of the main text (columns, see labels on top). Each row corresponds to one of the measured qubits, Q1 to Q7. 
Each panel illustrates the raw~$\langle Z \rangle$ expectation values obtained from performing zero-noise extrapolation (ZNE), but after implementing error suppression and readout error correction. 
The x-axes denote the ZNE stretch factor~$\lambda$ applied to a circuit.
Blue points represent experimental data. Associated error bars indicate the empirical statistical noise of the measurement. Orange solid lines are exponential fits to the data. To facilitate plotting on a logarithmic scale, all data and fits have the small fit offset subtracted and those curves with $\langle Z \rangle<0$ at $\lambda=1$ are multiplied by $-1$ so they can be plotted. Faint orange shaded regions represent the one-sigma confidence bands for the fit model, calculated from the uncertainties in the fit parameters.
}
\label{fig:si_fig2_zne}
\end{figure}

\begin{figure}[t]
    \centering 
\includegraphics[
]{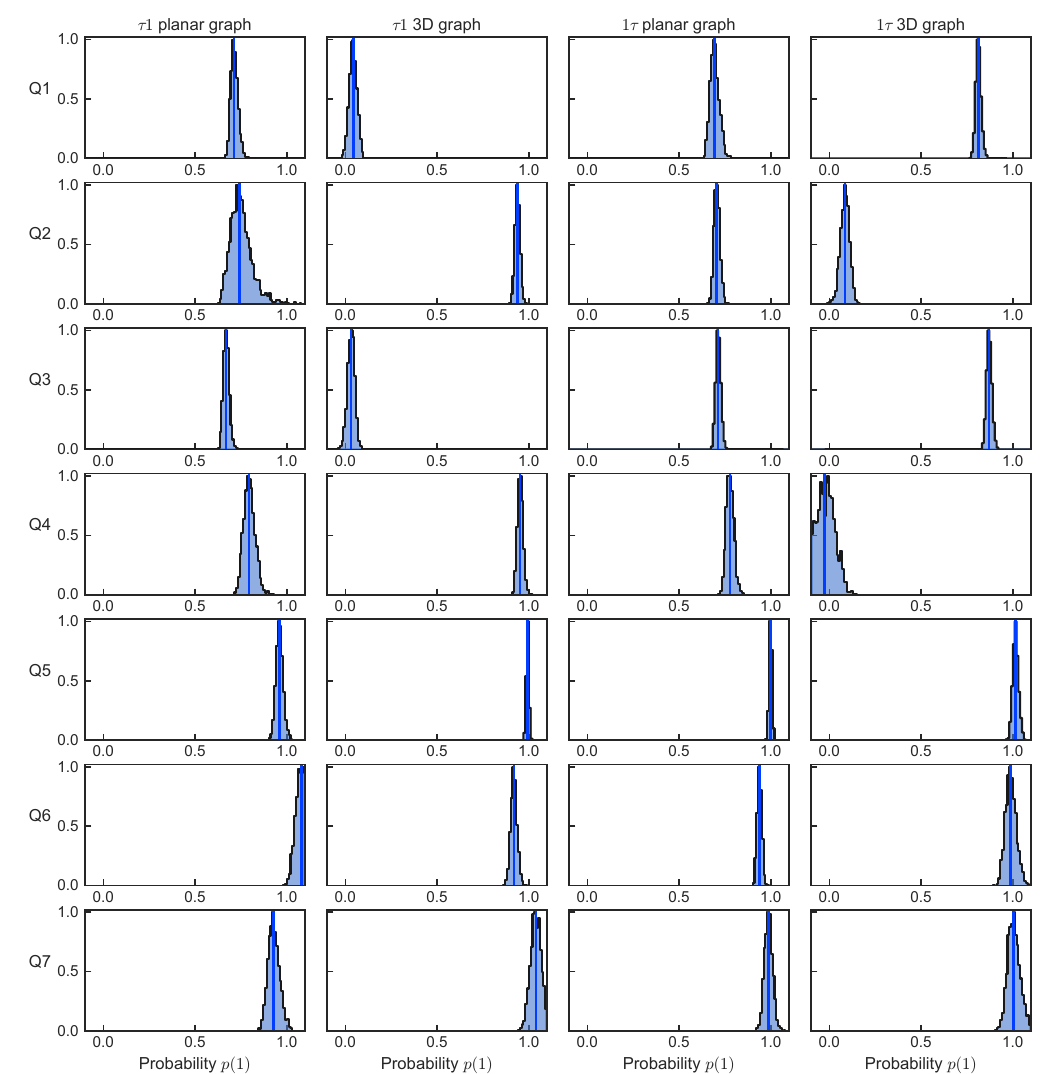}
\caption{
\textbf{Bootstrap resampled distribution of qubit state probability $p(1)$ for experiments in Fig. 2 of the main text.}
Bootstrap resampled distribution of the inferred probability $p(1)$ of measuring a qubit in the one state for the four experiments shown in Figure 2 of the main text (columns, see labels on top). Each row corresponds to one of the seven measured qubits (Q1 to Q7). Bootstrapping is performed on the raw sampled data to model the effect of statistical fluctuations, determined by the empirically measured noise on each data point.
Each panel shows the distribution of zero-noise extrapolation (ZNE) mitigated probabilities~$p(1)$ over different random instances of the resampled raw data. Vertical blue lines represent the extrapolated values of the raw data without any resampling. The standard deviation of these distributions represents a measure of the one-sigma confidence bands extracted from bootstrap resampling.
}
\label{fig:si_fig2_zne_bootstrap}
\end{figure}

To carefully understand the experiment results and the effect of noise and mitigation, we detail in the following several analyses of the raw data and the inferred probabilities from our measurements. 

\paragraph*{Figure 2 of the main text.}
First, we focus on the experiments presented in Fig.~2 of the main text and use them to describe the analyses. 
Figures~\ref{fig:si_fig2_zne} and~\ref{fig:si_fig2_zne_bootstrap} provide a detailed portrait of the experimental data, ZNE model, and the statistical robustness of the inferred results.

\paragraph*{Raw data.}
Figure \ref{fig:si_fig2_zne} shows the raw measured $\langle Z \rangle$ values from the four experimental configurations described in Figure 2 of the main text. Each column represents a distinct experiment involving the $\tau_1$ anyons and $1\tau$ anyons in both the planar and 3D graph configurations. Each row corresponds to one of the measured qubits, Q1 to Q7. Refer to Figure 2 of the main text for the definition of the these qubit labels.  The x-axes represent the stretch factor $\lambda$ used in the zero-noise extrapolation (ZNE). The blue points in each panel denote the raw experimental data, and the associated error bars indicate the empirical statistical noise. The orange solid lines are exponential fits to the data, with faint orange shaded regions representing the one-sigma confidence bands derived from the fit uncertainties. To enable plotting on a logarithmic scale, a small fit offset is subtracted from all data, and any $\langle Z \rangle$ values that are negative at $\lambda=1$ are multiplied by $-1$.  It is evident that some of the decays are fast, others are slower. Overall the signal is damped in a wide range. The largest damping factor (noisy value divided by ideal value) is 0.15. The mean and median damping factors across the entire data set are 0.40 and 0.38, respectively, with a standard deviation of 0.15. Thus on average, our raw data has lost 60\% of the initial signal.

\paragraph*{Error bars on the raw data.}
The mean and median error bar (standard deviation) on the raw $\langle Z \rangle$ data points are $5.6\times10^{-3}$ and $4.7\times10^{-3}$, respectively. We note that when measuring a spin with a probability $p$ of being in the one state, the theoretical standard deviation is given by $\sigma = \sqrt{p(1-p)/N_{\text{shots}}}$. For $N_{\text{shots}} = 10^5$ and $p=0.5$, we expect a standard deviation of approximately $3.1\times10^{-3}$. This theoretical value is slightly smaller than the empirical standard deviation observed in our data. This is expected due to the practical effects, such as the binning of the shots into different twirled circuits and the finite sampling effect of the twirling.

\paragraph*{Decay of the data.} 
Empirically, we observe conspicuous signatures of simple exponential decays in each of the panels in Fig.~\ref{fig:si_fig2_zne}. 
Under typical incoherent noise in the circuits, these decays are in practice expected to approach near-zero values at high stretch factors, nearing a mixed state. However, we do not expect them to strictly converge to zero. For example, measurement errors in our readout are mitigated only up to some finite precision due to finite sampling and time drifts. This leaves behind a very small but non-negligible offset. Empirically, we find that the median absolute value of the fit offset is only~$1.4\times10^{-2}$. The large and deep number of noise factors we use in our experiment allows us to better account for these offsets and to yield tighter model fits.

\paragraph*{Decay model.}  
By fitting the data to an exponential model \( A \exp(-k \lambda) + B \), where \( A \), \( B \), and \( k \) are the model parameters, we quantitatively assess the model's quality. 
The exponential model fit parameter $k$ for the data in Figure 2 has a mean value of $0.936$, a median value of $0.975$, a standard deviation of $0.365$, with values ranging from a minimum of $0.344$ to a maximum of $1.901$.

\paragraph*{Decay model: Coefficient of determination.}  
One standard measure of how well the data are replicated by the model is the coefficient of determination \( R^2 \). It ranges from 0 to 1. A zero value indicates that the model does not explain any of the variance in the data. A value of unity indicates that the model explains all of the variance.
In our data, the mean, median, and standard deviation of the coefficient of determination \( R^2 \) across the panels are \( 0.995 \), \( 0.996 \), and \( 0.003 \), respectively. These high \( R^2 \) values signal that, on average, 99.5\% of the variance in the data is explained by the model. 
While the \( R^2 \) value is a key measure, visual agreement with the data is equally important. The consistent fit across different datasets indicated by a simple visual inspection and the high \( R^2 \) values lends support to quality of the fit model for the experimental data.

\paragraph*{Second error analysis: Bootstrap resampling of the raw data.}
While the fit confidence intervals give us a first error analysis of the fit and extrapolation, we wish to perform a more stringent test.  We use the bootstrapping approach to model the effect of statistical fluctuations inherent in the raw sampled data to find an alternative robust estimate of the variability in our reported measurements. 

Figure \ref{fig:si_fig2_zne_bootstrap} complements this analysis by presenting the bootstrap resampled distributions of the inferred probability for the qubits to be in the one state, $p(1)$. Each panel shows the distribution of ZNE mitigated probabilities $p(1)$ over different random instances of the resampled data, with vertical blue lines representing the extrapolated values of the raw data without any resampling. The standard deviation of these distributions serves as a measure of the one-sigma confidence bands, offering a comprehensive view of the statistical robustness and reliability of our inferred probabilities.

Together, these figures and analyses provide a detailed examination of the experimental data and it's statistical confidence intervals. We emphasize the importance of statistical analyses for mitigated data and inferred results.

\begin{figure}[t]
    \centering 
\includegraphics[
]{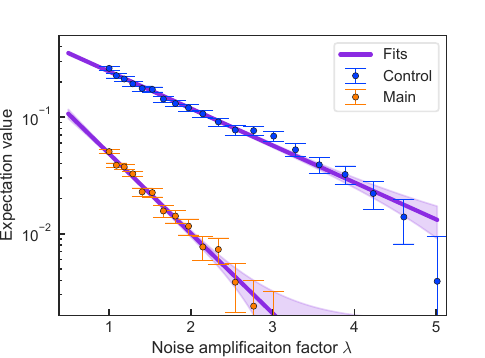}
\caption{
\textbf{Raw data and model fits for the experiments of Fig.~3 of the main text.}
Raw measured $\langle Z \rangle$ values from the two experiments shown in Fig.~3 of the main text. Each curve represents one of the experimental setups: control (blue) and the main braiding experiment (orange). The x-axis denotes the noise amplification factor~$\lambda$ applied to the circuit.
The blue and orange points represent the experimental data, shown with cap-style empirical error bars. The purple solid lines are exponential fits to the data. To facilitate plotting on a logarithmic scale, all data and fits have the small fit offset subtracted. Additionally, since the $\langle Z \rangle$ values fir the main experiment are all negative, they have been multiplied by $-1$ for log-scale plotting. The shaded regions around the fit lines represent the one-sigma confidence bands, calculated from the uncertainties in the fit parameters.
}
\label{fig:si_fig3_zne}
\end{figure}

\paragraph*{Figure 3 data analysis.}

Figure \ref{fig:si_fig3_zne} presents the raw $\langle Z \rangle$ data and corresponding exponential fits for the two experimental setups detailed in Fig.~3 of the main text: the control experiment (blue) and the main braiding experiment (orange). The x-axis indicates the noise amplification factor $\lambda$ applied to the circuit. Here, due to the even more challenging circuits, we use 20 noise factors, as discussed below, in order to improve the quality of the mitigated results.
The experimental data points, depicted by blue and orange markers, include empirical error bars shown as cap markers to reflect statistical noise. The purple solid lines represent exponential fits to the data, with shaded regions indicating the one-sigma confidence bands derived from the fit uncertainties.

The main braiding experiment presented in Fig.~3 of the main text faces significant challenges due to its noise-free expectation $\langle Z \rangle$ value being small and its noisy version very small. After noise amplification, the measured $\langle Z \rangle$ values of the circuits with amplified noise are further reduced to become extremely low, approaching the noise floor. We reduced the floor and increased the signal-to-noise ratio by increasing the number of shots, twirls, and averaging relative to the experiments so far presented. Data collection for Fig.~3 required approximately 7.27 times longer in quantum processor runtime per raw circuit compared to that of Fig.~2 of the main text. This extended duration increases susceptibility to noise fluctuations during data acquisition.

A comparison of the data in Fig.~\ref{fig:si_fig2_zne} and Fig.~\ref{fig:si_fig3_zne} supports this expectation. The Fig.~3 data for the control experiment shows a weak low-frequency fluctuation in the model residual, observable upon close examination of the blue points relative to the purple fit line. Such a fluctuation is not evident in the Fig.~2 data. We attribute this to noise and calibration drifts associated with the longer timescale of this particular experiment.
For the control experiment  (blue dots), the signal is relatively large and the suspected drift deviations are small relative to the signal, they appear to be within the measurement error. Their effect is largely accounted for in the fit error and the bootstrapped resampling of the data. 

In the main experiment (orange dots), the signal is much smaller, and the last few data points of our 20 noise factors approach the empirical noise floor. To provide our best estimate of the extrapolated value, we want to minimize the effect of time drift and we want to maximize the signal to noise ratio in the data. To this end, we exclude the last six stretch factors from the fit, which fall below the $2\times10^{-3}$ level in Fig.~\ref{fig:si_fig3_zne}, our empirical noise floor. What is the impact of including or excluding these points? We find their effect is small either way. The changes in the extrapolated value fall within within the reported uncertainty of the value. To minimize the effect of time drifts, we choose to report the shorter-time data sequence.

The bootstrap resampling distributions for both experiments is shown in the main text. It is based on resampling the raw data points here subject to the empirical error bars they have in order to account for the statistical fluctuations. For each bootstrap resample instance of the data, we perform a new model fit and extrapolation, calculate the ratio of probabilities for the qubit to be in the one versus the zero state $p(1)/p(0)$,  and report the distribution in the main text.  

Because of the smaller data values for the main experiment, we find a larger fluctuation in the extrapolation and data compared to the control experiment, which starts at a larger initial value and is hence less susceptible to the noise floor.  Moreover, as shown in the following paragraph, the error-propagation from the expectation value to the ratio is non-linear and leads to an positive skew in the bootstrap resampled data. 

\paragraph*{Figure 3: Further error propagation analysis for the probability ratio.}

In addition to the bootstrap error analysis, we can also write down a simple analytic expression to understand the propagation of uncertainties from the measured $\langle Z \rangle$ values to the derived ratios. This will use the standard error analysis based on standard error propagation. The error propagation for a function $f(x)$ given an uncertainty $\delta x$ in the variable $x$ is described by:
\[
\delta f = \left| \frac{df}{dx} \right| \delta x\;.
\]

In our specific case, the function $r$ represents the ratio, which is the probability of being in state 1 over the probability of being in state 0. It is calculated from the measured mitigated expectation value $z_{\text{exp}}$. The function $r(z_{\text{exp}})$ is hence
\[
r(z_{\text{exp}}) = \frac{1 - z_{\text{exp}}}{1 + z_{\text{exp}}}\;.
\]
Therefore, the error propagation formula becomes $\delta r = \left| \frac{dr}{dz_{\text{exp}}} \right| \delta z_{\text{exp}}$. 
To find the propagated error $\delta r$, we need to compute the derivative of $r$ with respect to $z_{\text{exp}}$, $
\frac{dr}{dz_{\text{exp}}} = \frac{-2}{(1 + z_{\text{exp}})^2}
$. Thus,
\[
\delta r = \left| \frac{dr}{dz_{\text{exp}}} \right| \delta z_{\text{exp}} = \left| \frac{-2}{(1 + z_{\text{exp}})^2} \right| \delta z_{\text{exp}}\;.
\]

This relationship shows how the uncertainty in the measured $z_{\text{exp}}$ values ($\delta z_{\text{exp}}$) propagates to the uncertainty in the derived function $r(z_{\text{exp}})$. The negative sign in the derivative indicates the direction of change, but for the magnitude of uncertainty, we consider the absolute value. 

It is evident from this that $z_{\text{exp}}$ values closer to +1 will result in lower error propagation, while values such as the main experiment $z_{\text{exp}}\approx-0.24$ will propagate with more uncertainty. Moreover, we note that the error propagation is non-linear in $z_{\text{exp}}$, since $r(z_{\text{exp}}) $ is itself non-linear and blows up to infinity at $z_{\text{exp}} = -1$.

\paragraph*{Figure 4 data analysis.}

Here, we describe the data taking, processing, and analysis of the experimental data reported in Fig.~4 of the main text. 
For the requisite background theory, see Sec.~\ref{subsec:string-net-sampling}.

Although the main circuit requires only 9 qubits for the $2 \times 2$ plaquettes depicted in Fig.~4a of the main text, we employed an equivalent form of the 5-qubit $F$-move using relative phase Toffoli gates to reduce the depth required for implementing the multiple control Toffoli gate, as discussed in Sec.~\ref{Circuits_Fmoves}. 
This method, by adding one ancilla qubit for each 5-qubit $F$-move, decreases the depth from 63 to 18 for the Toffoli gate, as reported in Ref.~\cite{Maslov_Fmov_2016}; see Table~1 within the reference. Thus, in our case, we added two ancilla qubits in the experiment for the 5-qubit $F$-moves depicted in Figs.~4b--c of the main text and implemented on (Q2, Q4, Q6, Q7, Q8) and (Q0, Q1, Q2, Q6, Q8). The significant reduction in depth led to an improvement in the experimental data.

In Fig.~4h of the main text, we report the raw experimental data obtained from a total of $8.8 \times 10^6$ experimental realizations gathered across 1,200 quantum circuit randomizations of the 11-qubit circuit for preparing and sampling the Fib-SNC ground state. 
In the main large panel, we report the measured probability distribution over all $2^{11} =2,048$ possible computational bitstring configurations. 
Experimental data is shown in red and theoretical prediction in blue. 
The theoretically valid bitstrings, those that satisfy the fusion (branching) rules, are grouped on the left. 
Illegal bitstrings, those that violate the trivalent fusion (branching) rules, are grouped on the right.  
Theoretically, all illegal bitstrings have zero probability. 
Experimentally,  a tail of illegal bitstrings is evident, though their individual amplitudes are generally much smaller than those of the legal bitstrings, showing a clear separation. 
To quantify this better, we compare the median probability of the smallest 15\% of the valid bitstrings (which is $0.015$) to the maximum value of the invalid bitstrings observed in the experiment (which is $0.006$). The two values indicate a good separation in the measured amplitudes.

In the inset of Fig.~4h, we focus only on the valid bitstrings. We note that it is easy to check if a bitstring is allowed or now. We truncate all bitstrings not allowed by the fusion (branching) rule and uniformly renormalize their probability into valid bitstrings. Thus, the sum of the probabilities over the valid bitstrings is now unity.
To compare the experiment against the theory, we first plot the experimental data on top of the theoretical prediction. The theory prediction here is possible due to the proof-of-principle small system size, but would be hard for a larger number of qubits. Valid bitstrings only require measurement outcomes over the subspace of the 9-qubits; hence, we trace out here the two ancilla qubits.
We observe reasonable agreement. 
To provide a clearer guideline, we also show the average value (thick lines) over each topologically equivalent class of graph configurations, denoted as $[G_0, G_1, G_{2A}, G_{2B}, G_{3A}, G_{3B}, G_4]$ from left to right, indicating that the average values manifest behavior similar to those predicted theoretically.

Finally, in Fig.~4i, we report the chromatic polynomials calculated from this distribution. For a class of isomorphic subgraphs, defined in the preceding parts of Fig.~4 of the main text, we report the mean value~$ \bar{\chi}$ over the equivalent subgraphs,
\begin{equation}\label{eq:meansample}
 \bar{\chi}= \frac{1}{m}\sum_{i=1}^{m} \chi_{i}\;, 
 \end{equation}
where multiplicity~$m$ of each graph class is reported in Fig.~4j of the main text and $\chi_{i}$ is the chromatic polynomial value computed over the $i$-th bitstring in the class.
In Fig.~4i, the error bars indicate the standard error of the sample means, defined in the usual way, as the sample standard deviation, accounting for Bessel's correction,  
 \begin{equation}\label{eq:stdsample}
\sigma_\chi=\sqrt{\frac{1}{m-1}\sum_{i}(\chi_i-\bar{\chi})^2}\;,
 \end{equation}
divided by the square root of the number of samples $m$; that is ${\sigma_\chi}/{\sqrt{m}}$.

In the case of two of the graphs, \([G_{\mathrm{2B}}]\) and \([G_{4}]\), the sample multiplicity \(m\) is unity. Thus, we cannot directly use the concept of repeated measurements, which should all yield the same outcome nominally, to estimate the experimental uncertainty. Therefore, we have to adopt a different approach to estimate the experimental uncertainty of the reported values.
First, we observe that the \(\chi\) and probability outcomes for the valid bitstrings are similar in broad magnitude across the six graph classes. Second, all the data were collected simultaneously using the same experimental setup and measurement. Assuming \([G_{\mathrm{2B}}]\) and \([G_{4}]\) are not fundamentally different regarding their experimental uncertainty behavior from that of graph classes with multiplicity that is near to but not unity, we can crudely infer an estimate of their uncertainty by proxy. 
To this end, we use the average measurement uncertainty of the graph classes that have a small \(m\) as a proxy for those with $m=1$. Specifically, we use the average uncertainty of the \([G_{\mathrm{3A}}]\) and \([G_{\mathrm{3B}}]\) graphs to provide an estimate of the uncertainty for \([G_{\mathrm{2B}}]\) and \([G_{4}]\). 
For related theory, see Sec.~\ref{subsec:string-net-sampling}.

\clearpage

\end{appendix}

\bibliography{Fib-simulation.bib, fibonacci.bib, Fibonacci2.bib}

\end{document}

%% file: zkm_math_common.tex
\DeclareOldFontCommand{\bf}{\normalfont\bfseries}{\mathbf}

\usepackage{nicematrix} 
\NiceMatrixOptions{renew-dots,renew-matrix}


%% file: util-macros.tex
\global\long\def\ket#1{\left|#1\right\rangle }%

\global\long\def\bra#1{\left\langle #1\right|}%

\global\long\def\kb#1#2{\left|#1\vphantom{#2}\right\rangle \left\langle \vphantom{#1}#2\right|}%